\documentclass[12pt]{article}
\usepackage{authblk}
\usepackage[bottom]{footmisc}

\usepackage[utf8]{inputenc}
\usepackage{epsfig}
\usepackage{amssymb}
\usepackage{amsfonts}
\usepackage{amsbsy}
\usepackage[all]{xy}
\usepackage{amsmath}
\usepackage{enumerate}
\usepackage{float}

 \usepackage{enumerate}
 \usepackage{hyperref}

\usepackage{amssymb,amscd}
\usepackage{mathrsfs}
\usepackage{amsmath,amsthm}
\usepackage{xspace}
\usepackage{url}
\newcommand{\del}{\partial}
\newcommand{\ov}{\overline}

\usepackage{epsfig}
\usepackage{amssymb}
\usepackage{amsfonts}
\usepackage{amsbsy}
\usepackage[all]{xy}
\usepackage{amsmath}
\usepackage{enumerate}

\usepackage{amssymb,amscd}
\usepackage{mathrsfs}
\usepackage{amsmath,amsthm}
\usepackage{xspace}
\usepackage{url}

\newcommand{\bea}{\begin{eqnarray}\displaystyle}
\newcommand{\eea}{\end{eqnarray}}

\newcommand{\mZ}{\mathbb{Z}}

\newcommand{\hp}{\text{Hop}}
\newcommand{\mft}{\mathfrak{T}}

\newcommand{\wR}{\widehat{R}}
\newcommand{\bphi}{\boldsymbol \phi}
\newcommand{\cwarrow}{\text{\Large$\curvearrowright$}}

\title{Categorical Wall-Crossing in Landau-Ginzburg Models} 
\author{Ahsan Z. Khan\footnote{ {\tt ahsan.z.khan@rutgers.edu	} } } 
\author{Gregory W. Moore\footnote{ {\tt gmoore@physics.rutgers.edu} }}
\affil{New High Energy Theory Center and Department of Physics, Rutgers University, Piscataway, NJ 08854}
\date{\today}
\begin{document}

\maketitle  

\begin{abstract}
We describe how categorical BPS data including chain complexes of solitons, CPT pairings, and interior amplitudes jump across a wall of marginal stability in two-dimensional $\mathcal{N}=(2,2)$ models. We show that our jump formulas hold if and only if the $A_{\infty}$-categories of $\frac{1}{2}$-BPS branes constructed on either side of the wall are homotopy equivalent. These results can be viewed as categorical enhancements of the Cecotti-Vafa wall-crossing formula.
\end{abstract}

\pagebreak
\tableofcontents

\section{Introduction and Outline}

\paragraph{} BPS states have played an important role in many aspects of physical mathematics. As is very well-known, the spaces of BPS states can jump discontinuously as physical parameters are varied, a phenomenon known as wall-crossing. Investigations of BPS wall-crossing have led to a wide variety of very interesting developments. For some reviews of BPS wall-crossing see \cite{CecottiLectures, Kontsevich:2009xt, KontsevichReview, MooreLectures, NeitzkeLectures, Pioline:2011gf}.

\paragraph{} BPS wall-crossing appears in two-dimensional quantum field theories with $\mathcal{N}=(2,2)$ supersymmetry, where it was first discovered \cite{Cecotti:1992qh, Cecotti:1992rm} as well as in four-dimensional supergravity and field theory with $\mathcal{N}=2$ supersymmetry \cite{Denef:2007vg, Dorey:1998yh, Gaiotto:2008cd, Lee:1998nv, Seiberg:1994rs}. It also appears in a more elaborate form in coupled 2d-4d systems \cite{Gaiotto:2011tf}.

\paragraph{}Indeed, there are quantitative formulae expressing how BPS \underline{indices} change across walls of marginal stability. It is natural to ask if one can obtain more refined information about the spaces of BPS states. For example, if BPS states are identified with the cohomology of some chain complexes one would like to know how the chain complexes themselves jump across walls of marginal stability. One cannot expect an answer at the level of chain complexes \textit{per se}, since homotopy equivalent chain complexes are also physically equivalent, but it is meaningful to ask how the equivalence class of the chain complexes (up to homotopy) changes\footnote{Note that the homotopy class of a chain complex contains more information than the index.  As a simple example, consider \bea C = (\mathbb{Z} \oplus \mathbb{Z}[1], d=0) \eea and \bea C' =(\mathbb{Z} \oplus \mathbb{Z}[1],d') \eea where $d'$ maps a generator of $\mathbb{Z}$ to a generator of $\mathbb{Z}[1]$.  Both have vanishing Euler characteristics \bea \chi(C) = \chi(C') =0,\eea but their cohomology is different so they are not homotopy equivalent.}. In particular, relating the homotopy equivalence class of chain complexes across a marginal stability wall allows us, by taking cohomology, to answer \textit{How do the BPS Hilbert spaces jump across a wall of marginal stability?}  This is the question a categorified wall-crossing formula is meant to answer. 

\paragraph{} The present paper addresses the categorification of the renowned Cecotti-Vafa wall-crossing formula for BPS indices in two-dimensional $\mathcal{N}=(2,2)$ quantum field theory. We have made use of a formalism developed in \cite{Gaiotto:2015aoa, Gaiotto:2015zna}, specifically for the purpose of carrying out the program of categorification of wall-crossing formulae. Indeed, in \cite{Gaiotto:2015aoa, Gaiotto:2015zna} it was explained how to categorify the so-called ``framed wall-crossing'' or ``S-wall-crossing'' formulae in the two-dimensional models. The present paper adds to the story with an improved understanding of how to phrase the categorification of the Cecotti-Vafa wall-crossing formula. 

\paragraph{} Much remains to be done in the program of the categorification of wall-crossing formulae. In particular, the categorification of the four-dimensional wall-crossing formula of Kontsevich-Soibelman is not known.\footnote{The \underline{change} in the 4d BPS state spaces is nicely understood using the halo formalism of \cite{Andriyash:2010qv, Denef:2007vg, Gaiotto:2010be}. In some sense, this answers the question of the categorification of wall-crossing formulae, but the categorification program is more ambitious, and seeks to describe the full set of BPS states on either side of the wall in homotopical algebra terms.} We believe an important step forward is to include twisted masses in two-dimensional Landau-Ginzburg models. This is work in progress and we hope to post a paper on the subject in the near future.

\paragraph{} In the remainder of this introduction we outline in more detail the difficulties which must be overcome to categorify the Cecotti-Vafa wall-crossing formula, and how we will achieve this. 

\subsection{A Failure of Naive Categorification}

\paragraph{} Supposing that $i,j,k$ denote distinct massive vacua of a two-dimensional $\mathcal{N}=(2,2)$ theory, recall that the Cecotti-Vafa wall-crossing formula states that across a wall of marginal stability of type $ijk$, the BPS indices $\mu$ and $\mu'$ on either side of the wall are related by \bea \mu'_{ij} &=& \mu_{ij}, \\ \mu'_{jk} &=& \mu_{jk}, \\ \label{cvwcf} \mu'_{ik} &=& \mu_{ik} \pm \mu_{ij} \mu_{jk},\eea the sign accounting for which way the wall-crossing occurred. As a first step in categorification, it's indeed encouraging, as we recall in section \ref{chain}, that for Landau-Ginzburg models one can formulate finite-dimensional chain complexes $(R_{ij}, d_{ij})$ such that the BPS index $\mu_{ij}$ is given by a graded trace \bea \mu_{ij} = \text{Tr}_{R_{ij}}(-1)^F .\eea The BPS Hilbert space \footnote{Throughout this paper, we have factored out the (super)translational mode of the soliton. With it included the chain complex will be \bea \widetilde{R}_{ij} = R_{ij}\otimes(\mathbb{Z}[-1] \oplus \mathbb{Z}),\eea and the BPS index would be the ``new index" $\text{Tr}_{\widetilde{R}_{ij}} \big( F(-1)^F \big)$ of \cite{Cecotti:1992qh}. The spectrum of $F$ on $R_{ij}$ lies in a $\mathbb{Z}$-torsor, so after a suitable phase redefinition, the $\mu_{ij}$ will be integers.} of type $ij$ is isomorphic to the $d_{ij}$-cohomology, \bea \mathcal{H}^{\text{BPS}}_{ij} = H^{\bullet}(R_{ij}, d_{ij}).\eea  A categorified wall-crossing formula should then relate the BPS chain complexes $(R'_{ij}, d'_{ij})$ upon crossing a wall of marginal stability to the original chain complexes $(R_{ij}, d_{ij})$. The simplest guess consistent with \eqref{cvwcf} is to say that the underlying vector spaces of the chain complexes are related by \bea   R'_{ij} &=& R_{ij} ,\\ R'_{jk} &=& R_{jk}, \\ \label{wcfcomplex} R'_{ik} &=& R_{ik} \oplus (R_{ij} \otimes R_{jk}),  \eea accompanied possibly with a degree shift on the $(R_{ij} \otimes R_{jk})$ summand to account for which way the wall-crossing occurred. The simplest differentials that one can guess on the primed spaces are \bea d'_{ij} &=& d_{ij}, \\ d'_{jk} &=& d_{jk} ,\\ \label{wcfdiff} d'_{ik} &=& d_{ik} \oplus( d_{ij} \otimes 1 + 1 \otimes d_{jk}). \eea Indeed, the Cecotti-Vafa statement \eqref{cvwcf} would follow as a corollary from this guess, simply by taking graded traces. Under this formula for the differentials, the primed BPS Hilbert spaces are simply \bea  (\mathcal{H}^{\text{BPS}}_{ij})' &\cong& \mathcal{H}^{\text{BPS}}_{ij} ,\\ (\mathcal{H}^{\text{BPS}}_{jk})' &\cong& \mathcal{H}^{\text{BPS}}_{jk}, \\ (\mathcal{H}^{\text{BPS}}_{ik})' &\cong& \mathcal{H}^{\text{BPS}}_{ik} \oplus \big( \mathcal{H}^{\text{BPS}}_{ij} \otimes \mathcal{H}^{\text{BPS}}_{jk} \big).  \eea Things are not so simple: it is very easy to construct counter-examples to this naive prediction of how BPS Hilbert spaces jump across a wall of marginal stability. Here is a simple one.

\paragraph{} Consider the quartic Landau-Ginzburg model, namely the theory of a chiral superfield $\Phi$ with superpotential \bea W = \frac{1}{4}\Phi^4 - \Phi. \eea Denote the three vacua $\Phi_1 = e^{-2\pi i /3 }$, $\Phi_2 = 1$, $\Phi_3 = e^{2\pi i /3}$ with corresponding critical values $W_1, W_2, W_3$. One can show that the \text{absolute} number of solitons is $1$ between each pair of distinct vacua. By taking into account the fermion degree we have that \bea R_{12} &=& \mathbb{Z} , \\ R_{23} &=& \mathbb{Z}, \\ R_{13} &=& \mathbb{Z},\eea with all differentials identically zero. We can vary the lower order terms of the superpotential (for instance we can turn on a quadratic term) so that $W_2$ passes through the line connecting $W_1$ and $W_3$. The naive guess implies that upon this wall-crossing the chain complex $R'_{13}$ is \bea R'_{13} &=& R_{13}\oplus (R_{12} \otimes R_{23}) [1], \\ &=& \mZ \oplus \mathbb{Z}[1]. \eea Because every differential in sight acts trivially, we conclude that $(\mathcal{H}^{\text{BPS}}_{13})'$ is two-dimensional. On the other hand, every Landau-Ginzburg model with target space $\mathbb{C}$ and a polynomial superpotential has an absolute number of solitons between each pair of critical points given by either $0$ or $1$ \footnote{For a proof see Appendix \ref{polynomialsolitons}.}. Thus the cohomology in such a model is either trivial or one-dimensional and we have found a contradiction. Our naive attempt at categorification has failed.

\subsection{Missing Instantons}

\begin{figure}%[here!]
\centering
\includegraphics[width=.6\textwidth]{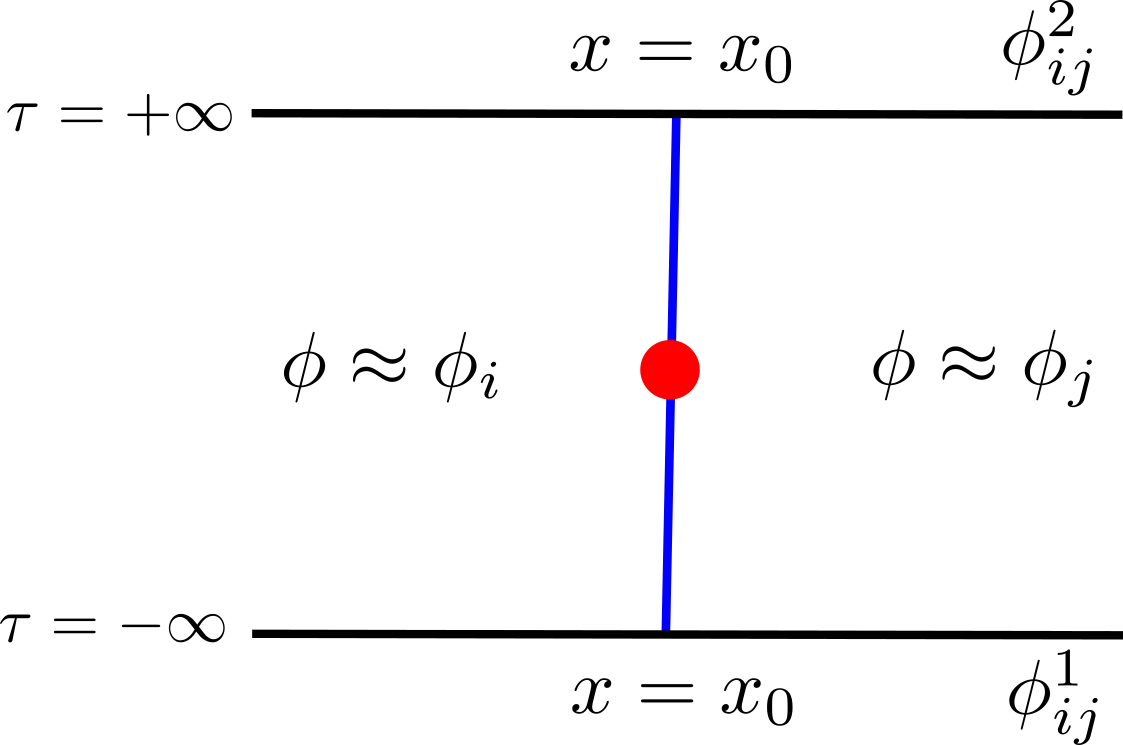}
\caption{An instanton interpolating between two-different $ij$-solitons}
\label{inst1}
\end{figure}

\paragraph{} The reason for the failure of the differential $d'_{ik}$ \eqref{wcfdiff} is simple, but also interesting: We have missed instantons.

\paragraph{} The spaces $R_{ij}$ are made of perturbative BPS states $|\phi_{ij} \rangle$ coming from quantizing around a classical soliton $\phi_{ij}$. The differentials $d_{ij}$ on $R_{ij}$ are meant to encode matrix elements \bea \langle \phi^b_{ij} | \mathcal{Q}_{ij} | \phi^a_{ij} \rangle,\eea where the superscripts $a,b$ label different classical solitons of type $ij$. When these are non-zero there is a difference between the exact ground states and the perturbative ones. We know from the relation between Morse theory and supersymmetry \cite{Witten:1982im}, that the former are computed by considering suitable instantons between these perturbative ground states. Now within a fixed sector, say the $ij$-sector, solutions of such an instanton on the plane look as in Figure \ref{inst1}: The soliton $\phi_{ij}^a$ is stationary, sitting at a fixed point $x_0$, whereas at an instant $\tau_{0}$, we transition from the $\phi_{ij}^a$ to $\phi_{ij}^b$. Such a process will contribute to the matrix element if the fermion numbers of $\phi_{ij}^a$ and $\phi_{ij}^b$ differ by $1$. 

%which probe non-perturbative corrections to the $\mathcal{Q}_{ij}$-closure of these perturbative states.
\begin{figure}%[here!]
\centering
\includegraphics[width=.6\textwidth]{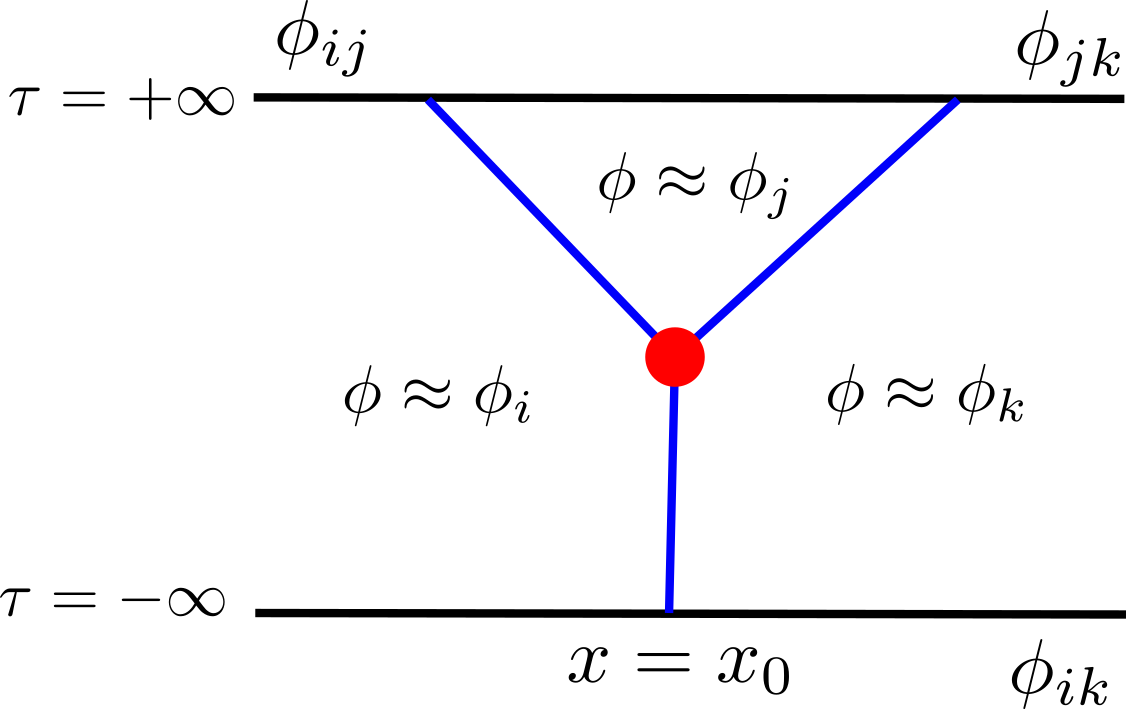}
\caption{An instanton which contributes to an off-diagonal element of $d'_{ik}$.}
\label{inst2}
\end{figure}
\paragraph{} Close to a wall of marginal stability, it is reasonable to postulate that bound states of $ij$ and $jk$-solitons give rise to an approximate $ik$-soliton, post wall-crossing, thus giving our guess \eqref{wcfcomplex}. Instantons of the sort depicted in Figure \ref{inst1}, contribute to matrix elements of the type \bea \langle \phi^b_{ik} | \mathcal{Q}_{ik} | \phi^a_{ik} \rangle \eea and \bea \langle \phi^a_{ij}, \phi^b_{jk} | \mathcal{Q}_{ik} | \phi^{a'}_{ij}, \phi^{b'}_{jk} \rangle .\eea Such contributions are indeed reflected in our guess for the differential \eqref{wcfdiff}. Our formula for $d'_{ik}$ has made an implicit assumption that the off-diagonal matrix element \bea \langle \phi_{ij}^a, \phi_{jk}^b| \mathcal{Q}_{ik} | \phi_{ik}^c \rangle \eea vanishes. However, it turns out, as we will explain in section \ref{intamp} that in addition to the familiar instanton of Figure \ref{inst1}, there can be a more interesting object, where a stationary $ik$-soliton can split into $ij$ and $jk$ solitons traveling at just the correct angles to preserve $\mathcal{Q}_{ik}$-supersymmetry. Such an instanton is depicted in Figure \ref{inst2}. Counting instantons of this type allows one to write down a corrected differential on $R'_{ik}$. This is the main new ingredient that enters the categorified wall-crossing formula.

\subsection{Wall-Crossing Invariants}

\begin{figure}%[here!]
\centering
\includegraphics[width=.55\textwidth]{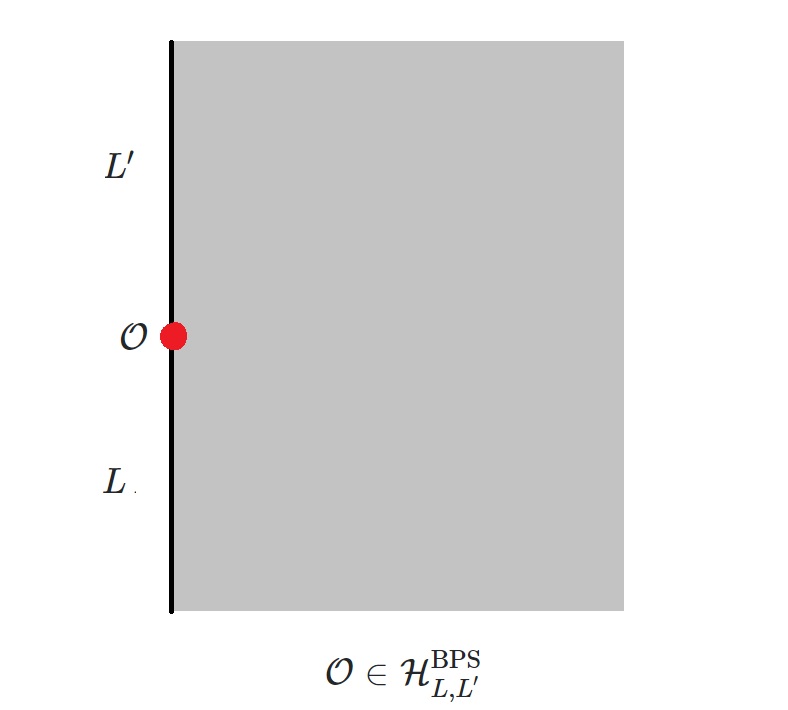}
\caption{A boundary local operator $\mathcal{O}$ between two branes $L$ and $L'$}
\label{bdry}
\end{figure}

\paragraph{} In order to derive wall-crossing formulas such as \eqref{cvwcf} it is extremely useful to introduce certain \textit{wall-crossing invariants}. For Cecotti-Vafa wall-crossing an example of such a wall-crossing invariant is the spectrum generator \footnote{Notation: $\mathbb{V}$ is the vacuum set, assumed to be finite in this paper. $\mathbb{H}$ is the upper-half plane, $Z_{ij}$ are central charges and $e_{ij}$ is the $ij$ elementary matrix. $\cwarrow$ is meant to indicate a clockwise ordered product with respect to the central charges. Implicit in the notation is that an ordering on $\mathbb{V}$ has been chosen. } \bea \label{spec} S = \prod^\cwarrow_{Z_{ij} \in \mathbb{H}} (\mathbf{1} +  \mu_{ij} e_{ij}) \in SL(|\mathbb{V}|, \mathbb{Z})\eea which must be invariant under crossing marginal stability walls \cite{Kontsevich:2008fj}, so long as no BPS rays enter of exit the half-plane $\mathbb{H}$. The wall-crossing invariant $S$ has a simple conceptual meaning. One can show that $S_{ij}$ is the Witten index of the space of boundary local operators at a junction of thimbles of type $i$ and $j$  \cite{Gaiotto:2015aoa} (a related interpretation appeared in \cite{Hori:2000ck}), see Figure \ref{bdry}. Such a space is insensitive to marginal stability walls. Nonetheless the BPS indices $S$ at a given point in parameter space allow the computation of the boundary Witten indices $S$. Comparing $S$ on different sides of the wall of marginal stability leads to \eqref{cvwcf}.  

\paragraph{} It is natural then to expect that a categorical wall-crossing invariant can also be constructed. The invariance of $S$ is categorically enhanced as follows.  The BPS chain complexes $(R_{ij}, d_{ij})$, along with counts of $\zeta$-instantons of the type depicted in Figure \ref{inst2}, allow for the construction of an $A_{\infty}$-category $\widehat{R}[X,W]$ whose objects can be thought of thimble branes \footnote{Note that considering a category with only thimble objects is not restrictive. $\widehat{R}[X,W]$ can be enlarged to a triangulated $A_{\infty}$ category for which the thimble objects provide a semi-orthogonal decomposition.} and morphisms are vector spaces of boundary local operators at brane junctions \cite{Gaiotto:2015aoa} \footnote{ The $A_{\infty}$ category of \cite{Gaiotto:2015aoa} can be viewed as an infrared construction of the category of A-branes in a Landau-Ginzburg model, which to mathematicians is known as the Fukaya-Seidel category \cite{Seidel} of $(X,W)$, and is denoted by $FS[X,W]$. It is expected that $FS[X,W]$ and $\widehat{R}[X,W]$ are quasi-isomorphic as $A_{\infty}$-categories. An outline of a proof of this expectation was given in \cite{Gaiotto:2015aoa}.}. The categorical wall-crossing constraint is then formulated as follows.

\paragraph{} \textit{The homotopy class of $\widehat{R}[X,W]$ is a wall-crossing invariant.}

\paragraph{}In the above statement \textit{homotopy class} refers to the homotopy equivalence of $A_{\infty}$-categories which is defined in Appendix \ref{app}. We show how our categorical wall-crossing formula can be derived from this wall-crossing constraint in section \ref{homotopy}. 

\paragraph{Remark} Note that instead of $\widehat{R}[X,W]$, there are other wall-crossing invariants one could have used as a starting point. For instance instead of imposing $A_{\infty}$-equivalence of the ``open string algebra" $\widehat{R}[X, W]$ across a marginal stability wall like we do in this paper, one could have imposed $L_{\infty}$-equivalence of the closed string algebra $R_c$, defined in \cite{Gaiotto:2015aoa}. Another way of describing the categorical wall-crossing formula makes
use of half-BPS interfaces. These can be used to construct a categorical notion of a flat parallel transport on a bundle of categories of boundary conditions over the space of Morse superpotentials \cite{Gaiotto:2015aoa}. The absence of monodromy around contractible cycles that intersect walls of marginal stability implies a categorified version of the invariance of $S$ defined in equation \eqref{spec}. This categorical equation can  in  turn can be reduced to categorified braid relations. For details see \cite{Gaiotto:2015aoa, GMSlides}. These superficially distinct starting points are all expected to lead to the same eventual result. 

%\subsection{Towards Four Dimensions via Twisted Masses} It is natural to ask the extent to which one can apply the lessons learned from the two-dimensional wall-crossing case to the case of four-dimensional wall-crossing. Interestingly, there is an intermediate case which lives in the two-dimensional world but has some properties in common with four dimensions. This is when so called ``twisted mass deformations" are turned on in the $\mathcal{N} = (2,2)$ theory. The main point is that the twisted mass deformation changes the formula for the central charge \bea Z_{ij} = W_i - W_j\eea by giving rise a natural charge lattice $\Gamma$ such that \bea Z_{ij}(\gamma) = W_i - W_j + M \cdot \gamma .\eea The modified formula now allows the existence of massive BPS states supported in a fixed vacuum $i$. Such a ``periodic soliton" obeys conceptually similar properties to BPS particles in four-dimensions. 

%\paragraph{} We discuss the wall-crossing formula and BPS branes for theories with twisted masses in section []. We then proceed to apply the categorical formalism to the case of the $\mathbb{CP}^1$-model, and construct brane categories on both sides of the wall of marginal stability. We construct differentials by solving the Maurer-Cartan equation and show that the deformed brane categories on both sides of the wall are quasi-isomorphic. We then make some remarks about the general deformation theory when twisted masses are included.

\subsection{Outline of the Paper} The outline of this paper is as follows. In section \ref{review}, we recall the standard discussion of wall-crossing at the level of BPS indices. This is followed in section \ref{chain} by a discussion of how to formulate chain complexes that categorify the BPS indices. The crucial concept of a $\zeta$-instanton with fan boundary conditions is discussed and we formulate the statement of categorical wall-crossing by using counts of certain trivalent instantons in section \ref{statement}. After reviewing the construction of the $A_{\infty}$ category of half-BPS branes associated to a Landau-Ginzburg model in section \ref{brane}, we show the equivalence of the categorical wall-crossing formula to the homotopy equivalence of $A_{\infty}$ categories constructed on either side of a marginal stability wall in section \ref{homotopy}. After a brief digression on fermion degrees of a $\zeta$-instanton in section \ref{maslov}, we turn our attention to some examples that illustrate our formulas in section \ref{examples}. We conclude with some speculations in section \ref{conclusion} and review some aspects of $A_{\infty}$-theory and homological algebra in Appendices \ref{ainf} and \ref{hom}. 

%\paragraph{Language and Prerequisites} In an effort to make the paper self-contained, we have included appendices that review some aspects of homotopical and homological algebra. In particular, the mapping cone construction, discussed in Appendix \ref{hom} is used for expressing the categorical wall-crossing formula. Some aspects of the web formalism of \cite{Gaiotto:2015aoa} are also used and we review the relevant points as they come up at various points in the paper. 

\section{Wall-Crossing of BPS Indices} \label{review} While our formulas are expected to hold for arbitrary massive two-dimensional $\mathcal{N}=(2,2)$ theories (with a non-anomalous $U(1)_R$-symmetry), it is simplest to work in the setting of Landau-Ginzburg models. A Landau-Ginzburg model is a supersymmetric sigma model with a K\"{a}hler manifold target $X$ and a potential of the form \bea V = |dW|^2,\eea where $W: X \rightarrow \mathbb{C}$ is a holomorphic function known as the superpotential. More precisely, working in two-dimensional $\mathcal{N}=2$-superspace, we can use the K\"{a}hler structure on $X$ to write D-terms \bea L_D = \int d^4 \theta \, K(\Phi, \ov{\Phi}),\eea and the holomorphicity of $W$ to write F-terms \bea L_F = \int d^2 \theta \, W(\Phi) + \int d^2 \ov{\theta}\, \ov{W}(\ov{\Phi}) ,\eea to get a Lagrangian \bea L = L_D + L_F,\eea invariant under two-dimensional $\mathcal{N}=(2,2)$ Poincar\'e supersymmetry. The reader is encouraged to consult \cite{Hori:2003ic}, whose notation we adopt, for more details. Various non-renormalization theorems \cite{Seiberg:1994bp} of $W$ tell us that one can get great mileage simply by studying the superpotential and its various properties. One use of the superpotential $W$ is that it is sufficient to study many aspects of BPS states.

\paragraph{}  Supposing that $W$ only has a finite number of isolated singularities, a familiar argument shows that the classical energy in such a theory obeys the BPS bound, \bea E \geq |Z_{ij}|\eea where \bea Z_{ij} = W_i - W_j\eea and $W_i$ denotes the critical value $W(\phi_i)$ of the critical point $\phi_i$.  Denoting the bosonic fields of the LG model as $\phi$, the standard Bogomolny trick leads to the BPS equation \bea \frac{d\phi}{dx} = \nabla \text{Re}(\zeta^{-1}W), \eea known as the $\zeta$-soliton equation, $\zeta$ being an arbitrary phase. Solutions on $\mathbb{R}$ with prescribed vacua $\phi_i$ and $\phi_j$ at the ends of $\mathbb{R}$ can only exist if \bea \zeta =  \zeta_{ji} := \frac{W_j - W_i}{|W_j - W_i|}. \eea Using intersection theory of vanishing cycles, it is possible to get a well-defined \textit{signed} count of the number of BPS solitons in the $ij$-sector. Let \bea L_{i}(\zeta) &=& \{p \in X | \text{lim}_{x \rightarrow -\infty} f^{\zeta}_x(p) = \phi_i\} ,\\ R_i(\zeta) &=& \{p \in X | \text{lim}_{x \rightarrow +\infty} f^{\zeta}_x(p) = \phi_i\} \eea be the ascending and descending manifolds respectively, emanating from the critical point $\phi_i$ of the Morse function $\text{Re}(\zeta^{-1}W)$. $f^{\zeta}_x$ denotes the one-parameter map $f^{\zeta}_x: X \rightarrow X$ defined by the gradient vector field of $\text{Re}(\zeta^{-1}W)$. We then set \bea \label{indexintersection} \mu_{ij} = L_i^{-} \circ R_j^{+} \eea where $L_i^- = L_i(\zeta_{ji}e^{-i \epsilon})$ and $R_j^+ = R_j(\zeta_{ji}e^{+i\epsilon})$ and $\epsilon$ is a small positive number. The infinitesimal rotation ensures that the intersection is transversal. 

\paragraph{} The significance of $\mu_{ij}$ from the perspective of the $\mathcal{N}=2$ field theory defined by $(X,W)$ is that one can show \cite{Cecotti:1992qh, Cecotti:1992rm} that \bea \mu_{ij} = \text{Tr}_{\mathcal{H}^{\text{BPS}}_{ij}} (-1)^F F \eea where $F$ is the fermion number and \bea \mathcal{H}_{ij}^{\text{BPS}} = \text{ker}(\mathcal{Q}_{ij}) \cap \text{ker}(\ov{\mathcal{Q}}_{ij}), \eea where \bea \mathcal{Q}_{ij} =  Q_- - \zeta_{ij}^{-1} \ov{Q}_+.\eea $\mu_{ij}$ is thus a supersymmetry protected index that counts the degeneracy of BPS states of type $ij$. Some of its elementary properties are as follows. 

\paragraph{Metric Independence} While the BPS soliton equation does depend on the K\"{a}hler metric on $X$, the BPS index $\mu_{ij}$ is metric-independent. 
\paragraph{CPT} Reversing $x \rightarrow -x$ takes $F \rightarrow -F$ so that $\mu_{ij} = -\mu_{ji}$.

%\paragraph{}The Cecotti-Vafa formula relates to the study of a two-dimensional quantum field theory, having Poincare generators $H,P,M$ along with odd generators $Q_+, Q_-, \ov{Q}_+, \ov{Q}_-$, and a $U(1)$ fermion number $F$ as part of its symmetry algebra. The algebra these generators obey along with some of its properties can be found in ref. Together it's known as the $\mathcal{N}=(2,2)$ algebra. We are interested in a theory having a finite number of massive vacua, a typical vacuum denoted by a lowercase Latin letter such as $i$ and $j$. Because we consider centrally extended supersymmetry algebra each vacuum gives rise to a complex number $W_i$ so that the central charge in the $ij$-sector is given by \bea Z_{ij} = W_i - W_j .\eea We study the Hilbert space on $\mathbb{R}$ with asymptotic boundary conditions going to $i$ at $-\infty$ and $j$ at $+\infty$ and we are interested in states which are annihilated by the two supercharges \bea Q_A^{ij} = \ov{Q}_+ + \zeta_{ij} Q_- , \\ \ov{Q}_A^{ij} = Q_{+} + \zeta_{ij}^{-1} \ov{Q}_- \eea where \bea \zeta_{ij} = \frac{W_i - W_j}{|W_i - W_j|}. \eea The BPS Hilbert space $\mathcal{H}_{ij}^{\text{BPS}}$ is the simultaneous kernel of $Q^{ij}_{A}, \ov{Q}^{ij}_{A}$. The BPS index is given by \bea \label{index} \mu_{ij} := \text{Tr}_{\mathcal{H}^{\text{BPS}}_{ij}}(-1)^F F \eea where we have to shift $F$ by a universal sector-dependent constant to make it integral. Later we give a construction of $\mathcal{H}^{\text{BPS}}_{ij}$ involving finite-dimensional vector spaces only, for the special case of Landau-Ginzburg models.

\paragraph{} It is familiar that supersymmetric indices such as the Witten index are quantities that are piecewise constant in parameter space.  For instance, we can consider the one-dimensional system given by the real superpotential \bea h = x^4 + \alpha x^2 + \beta x .\eea While the conventional partition function $Z = \text{Tr}(e^{-\beta H})$ of the system will be a very non-trivial function of $\alpha$ and $\beta$, the Witten index $I = \text{Tr}(-1)^F e^{-\beta H}$ is simply equal to $+1$, \bea I = 1,\eea irrespective of $\alpha$ and $\beta$. In contrast the behavior of the BPS index is more subtle.

\paragraph{} Historically\footnote{We thank S. Cecotti for narrating this story.} wall-crossing was first noticed by considering points in the parameter space of the Landau-Ginzburg model with \bea W = X^4 + t_1 X^2 + t_2 X \eea with distinct symmetry groups.   Supposing we start out at $(t_1, t_2) = (0,1),$ where the model is $\mathbb{Z}_3$-symmetric, the latter permuting the three vacua. We can show that there is indeed a single soliton between each pair of distinct critical points, \bea \mu_{12} &=& 1, \\ \mu_{23} &=& 1,\\ \mu_{13} &=& 1,\eea a spectrum consistent with the $\mathbb{Z}_3$ symmetry.  If we move slightly away from this point, the collection of numbers doesn't change. On the other hand at $(t_1, t_2) = (1,0)$, the superpotential has $\mathbb{Z}_2$-symmetry. Requiring a $\mathbb{Z}_2$-symmetric spectrum requires that one of the solitons disappears and the BPS indices are \bea  \mu'_{12} &=& 1, \\ \mu'_{23} &=& 1, \\ \mu'_{13} &=& 0.\eea Thus BPS indices are examples of indices that are not constant but only piecewise constant.

\begin{figure}%[here!]
\centering
\includegraphics[width=.9\textwidth]{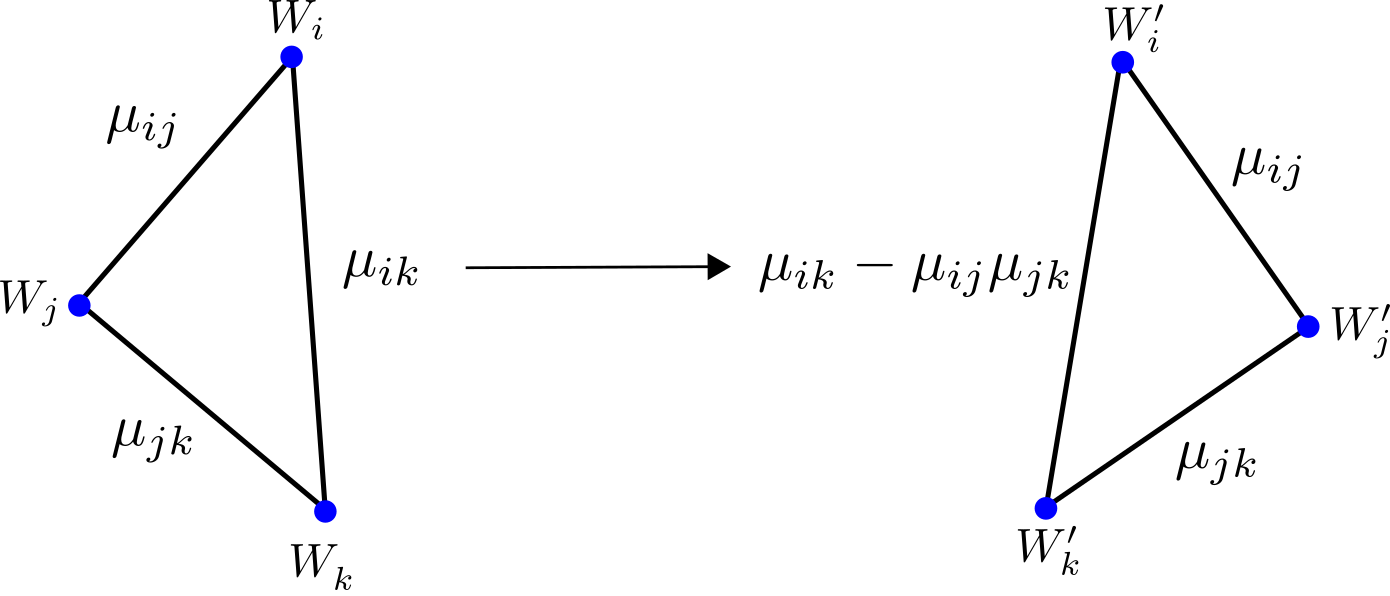}
\caption{Wall-crossing summarized in the $W$-plane.}
\label{cvform}
\end{figure}

\paragraph{} The content of the Cecotti-Vafa formula is as follows. It first states that potential discontinuous jumps in the BPS spectrum can occur when three critical values $W_i, W_j, W_k$ become co-linear as we vary parameters. This is the locus where $\text{Im}(Z_{ij} \ov{Z}_{jk}) = 0$. Next it gives an explicit formula for the quantitative nature of this jump: If $\mu$ and $\mu'$ denote BPS degeneracies on different sides of the wall of marginal stability, they must be related by \bea \label{wcf} \mu'_{ij} &=& \mu_{ij}, \\ \mu'_{jk} &=& \mu_{jk}, \\ \mu'_{ik} &=& \mu_{ik} \pm \mu_{ij} \mu_{jk},\eea where the sign $-$ is picked in going from the negative side, where $\text{Im}(Z_{ij} \ov{Z}_{jk})<0$ to the positive side, where $\text{Im}(Z_{ij} \ov{Z}_{jk})>0$ and the $+$ is picked in the reverse move.  We summarize the formula from the perspective of the $W$-plane in Figure \ref{cvform}. 

\begin{figure}%[here!]
\centering
\includegraphics[width=.9\textwidth]{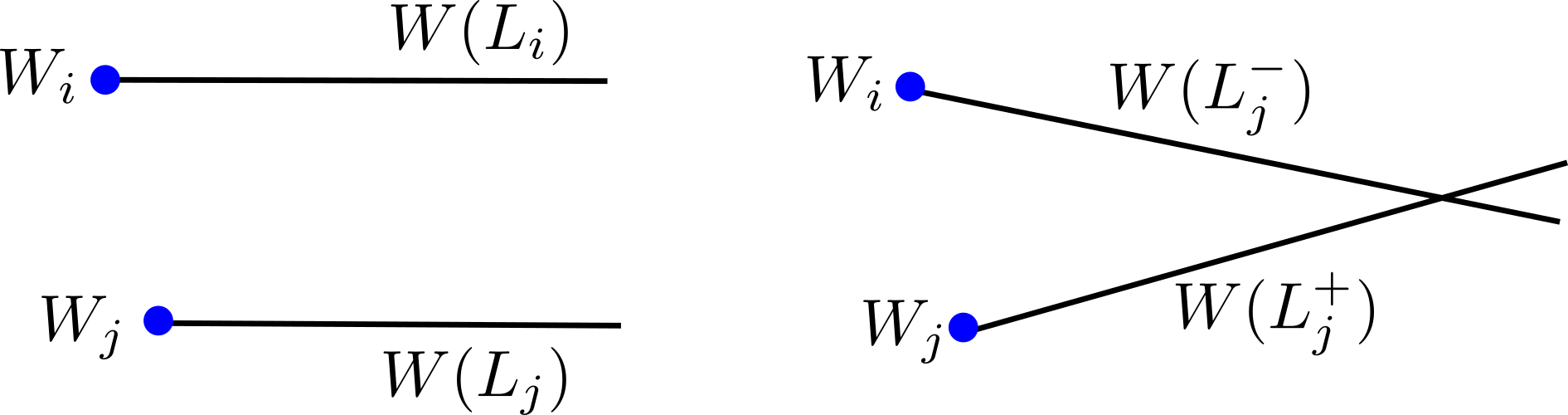}
\caption{The topological intersection numbers $\widehat{\mu}_{ij}$ obtained by looking at intersection numbers of slightly rotated thimbles.}
\label{FS}
\end{figure}

\paragraph{} The trick in arguing for this is to consider not just BPS states, but rather to look at \bea \mathcal{Q}(\zeta) = Q_- - \zeta^{-1} \ov{Q}_+\eea preserving boundary conditions of our Landau-Ginzburg model when the latter is formulated on a half-space such as $(-\infty, 0] \times \mathbb{R}_t$. Such branes have been analyzed in great detail in references, \cite{Gaiotto:2015aoa, Hori:2000ck}. One finds that the homology class of the support of these branes lives in the finite rank $\mathbb{Z}$-module \bea B(\zeta) := H_{\frac{1}{2}\text{dim}(X)}(X, \text{Re}(\zeta^{-1}W) \rightarrow \infty ; \mathbb{Z}).\eea We can equip $B(\zeta)$ with a natural bilinear form \bea \widehat{\mu}^{\zeta}: B(\zeta) \times B(\zeta) \rightarrow \mathbb{Z},\eea defined as follows. When $W$ is Morse, there is a natural $\mathbb{Z}$-module basis for $B(\zeta)$ given by the homology class of Lefschetz thimbles $\{[L_i(\zeta)]\}_{i \in \mathbb{V}}$. The thimble $L_i(\zeta)$ projects to half-infinite rays emanating from the critical value $W_i$ in the $\zeta$-direction. We then define \bea \label{leftleft}  \widehat{\mu}^{\zeta}_{ij} := \widehat{\mu}(L_i, L_j) = L^{-}_i \circ L^{+}_j, \eea where $L^{\pm}$ denote thimbles with phases slightly rotated by a small positive or negative angle respectively, as in Figure \ref{FS}. 

\paragraph{} Some basic properties of $\widehat{\mu}^{\zeta}$ are as follows. First: if $i$ and $j$ are distinct vacua, $\widehat{\mu}^{\zeta}_{ij}$ and $\widehat{\mu}^{\zeta}_{ji}$ cannot both be non-zero. In the case they are equal, \bea \widehat{\mu}_{ii} = 1 .\eea Finally, if the vacuum weights are $\zeta$-generic \footnote{A set of critical values is called $\zeta$-generic, following the terminology in \cite{Kapranov:2014uwa}, if none of the relative phases $\zeta_{ij}$ are equal to $\zeta$.}, we can order the thimble basis in decreasing order of $\text{Im}(\zeta^{-1}W_i)$. Making this choice of ordering, we find that $\widehat{\mu}^{\zeta}$ is an upper-triangular $|\mathbb{V}| \times |\mathbb{V}|$ matrix with $+1$ on the diagonal.

%\paragraph{} Perhaps unsurprisingly, the bilinear form $\widehat{\mu}$ can be determined from the BPS indices $\mu_{ij}$. $\widehat{\mu}_{ij}$ can be recovered from the BPS indices $\mu_{ij}$ by taking certain phase ordered products of elementary factors associated with each BPS ray $Z_{ij}$, as summarized in \eqref{spec}. We will find a slight rephrasing in terms of \textbf{half-plane fans} quite useful in what follows.  

\paragraph{} For definiteness and to avoid notational clutter we set $\zeta=1$ and set $\widehat{\mu} = \widehat{\mu}^{\zeta = 1}$. This is equivalent to choosing the half-plane in which we take phase ordered products to be the upper-half plane, as was done in \eqref{spec}. 

\paragraph{} The matrix representation $\widehat{\mu}_{ij}$ for the bilinear form can be calculated from the BPS indices $\mu_{ij}$ by a nice rule expressed in terms of convex geometry. 

\begin{figure}%[here!]
\centering
\includegraphics[width=.55\textwidth]{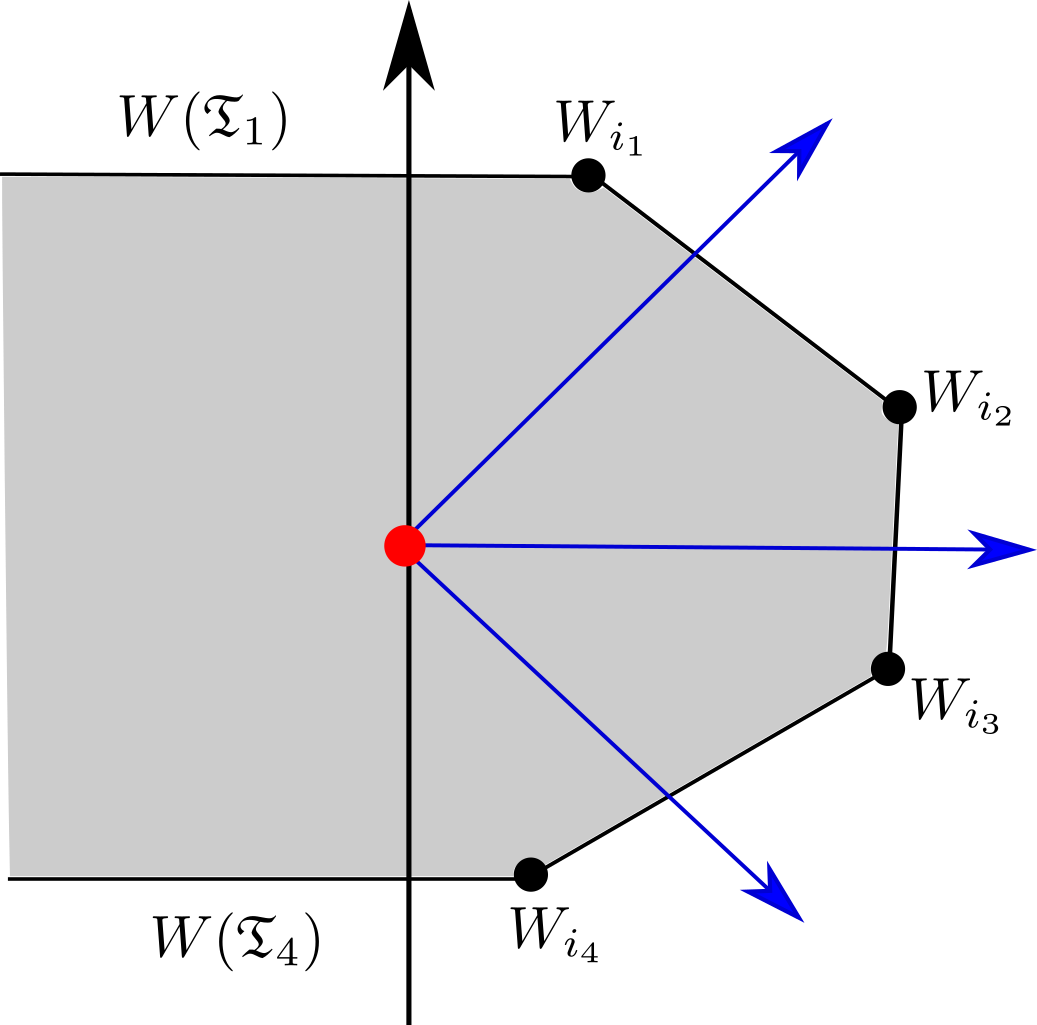}
\caption{A half-plane fan $F_{i_1 i_4} = \{i_1, i_2, i_3, i_4\}$ for $\zeta=1$ and the semi-infinite polygon it forms in the $W$-plane. }
\label{infinitepolygon}
\end{figure}

\paragraph{Definition:} A half-plane fan $F$ of phase $\zeta$ is a collection of vacua $F = \{i_1, \dots, i_n\}$ such that $W(F) = \{W_{i_1}, \dots, W_{i_n} \}$ are the clockwise-ordered vertices of a semi-infinite convex polygon going off to infinity in the $-\zeta$-direction.  See Figure \ref{infinitepolygon} for an example with $n=4$. The dual graph looks like a half-plane fan (and indeed has a space-time interpretation), hence the terminology.

\paragraph{}  To a given half-plane fan $F = \{i_{1}, i_2. i_3, \dots, i_n \}$ assign the number \bea \mu_{F} = \mu_{i_1 i_2} \mu_{i_2 i_3} \dots \mu_{i_{n-1} i_n}.\eea We then make the 

\paragraph{Claim} \bea \label{polygonrule} \widehat{\mu}_{ij} = \sum_{\substack{ F_{ij} = \{i, i_1, \dots, i_k, j\} \\ F_{ij} \text{ half-plane fan}}} \mu_{i i_1} \dots \mu_{i_k j} .\eea  

\paragraph{Proof} The proof is a straightforward inductive argument, where we induct on distance between $i$ and $j$. To show the base case, for two neighboring vacua $i<j$, one has $\widehat{\mu}_{ij} = \mu_{ij}$ due to \eqref{indexintersection}\footnote{Note that for $\zeta$ being the phase of an $ij$-soliton left-right intersection number of \eqref{indexintersection} agrees with the left-left intersection number of \eqref{leftleft}}. On the other hand there's only one polygon between two neighboring vacua, whose finite segment is given by the segment connecting them, to which we also assign $\mu_{ij}$. For the inductive step, assume that the polygon rule \eqref{polygonrule} holds for vacua that are up to $n$ units apart and consider a pair of vacua $\{i,j\}$ that are $n+1$ units apart. We know that \bea L_i \circ \widetilde{L}_{j} = \mu_{ij}\eea where $\widetilde{L}_j := L_j(\zeta_{ij} e^{-i \epsilon})$. We thus want to compare $\widetilde{L}_j$ with $L_j^+$ namely we must rotate this thimble in a clockwise direction by the phase of $\zeta_{ij}$. In doing this rotation we pick up Picard-Lefschetz discontinuities: For each critical value $W_k$ such that $\{i,k,j\}$ forms half-plane fan, we pick up a contribution of $L^+_k \mu_{ki}$. Summing these up we get \bea L^+_j = \widetilde{L}_j + \sum_{\substack{k \\ \{i,k,j\} \text{ is a fan}} } L^+_k \mu_{ki}. \eea Thus we can compute that \bea \widehat{\mu}_{ij} = \mu_{ij} + \sum_{\substack{k \text{ s.t.} \\ \{i,k,j \} \text{ is a fan}}} \widehat{\mu}_{ik} \mu_{kj}.\eea The polygon rule applies to $\widehat{\mu}_{ik}$ so that \bea \widehat{\mu}_{ik} \mu_{kj} = \sum_{F_{ik}} \mu_{F_{ik}} \mu_{kj}. \eea On the other hand if $\{i,k,j\}$ is a fan, we can form an $ij$ half-plane fan by taking the fan $F_{ik} = \{i, \dots, k \}$ and putting $j$ at the end $\{i, \dots, k, j\}$. To this one precisely assigns $\mu_{F_{ik}} \mu_{kj}$. Conversely, every $ij$ fan can be obtained in this way. $\qed$

\paragraph{} To see that this implies the wall-crossing formula, consider $\widehat{\mu}$ restricted to the three-dimensional $\{i,j,k\}$ space and note that if we are on the left side of Figure \ref{cvform} then there is only one half-plane of type $ij$ $ik$ $jk$ respectively, so that \bea \widehat{\mu} = \begin{pmatrix}
1 && \mu_{ij} && \mu_{ik} \\
0 && 1 && \mu_{jk} \\
0 && 0 && 1
\end{pmatrix}. \eea On the other side of the wall we have two half-plane fans of type $ik$, depicted in Figure \ref{hpfans}, leading us to write \bea \widehat{\mu} = \begin{pmatrix}
1 && \mu'_{ij} && \mu'_{ik} + \mu'_{ij} \mu'_{jk} \\
0 && 1 && \mu'_{jk} \\
0 && 0 && 1
\end{pmatrix}. \eea The two expressions for $\widehat{\mu}$ are equal if and only if the wall-crossing formula holds.  

\paragraph{} More generally suppose that $\{l,m\}$ is any pair of vacua such that there is a fan \bea F_{lm} = \{l,i_1, \dots, i,k, \dots, i_n, m\} \eea in which $\{i,k\}$ appears as a subset of consecutive vacua. Then on the other side of the wall, for every such fan, the set of $lm$-fans gains an additional fan obtained by taking $F_{lm}$ and inserting $j$ in between $i$ and $k$. Moreover these are the only additional fans we gain, assuming we cross no other marginal stability walls in the move. Thus we compare \bea \mu_{li_1} \dots \mu_{ik} \dots \mu_{i_n m} \eea with \bea \mu_{l i_1} \dots \mu'_{ik} \dots \mu_{i_n m} + \mu_{l i_1} \dots \mu'_{ij} \mu'_{jk} \dots \mu_{i_n m}\eea and the two are equal if and only if the wall-crossing formula holds. Therefore we conclude that the wall-crossing formula is equivalent to the invariance of the bilinear form $\widehat{\mu}$ across a wall of marginal stability.

\section{BPS Chain Complexes and $\zeta$-instantons} \label{chain} \subsection{BPS Chain Complexes}
 
\paragraph{} The chain complexes $R_{ij}$ that categorify $\mu_{ij}$ can be formulated by using an infinite-dimensional version of Morse theory. Suppose that the symplectic form $\omega$ on $X$ is exact and choose a Liouville form $\lambda$ so that $\omega = d\lambda$.  We consider the (family of) ``Morse" functions \bea h_{\zeta}[\phi] = \int_{\mathbb{R}} \phi^*(\lambda) + \text{Im}\big(\zeta^{-1}W(\phi) \big) dx \eea acting on the space \bea \mathscr{X}_{ij} =\{ \phi: \mathbb{R} \rightarrow X | \text{lim}_{x \rightarrow -\infty}\phi(x) = \phi_i \,\,\,\, \text{lim}_{x \rightarrow \infty } \phi(x) = \phi_j \}. \eea \paragraph{Generators} The critical points are the points where $\delta h_{\zeta} = 0$ which are solutions of the $\zeta$-soliton equation \bea \label{solitoneqn} \frac{d \phi^{I}}{dx} = \frac{ \zeta}{2} g^{I \bar{J}}\frac{\del \ov{W}}{\del \ov{\phi}^{\bar{J}}},\eea and so the critical point set is non-empty only for $\zeta = \zeta_{ji}$. The Morse function is actually not Morse because of the translational invariance of the soliton equation but we can mod out the solution space by this $\mathbb{R}$-action to obtain a (generically) finite set of critical points, in one-to-one correspondence with intersection points \bea L_i(\zeta_{ji} e^{-i \epsilon} )\cap R_j(\zeta_{ji} e^{i \epsilon} ). \eea Thus we look to the pair \bea (\mathscr{X}_{ij}, h_{-\zeta_{ij}}) \eea and assign a $\mathbb{Z}$-module $R_{ij}$ with one generator for each solution of the $\zeta_{ji}$-soliton equation \bea R_{ij} = \bigoplus_{p \in L_i^- \cap R_j^+} \mathbb{Z} \langle \phi_{ij}^p \rangle .\eea 

\paragraph{Gradations} Next we come to the subtle business of defining gradations on $R_{ij}$. The Fermion number, or homological degree of a generator in the Morse complex for a Morse function $f$ as reviewed in \cite{Gaiotto:2015aoa,Hori:2003ic} is given by \bea \label{homdegree} -\frac{1}{2} \sum_{\lambda \in \text{Spec } \text{Hess}(f(p))} \text{sign}(\lambda) ,\eea where $p$ is the critical point of $f$ whose degree we're computing.   To assign a degree to a $\zeta$-soliton we must therefore compute the second derivative $\delta^2 h_{\zeta}$. Equivalently we may linearize the $\zeta$-soliton equation \eqref{solitoneqn} which leads to \bea  \label{linearsolitoneqn} D^{(1,0)}_{x} \delta \phi^{I} = \frac{\zeta}{2} g^{I \bar{J}} D_{\bar{J}} \del_{\bar{K}} \ov{W} \delta \ov{\phi}^{\bar{K}} \eea where \bea D^{(1,0)}_x\delta \phi^I = \frac{\del}{\del x} \delta \phi^I + \Gamma^I_{JK} \frac{\del \phi^J}{\del x} \delta \phi^K \eea is the pullback connection on $\phi^*(T^{(1,0)}X)$. By considering also the complex conjugate of \eqref{linearsolitoneqn}, we can write the linearized soliton equation as \bea D_{\zeta} \delta \phi = 0  \eea where $D_{\zeta}$ is a Dirac type operator \bea D_{\zeta} : \Gamma\big(\phi^*(T^{(1,0)}X) \oplus \phi^*(T^{(0,1)} X)\big) \rightarrow \Gamma\big(\phi^*(T^{(1,0)}X) \oplus \phi^*(T^{(0,1)}X) \big) .\eea Writing \bea \delta \phi \in \Gamma\big(\phi^*(T^{(1,0)} X) \oplus \phi^*(T^{(0,1)}X) \big)\eea as a column vector \bea \delta \phi = \begin{pmatrix}
\delta \phi^I \\ 
\delta  \ov{\phi}^{\bar{I}}
\end{pmatrix} \eea the operator $D_{\zeta}$ reads\footnote{Note that the operator \eqref{diracop} differs from that given in equation 12.6 of \cite{Gaiotto:2015aoa}, v1. The authors of \cite{Gaiotto:2015aoa} forgot to include covariant derivatives.} \bea \label{diracop} \begin{split} D_{\zeta} = \begin{pmatrix}
\delta^{I}_J \del_x + \Gamma^{I}_{JK} \del_x \phi^K && 0 \\
0 && \delta^{\bar{I}}_{\bar{J}} \del_x + \Gamma^{\bar{I}}_{\bar{J} \bar{K}} \del_x \ov{\phi}^{\bar{K}} \end{pmatrix} \,\,\,\,\,\,\,\,\,\,\,\,\,\,\,\,\,\,\,\,\,\,\,\,\,\,\,\,\,\,\,\,\,\,\,\,\,\,\,\,\,\,\,\,\,\,\,\, \\ - \begin{pmatrix}
0 && \frac{\zeta}{2} g^{I \bar{K}} D_{\bar{K}} \del_{\bar{J}} \ov{W} \\
\frac{\zeta^{-1}}{2} g^{\bar{I}K} D_K \del_{J} W && 0 
\end{pmatrix}. \end{split} \eea The operator $D_{\zeta}$ is expressed a little more compactly by identifying \bea \phi^*(T^{(1,0)}X) \oplus \phi^*(T^{(0,1)}X) \cong \phi^*(TX),\eea where $TX$ denotes the complexified tangent bundle. Choosing real coordinates indexed by $a = 1, \dots, \text{dim}_{\mathbb{R}}(X)$,  we can write \bea D_{\zeta} = \delta^a_b D_x  - g^{ac} D_b \del_c\, \text{Re}(\zeta^{-1}W), \eea where \bea D_x \delta \phi^a = \del_x \delta \phi^a + \Gamma^a_{bc} \del_x \phi^b \,\delta \phi^c \eea is now the pullback connection on $\phi^*(TX)$. The Fermion number of an $ij$-soliton $\phi$ should thus be given by a regularized version of \eqref{homdegree}: \bea F(\phi) &=& -\text{lim}_{\epsilon \rightarrow 0} \frac{1}{2} \sum_{\lambda \in \text{Eigenvalues} (D_{\zeta_{ji}}(\phi))} \text{sign}(\lambda) e^{- \epsilon|\lambda|}  \\ &=& -\frac{1}{2} \eta \big(D_{\zeta_{ij}} (\phi) \big).\eea One wants chain complexes $R^{(1)}_{ij}, R^{(2)}_{ij}$ constructed from two different choices of K\"{a}hler metrics $g^{(1)}, g^{(2)}$ (namely by a different choice of D-terms) to be homotopy equivalent \bea R^{(1)}_{ij} \simeq R^{(2)}_{ij} .\eea A necessary condition for is this that if we continuously interpolate between the metrics $g^{(1)}$ and $g^{(2)}$ and evolve the soliton $\phi^{(1)}$ solving the $\zeta$-soliton equation for $g^{(1)}$ to $\phi^{(2)}$ a soliton for $g^{(2)}$ then their Fermion degrees must match. However the variational formula for the $\eta$-invariant says that \bea \label{anomaly} \frac{1}{2} \eta \big( D(\phi^{(1)}, g^{(1)}) \big) - \frac{1}{2} \eta \big(D(\phi^{(2)}, g^{(2)}) \big) = 2 \int_{\mathbb{R} \times [0,1]} \widetilde{\phi}^* \Big( \frac{1}{2\pi} \text{Tr} \, \mathcal{R} \Big), \eea where \bea \widetilde{\phi}: \mathbb{R} \times [0,1] \rightarrow X \eea is a path in $\mathscr{X}_{ij}$ interpolating between $\phi^{(1)}$ and $\phi^{(2)}$, and \bea \frac{1}{2\pi} \text{Tr} \,\mathcal{R} \eea is the Chern-Weil representative of $c_1(TX)$. This is nothing but a reminder that the LG model has an axial anomaly for arbitrary K\"{a}hler target. The axial anomaly is traditionally expressed as the statement that the right hand side of \eqref{anomaly} measures the net violation of Fermion number. The factor of two comes from taking into account the individual violations of both left and right moving fermions. Thus gradations are unchanged under metric variations only if $X$ is Calabi-Yau. Otherwise to ensure this property we must grade $R_{ij}$ by a cyclic group $\mathbb{Z}_N$ such that the image of $2c_1(X)$ in $H^2(X, \mathbb{Z}_N)$ vanishes. 

\paragraph{Differential} The differential $d_{ij}$ is provided by counting (with signs) solutions of the $\zeta_{ji}$-instanton equation \bea \del_{\ov{s}} \phi^I =  \frac{\zeta_{ji}}{2} g^{I \bar{J}}\frac{\del \ov{W}}{\del \ov{\phi}^{\bar{J}}}\eea interpolating between solitons of fermion number differing by a unit. Here $s=x+i\tau$, where $\tau$ is the Euclidean time. Thus we get well-defined chain complexes $(R_{ij}, d_{ij})$ from which we can construct $\mathcal{H}^{\text{BPS}}_{ij}$ by taking cohomology \bea \mathcal{H}^{\text{BPS}}_{ij} \cong H^{\bullet}(R_{ij}, d_{ij}).\eea

\paragraph{} A $\zeta$-instanton which contributes to the differential $d_{ij}$ in spacetime looks like Figure \ref{inst1}. Physically we expect the following properties.

\paragraph{Metric Dependence} BPS chain complexes constructed from two different choices of K\"{a}hler metrics should be homotopy equivalent.

\paragraph{CPT} Reversing the spatial coordinate, i.e the path $\phi^p_{ij}(-x)$ says that for every basis element $\phi_{ij}^p$ of $R_{ij}$ we get an element $\phi_{ji}^p$ such that \bea \text{deg}(\phi^p_{ji}) = 1-\text{deg}(\phi_{ij}^p) . \eea The shift in degree by $+1$ is a technical consequence of factoring out the translational mode of the soliton. For more details on this point see the discussion in section 12.3 in \cite{Gaiotto:2015aoa}. In basis independent terms, CPT says that we have a degree $-1$ non-degenerate pairing \bea K_{ij}: R_{ij} \otimes R_{ji} \rightarrow \mathbb{Z}.\eea 

\subsection{$\zeta$-instantons and Interior Amplitudes} \label{intamp} As alluded to in the introduction, a categorified wall-crossing formula will involve certain ``off-diagonal" maps \bea M[\beta_{ikj}]: R_{ij} \otimes R_{jk} \rightarrow R_{ik}\eea which allow construction of the correct differential. The construction of this map involves counting $\zeta$-instantons with fan boundary conditions, which we now discuss.

\begin{figure}%[here!]
\centering
\includegraphics[width=1.0\textwidth]{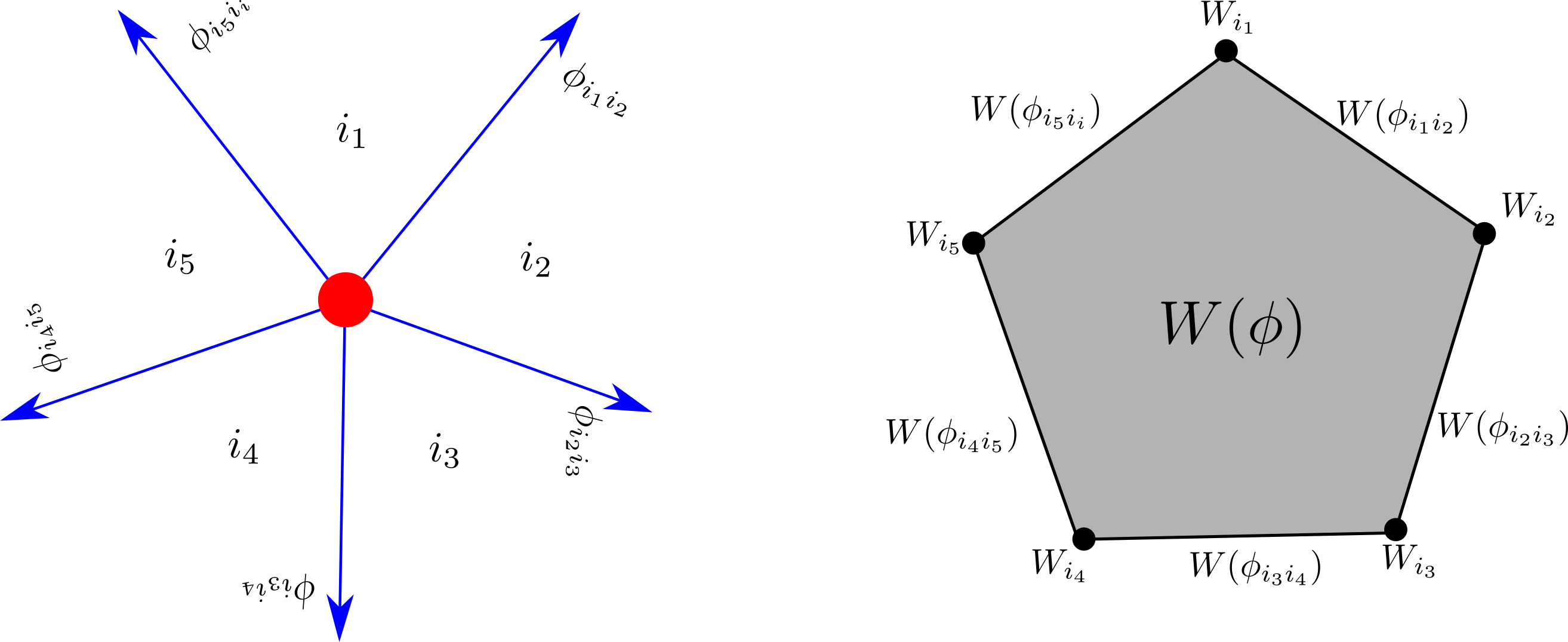}
\caption{Boundary conditions for a $\zeta$-instanton in the $(x,\tau)$-plane on the left, and the image of such a $\zeta$-instanton in the $W$-plane on the right.}
\label{zetainst}
\end{figure}

\paragraph{} We consider solutions of the $\zeta$-instanton equation \bea \ov{\del}_{\ov{s}} \phi^I =  \frac{\zeta}{4} g^{I \bar{J}}\frac{\del \ov{W}}{\del \ov{\phi}^{\bar{J}}}\eea which look like a collection of ``boosted solitons" at infinity. See \cite{Gaiotto:2015aoa} sections 14.1-14.2 and Appendix E for more details on such boundary conditions. Let \bea I= \{i_1, \dots, i_n\}\eea be a cyclic fan of vacua and \bea \bphi = \{\phi_{i_1 i_2}, \dots, \phi_{i_n i_1} \}\eea be a fan of solitons. We want to consider $\zeta$-instantons which support these particular solitons on the edges. $I$ is a fan if and only if the critical values \bea W_I = \{W_{i_1}, \dots, W_{i_n}\} \eea are the \underline{clockwise ordered} vertices of a convex polygon in the $W$-plane. Solutions of the $\zeta$-instanton equation with fan boundary conditions are known as a domain-wall junctions and have been studied in \cite{Carroll:1999wr, Gibbons:1999np, Ito:2000mt}, and elsewhere. In particular, it was noted in \cite{Carroll:1999wr}, that just the way a $\zeta_{ij}$-soliton maps to a line connecting $W_i$ and $W_j$ in the $W$-plane, a $\zeta$-instanton maps to the interior of the convex polygon with $W_I$ as vertices. See Figure \ref{zetainst} for an example with $n=5$. This fact motivates the terminology BPS or gradient polygon for $\bphi$, as was introduced in \cite{Kapranov:2014uwa}.

\paragraph{} Solutions of the $\zeta$-instanton equation modulo translations with a fixed fan and fixed soliton collection $\bphi$ supported on edges form a moduli space $\mathcal{M}_{\zeta}(\bphi)$. Its dimension is given by forming the vector \bea e_{\boldsymbol \phi} := \phi_{i_1 i_2} \otimes \dots \otimes \phi_{i_n i_1} \eea in the cyclic tensor product \bea R_{I} = R_{i_1 i_2} \otimes \dots \otimes R_{i_n i_1} \eea and considering its degree \bea F(\boldsymbol \phi) := \text{deg}(e_{\boldsymbol \phi}).\eea The (virtual) dimension of these moduli spaces is \cite{Gaiotto:2015aoa} \bea \text{dim}(\mathcal{M}_{\zeta}(\boldsymbol \phi)) = F(\boldsymbol \phi)-2. \eea Moreover $\mathcal{M}_{\zeta}(\bphi)$ can be oriented. In particular if $F({\boldsymbol{\phi}}) = 2$, we learn that the moduli space $\mathcal{M}_{\zeta}(\boldsymbol \phi)$ is a collection of oriented points and thus we can get a well-defined signed count of $\zeta$-instantons \bea N_{\zeta}(\boldsymbol \phi) := \# \mathcal{M}_{\zeta}(\bphi) .\eea 

\paragraph{} The integers $N(\bphi)$ \footnote{We can safely drop the $\zeta$-subscript from the notation because the integers $N_{\zeta}(\bphi)$ are $\zeta$-independent} satisfy some miraculous identities . There is an identity corresponding to each cyclic fan. \paragraph{}For a cyclic fan of length two, $\{i,j\}$ we have \bea \label{nilpotence} \sum_{\substack{\chi_{ij} \in L_i^- \cap R_j^+ \\ F(\phi_{ij}, \chi_{ji}) = 2\\ F(\chi_{ij}, \psi_{ji}) = 2 }} N\big(\phi_{ij}, \chi_{ji}\big) N\big(\chi_{ij}, \psi_{ji}\big) = 0.\eea This is nothing but the identity that the differential $d_{ij}$ counting $\zeta$-instantons between $ij$-solitons is nilpotent, which is a familiar fact from Morse theory. It involves the fact that the moduli space $\mathcal{M}(\phi_{ij}, \psi_{ji})$ such that $d(\phi_{ij}, \psi_{ji})=3$ has ends corresponding to broken flow lines gluing intermediate instantons.

\begin{figure}%[here!]
\centering
\includegraphics[width=.9\textwidth]{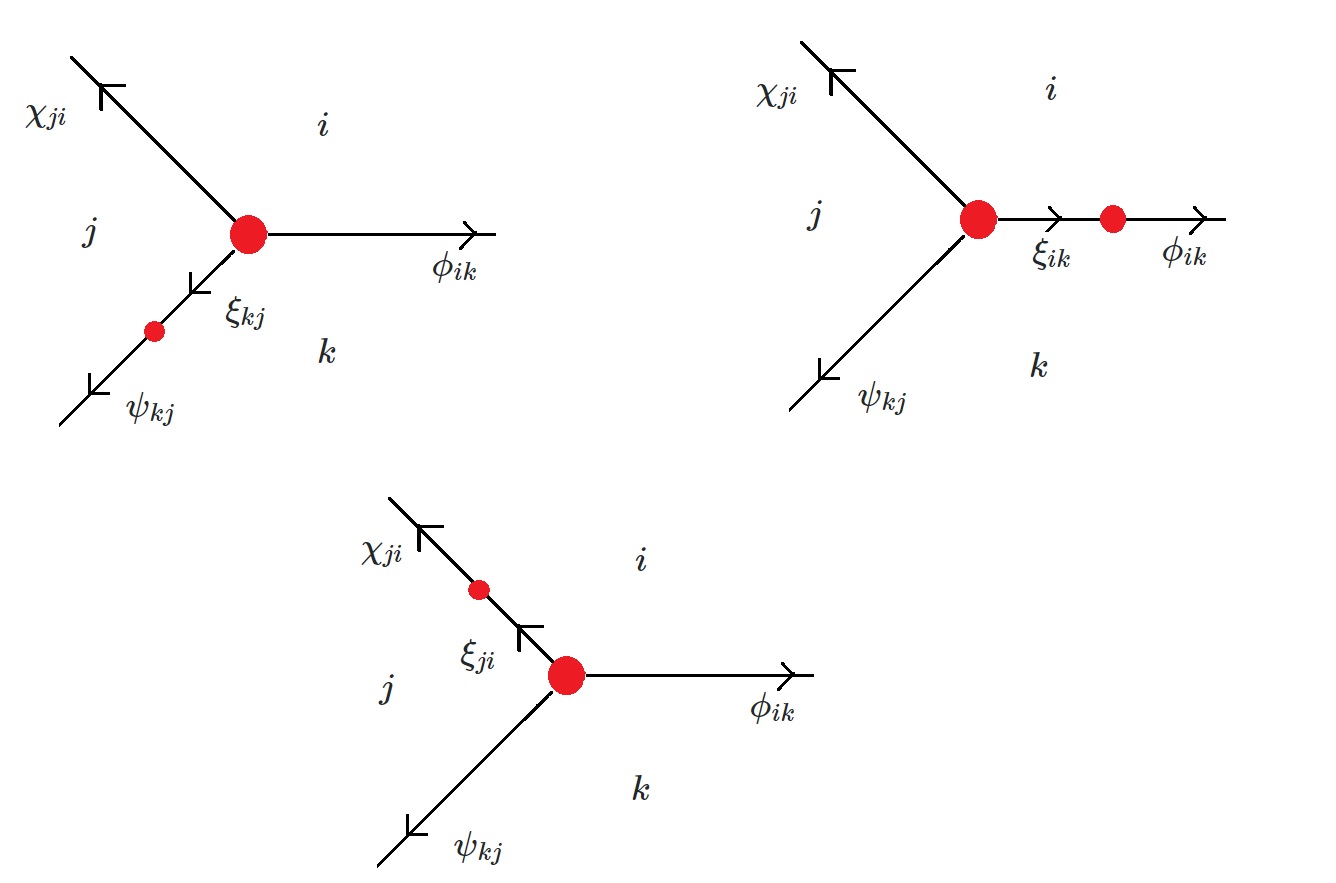}
\caption{The various ends of $\mathcal{M}\big(\phi_{ik}, \psi_{kj}, \chi_{ji})$ where $F(\phi_{ik}, \psi_{kj}, \chi_{ji})=3.$ }
\label{ends}
\end{figure}

\paragraph{}  For $\{i,k,j\}$ a cyclic fan of vacua of length three, we have the identity \bea \label{chainmap} \begin{split} \sum_{\substack{\xi_{ij} \in L_i^- \cap R_j^+ \\ F(\phi_{ik},\psi_{kj}, \xi_{ji}) = 2 \\ F(\xi_{ij}, \chi_{ji})=2}} N(\phi_{ik}, \psi_{kj}, \xi_{ji} \big) N\big(\xi_{ij}, \chi_{ji} \big) \\ + \sum_{\substack{\xi_{jk} \in L_j^- \cap R_k^+ \\ F(\chi_{ji}, \phi_{ik}, \xi_{kj}) = 2 \\ F(\xi_{jk}, \psi_{kj}) = 2}} N\big(\chi_{ji}, \phi_{ik}, \xi_{kj} \big) N\big(\xi_{jk}, \psi_{kj} \big) \\ + \sum_{\substack{\xi_{ik} \in L_i^- \cap R_k^+ \\ F(\xi_{ik}, \psi_{kj}, \chi_{ji}) = 2 \\ F(\phi_{ik}, \xi_{ki}) = 2} }N\big(\phi_{ik}, \xi_{ki}\big) N\big(\xi_{ik}, \psi_{kj}, \chi_{ji}\big) = 0. \end{split}\eea The argument for this involves looking at the ends of the moduli space \bea \mathcal{M}\big(\phi_{ik}, \psi_{kj}, \chi_{ji}) \eea of a fan of solitons such that $F(\phi_{ik}, \psi_{kj}, \chi_{ji})=3.$ There are three types of ends, where a rigid instanton of type $\{i,k\}$ is glued to a rigid instanton of type $\{i,k,j\}$, similarly for $\{i,j\}$ and $\{j,k\}$. See Figure \ref{ends}. Such ``broken flows" give \bea \begin{split} \del \mathcal{M}\big(\phi_{ik}, \psi_{kj}, \chi_{ji}) =\bigsqcup_{\substack{\xi_{ij} \in L_i^- \cap R_j^+ \\ F(\phi_{ik},\psi_{kj}, \xi_{ji}) = 2 \\ F(\xi_{ij}, \chi_{ji})=2}} \mathcal{M}\big(\phi_{ik}, \psi_{kj}, \xi_{ji} \big) \times \mathcal{M}\big(\xi_{ij}, \chi_{ji} \big) \\ \sqcup  \bigsqcup_{\substack{\xi_{jk} \in L_j^- \cap R_k^+ \\ F(\chi_{ji}, \phi_{ik}, \xi_{kj}) = 2 \\ F(\xi_{jk}, \psi_{kj}) = 2}} \mathcal{M}\big(\chi_{ji}, \phi_{ik}, \xi_{kj} \big) \times \mathcal{M}\big(\xi_{jk}, \psi_{kj} \big) \\ \sqcup \bigsqcup_{\substack{\xi_{ik} \in L_i^- \cap R_k^+ \\ F(\xi_{ik}, \psi_{kj}, \chi_{ji}) = 2 \\ F(\phi_{ik}, \xi_{ki}) = 2} }\mathcal{M}\big(\phi_{ik}, \xi_{ki}\big) \times \mathcal{M}\big(\xi_{ik}, \psi_{kj}, \chi_{ji}\big) . \end{split}\eea \eqref{chainmap} then follows from \bea \# \del \mathcal{M}\big(\phi_{ik}, \psi_{kj}, \chi_{ji}) =0. \eea More generally, one expects that the moduli spaces $\mathcal{M}_{\zeta}(\bphi)$ can be compactified, such that the compactified moduli space $\ov{\mathcal{M}}_{\zeta}(\bphi)$ has strata labeled by web diagrams of the type in Figure \ref{ends}.

\paragraph{}Although the identities \eqref{nilpotence} and \eqref{chainmap} are all we need for categorical wall-crossing, we should mention for completeness that there are more complicated identities involving fans of longer length which can be deduced from the web combinatorics of \cite{Gaiotto:2015aoa}. The summary is that all identities follow from a single $L_{\infty}$-Maurer-Cartan equation. Form the vector space\footnote{$R_i \cong \mathbb{Z}$} \bea R_c &=& \oplus_{I} R_I \\ &=& \oplus_{i \in \mathbb{V}} R_i \oplus_{i \neq j}(R_{ij} \otimes R_{ji}) \oplus \dots \eea corresponding to taking all possible cyclic tensor products. $R_c$ has the structure of an $L_{\infty}$-algebra. Namely there are maps \bea \rho(\mathfrak{t}): S_+R_c \rightarrow R_c,\eea where $S_+R_c$ denotes (the positive part of) the symmetric algebra, satisfying $L_{\infty}$-axioms. $\rho(\mathfrak{t})$ is defined through \textbf{taut webs} as in \cite{Gaiotto:2015aoa}. Define \bea \label{mc} \beta_{I} := \sum_{\substack{\boldsymbol \phi \text{ gradient polygons for } I \\ F(\boldsymbol \phi) = 2}} N(\boldsymbol \phi) e_{\boldsymbol \phi},\eea and let \bea \beta := \sum_{I} \beta_{I} \in R_c.\eea One of the main results of \cite{Gaiotto:2015aoa} is that analysis of various moduli spaces leads one to conclude that $\beta$ is a Maurer-Cartan element for the $L_{\infty}$-structure. Namely it satisfies the $L_{\infty}$ Maurer-Cartan equation \bea \rho(\mathfrak{t})(e^{\beta}) = 0.\eea $\beta$ was called the \textbf{ interior amplitude} in \cite{Gaiotto:2015aoa}. The identities \eqref{nilpotence}, \eqref{chainmap} are some simple equations that come from unpacking the $L_{\infty}$ Maurer-Cartan equation.

\paragraph{Remark} In general interior amplitudes will have components associated to arbitrary fans \bea \beta_{i_1  i_2 \dots i_n} \in R_{i_1 i_2} \otimes R_{i_2 i_3} \otimes \dots \otimes R_{i_n i_1}.  \eea However, only the trivalent components associated to the ``wall-crossing triangle"; $\beta_{ikj}$ on one side and $\beta'_{ijk}$ on the other, enter the discussion in categorical wall-crossing.

\subsection{Homotopy Equivalence of BPS Data} We have discussed the construction of the BPS chain complexes \bea \{(R_{ij}, d_{ij}) \},\eea the contraction maps \bea \{K_{ij} \} \eea and the important vector encoding counts of rigid $\zeta$-instantons \bea \beta \in R_c.\eea We have noted however that the BPS complexes by themselves are not physical observables, only their homotopy equivalence class is. It is natural to try to extend the notion of homotopy equivalence from the BPS complexes, to the full categorical BPS data, namely to introduce a natural notion of homotopy equivalence for the contraction pairings and interior amplitudes. We briefly formulate such a notion in this sub-section.

\paragraph{} Suppose we are given another collection of BPS data $(\{S_{ij} \}, \{L_{ij} \}, \gamma) $ where $S_{ij}$ denote complexes $L_{ij}$ contaction maps, and $\gamma$ is now a Maurer-Cartan element of the $L_{\infty}$-algebra $S_c$, constructed from $S_{ij}$ and $L_{ij}$. We say that the BPS data \bea \big( \{R_{ij} \},  \{K_{ij} \}, \beta \big) \text{ and } \big( \{S_{ij} \},  \{L_{ij} \}, \gamma)\eea are \textbf{homotopy equivalent} if there are homotopy equivalences of chain complexes \bea f_{ij}: R_{ij} \rightarrow S_{ij} \eea that fit into a collection of maps \bea f_n: R_c^{\otimes n} \rightarrow S_c \eea with $f_1 $ being induced canonically from the collection $\{f_{ij}\}$ that together define an $L_{\infty}$-equivalence from $R_c$ to $S_c$. The maps $\{f_{ij}\}$ and the $L_{\infty}$-morphism $\{f_n\}$ must be such that the diagram \bea \xymatrixrowsep{4pc}\xymatrixcolsep{5pc}\xymatrix{ R_{ij} \otimes R_{ji} \ar[r]^{f_{ij} \otimes f_{ji}} \ar[rd]_{K_{ij}} & 
S_{ij} \otimes S_{ji} \ar[d]^{L_{ij}} 
\\
&\mathbb{Z}   }\eea commutes up to homotopy, and the Maurer-Cartan element transports naturally: \bea f(e^{\beta}) \sim \gamma,\eea where $\sim$ denotes gauge equivalence of Maurer-Cartan elements, defined in Appendix \ref{app}.

\paragraph{} The general philosophy of this paper is that we should only consider homotopy equivalence classes of the categorical BPS data. For example a D-term variation will only result in homotopy equivalent BPS data. The equivalence in this section can be viewed as a relaxation of the notion of strict isomorphism of categorical BPS data as defined in \cite{Gaiotto:2015aoa} section 4.1.1.

\section{Statement of Categorical Wall-Crossing}\label{statement}

\paragraph{Notation}Given an element $r_{ik} \otimes  r_{kj} \otimes r_{ji} \in R_{ik} \otimes R_{kj} \otimes R_{ji}$ we can define \bea M[r_{ik} \otimes  r_{kj} \otimes r_{ji}]: R_{ij} \otimes R_{jk} \rightarrow R_{ik} \eea by using the contraction maps \bea \label{cont} M[r_{ik} \otimes  r_{kj} \otimes r_{ji}](r'_{ij} \otimes r'_{jk}) = K_{ji}(r_{ji}, r'_{ij}) K_{kj}(r_{kj}, r'_{jk}) r_{ik}. \eea Similarly we define \bea M'[r_{ik} \otimes r_{kj} \otimes r_{ji}]:R_{ki} \rightarrow R_{kj} \otimes R_{ji} \eea by contracting the $ik$ factor using $K_{ik}$, and using the Koszul sign rule. Finally the natural product rule differential on a tensor product chain complex of the form as $R_{ij} \otimes R_{jk}$ is denoted as $d_{ijk}$: \bea d_{ijk} = d_{ij} \otimes 1 \pm 1 \otimes d_{jk}.\eea When we write $\pm$ it means we are not being precise about the exact sign.

\begin{figure}%[here!]
\centering
\includegraphics[width=1.0\textwidth]{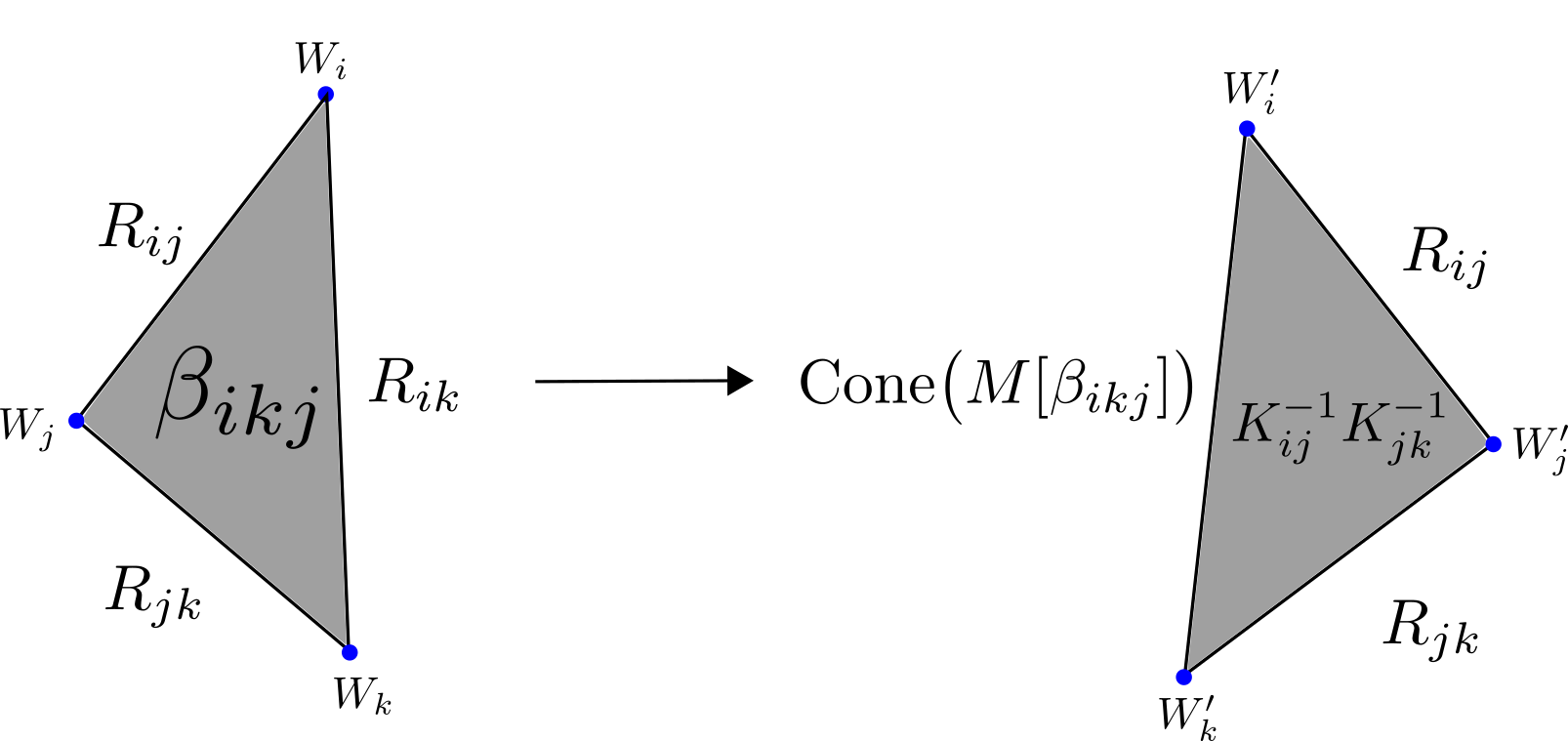}
\caption{Categorical wall-crossing summarized in the $W$-plane.}
\label{catcv}
\end{figure}

\paragraph{Marginal Stability Wall} Recall an $ijk$ wall of marginal stability is the locus where \bea \text{Im}(Z_{ij} \ov{Z}_{jk}) = 0. \eea See Figure \ref{catcv}.

\paragraph{Main Statement} Let \bea (R_{ij}, R_{jk}, R_{ik}, \beta_{ikj}) \eea be the chain complexes and interior amplitude component in a region where \bea \text{Im}(Z_{ij} \ov{Z}_{jk}) <0,\eea and \bea (R'_{ij}, R'_{jk}, R'_{ik}, \beta'_{ijk}) \eea be the chain complexes and interior amplitude component in a region where \bea \text{Im}(Z'_{ij}\ov{Z}'_{jk})>0.\eea Note that $\beta_{ikj}$ defines a \underline{chain}\footnote{This follows from $\beta$ being an interior amplitude, or equivalently, identity \eqref{chainmap}. The taut webs involved in this identity are the ones in Figure \ref{ends}.} map \bea M[\beta_{ikj}] : R_{ij} \otimes R_{jk} \rightarrow R_{ik},\eea and $\beta'_{ijk}$ defines a \underline{chain} map \bea M'[\beta'_{ijk}]: R'_{ik}[1] \rightarrow R'_{ij} \otimes R'_{jk}. \eea The categorical wall-crossing formula states that \bea R'_{ij} &\simeq& R_{ij}, \\ R'_{jk} &\simeq& R_{jk}, \\ R'_{ik} &\simeq& \text{Cone} \big(M[\beta_{ikj}]: R_{ij} \otimes R_{jk} \rightarrow R_{ik} \big). \eea Furthermore, letting $(P, Q)$ be the chain maps that implement the homotopy equivalence between the primed and unprimed sides, it states that the diagrams \bea \label{diagram1} \begin{CD} R'_{ik}[1] @>M'[\beta'_{ijk}]>> R'_{ij} \otimes R'_{jk} \\ @V P VV @VV P \otimes P V \\
\text{Cone} \big(M[\beta_{ikj}] \big)[1] @>>\pi> R_{ij} \otimes R_{jk} \end{CD}\eea and \bea \label{diagram2} \begin{CD} R'_{ik}[1] @>M'[\beta'_{ijk}]>> R'_{ij} \otimes R'_{jk} \\ @A Q AA @AA Q \otimes Q A \\
\text{Cone} \big(M[\beta_{ikj}] \big)[1] @>>\pi> R_{ij} \otimes R_{jk} \end{CD} \eea commute up to homotopy. 

\paragraph{} Equivalently, \bea \label{cat4} R_{ij} &\simeq& R'_{ij}, \\ \label{cat5} R_{jk} &\simeq& R'_{jk}, \\ \label{cat6} R_{ik} &\simeq& \text{Cone}(M'[\beta'_{ijk}]: R'_{ik}[1] \rightarrow R'_{ij} \otimes R'_{jk} ), \eea and letting $(S,T)$ be the chain maps implementing homotopy equivalence between the two sides, the diagrams \bea \label{diagram3} \begin{CD} R_{ij} \otimes R_{jk} @>M[\beta_{ikj}]>> R_{ik}, \\ @V T \otimes T VV @VV T V \\
R'_{ij} \otimes R'_{jk} @>>i> \text{Cone}\big(M'[\beta'_{ijk}] \big)  \end{CD}\eea and \bea \label{diagram4} \begin{CD} R_{ij} \otimes R_{jk} @>M[\beta_{ikj}]>> R_{ik}, \\ @A S \otimes S AA @AA S A \\
R'_{ij} \otimes R'_{jk} @>>i> \text{Cone}\big(M'[\beta'_{ijk}] \big)  \end{CD}\eea commute up to homotopy.

\paragraph{} These formulas are also sufficient to relate the contraction maps. Given chain complexes \bea (R_{ij}, R_{jk}, R_{ik}, \beta_{ikj}) \eea such that \bea \text{Im}(Z_{ij} \ov{Z}_{jk}) <0,\eea the dual complexes $(R_{kj}, R_{ji}, R_{ki})$ will be a triple such that \bea \text{Im}(Z_{kj} \ov{Z}_{ji})>0. \eea Therefore the formulas for going from $ \text{Im}(\cdots) >0 $ to $ \text{Im}(\cdots) <0 $ imply that \bea R'_{kj} &\simeq& R_{kj}, \\ R'_{ji} &\simeq& R_{ji}, \\ R'_{ki} &\simeq& \text{Cone} \big( M'[\beta_{ikj}]: R_{ki}[1] \rightarrow R_{kj} \otimes R_{ji} \big).\eea Note that there is a canonical degree $-1$ map \bea L : \text{Cone}\big(M[\beta_{ikj}] \big) \otimes \text{Cone} \big(M'[\beta_{ijk}]) \rightarrow \mathbb{Z} \eea given by \bea L = \begin{pmatrix}
0 && K_{ik} \\ 
K_{ij}\otimes K_{jk} && 0
\end{pmatrix}.\eea Denote the chain maps implementing the homotopy equivalence as $\widetilde{P}, \widetilde{Q}$. With this, the final part of categorical wall-crossing also determines the homotopy class of the contraction maps, by stating that the diagrams 

\bea \xymatrixrowsep{4pc}\xymatrixcolsep{5pc}\xymatrix{R_{ij} \otimes R_{ji} \ar[r]^{K_{ij}} & \mathbb{Z} \\
R'_{ij} \otimes R'_{ji} \ar[u]^{P \otimes \widetilde{P}} \ar[ur]_{K'_{ij}}} \eea

\bea \xymatrixrowsep{4pc}\xymatrixcolsep{5pc}\xymatrix{R_{jk} \otimes R_{kj} \ar[r]^{K_{jk}} & \mathbb{Z} \\
R'_{jk} \otimes R'_{kj} \ar[u]^{P \otimes \widetilde{P}} \ar[ur]_{K'_{jk}}} \eea

\bea \xymatrixrowsep{4pc}\xymatrixcolsep{5pc}\xymatrix{\text{C}(M)\otimes \text{C}(M') \ar[r]^{L} & \mathbb{Z} \\
R'_{ik} \otimes R'_{ki} \ar[u]^{P \otimes \widetilde{P}} \ar[ur]_{K'_{ik}}} \eea commute up to homotopy. In the above we have abbreviated $\text{Cone}\big(M[\beta_{ikj}]\big)$ and $\text{Cone}\big(M'[\beta'_{ijk}]\big)$ as $\text{C}(M)$ and $\text{C}(M')$ respectively. There will be similar diagrams with $(Q,\widetilde{Q})$.

\paragraph{Canonical Representatives} In practice given the chain complexes on one side, one wants to work with specific representatives within the homotopy equivalence class of chain complexes (and chain maps) for the other. There is a canonical choice for this. Suppose we treat the primed side as unknown. Then the canonical representatives for the primed complexes are \bea \label{cat1} R'_{ij} &=& R_{ij}, \\ \label{cat2} R'_{jk} &=& R_{jk}, \\ \label{cat3} R'_{ik} &=& \text{Cone} \big(M[\beta_{ikj}] \big).\eea By letting $P,Q$ to be identity maps, we can then make the diagrams \eqref{diagram1},\eqref{diagram2}, strictly commute by letting \bea M'[\beta'_{ijk}] = \pi,  \eea which is equivalent to saying that \bea \label{intprime} \beta'_{ijk} = K^{-1}_{ij} K^{-1}_{jk}.\eea The canonical representatives for the dual complexes are \bea R'_{kj} &=& R_{kj}, \\ R'_{ji} &=& R_{ji}, \\ R'_{ki} &=& \text{Cone}\big(M'[\beta_{ikj}] \big) \eea and one can then set the contraction maps to be \bea K'_{ij} &=& K_{ij}, \\ K'_{jk} &=& K_{jk}, \\ K'_{ik} &=& \begin{pmatrix}
0 && K_{ik} \\ 
K_{ij} \otimes K_{jk} && 0
\end{pmatrix}.\eea 

\paragraph{} Figure \ref{catcv} summarizes the categorical wall-crossing formula for going from a point in parameter space with $\text{Im}(Z_{ij} \ov{Z}_{jk})<0$ to a point where $\text{Im}(Z_{ij} \ov{Z}_{jk}) >0$ from the perspective of the $W$-plane. The formulas and the figure summarizing the specific representatives in the inverse move would look similar. These straightforward details are left for the reader.

\paragraph{Remark: Consistency Check} A consistency check our formulas must pass is whether jumping from the negative side of the wall of marginal stability where $\text{Im}(Z_{ij} \ov{Z}_{jk})<0$ to the positive side where $\text{Im}(Z_{ij} \ov{Z}_{jk})>0$ and then jumping back to the negative side is equivalent to doing nothing. We work with the canonical representatives. Starting from the complex $R_{ik}$ the wall-crossing formula says that \bea R'_{ik} = \text{Cone} \big(M[\beta_{ijk}]: R_{ij} \otimes R_{jk} \rightarrow R_{ik} \big),\eea and \bea\beta'_{ijk} = K_{ij}^{-1} K_{jk}^{-1}.\eea Jumping back to the right side, gives us \bea R''_{ik} = \text{Cone}\big(M'[K^{-1}_{ij} K^{-1}_{jk}] : \text{Cone}\big(M[\beta_{ikj}] \big)[1] \rightarrow R_{ij} \otimes R_{jk} \big).\eea But  \bea M'[\beta'_{ijk}] = \pi \eea and therefore we have \bea R''_{ik} &=& \text{Cone} \big(\pi: \text{Cone}\big(M[\beta]: R_{ij} \otimes R_{jk} \rightarrow R_{ik})[1] \rightarrow R_{ij} \otimes R_{jk} \big) \\ &=& \text{Cyl} \big(M[\beta_{ikj}]:R_{ij} \otimes R_{jk} \rightarrow R_{ik} \big) \\ &\simeq& R_{ik}.\eea The cylinder construction of homological algebra, used above is described in Appendix \ref{hom}. Therefore we end up with a complex canonically homotopy equivalent to the original complex. A similar check can be performed for $\beta''_{ikj}$. One shows that the diagram \bea 
\xymatrix{
    R_{ij} \otimes R_{jk}  \ar[dr]_{M[\beta''_{ikj}]} \ar[r]^{M[\beta_{ikj}]} & R_{ik} \ar[d]^{i} \\
                          & \text{Cyl}\big(M[\beta_{ikj}] \big) } \eea commutes up to homotopy. This shows clearly the need to work at the level of homotopy equivalence.
    
\paragraph{} In the next two sections we show how these conditions word-for-word are the homotopy equivalence of $A_{\infty}$ categories constructed at a point where $\text{Im}(Z_{ij} \ov{Z}_{jk})>0$ compared to a point where $\text{Im}(Z_{ij} \ov{Z}_{jk})<0$.

\section{$\zeta$-instantons and Brane Categories} \label{brane}

\subsection{Bare Thimble Category} While the chain complex $R_{ij}$ categorifies $\mu_{ij}$, categorification of $\widehat{\mu}_{ij}$ leads to more interesting structure. The correct viewpoint will be that $B$ must be upgraded to a category, and $\widehat{\mu}_{ij}$ will be categorified to vector spaces of morphisms. 

\paragraph{} The construction of the ``bare" thimble category $\widehat{R}^{\text{bare}}$ proceeds as follows.

\paragraph{Objects} The objects are an ordered collection of thimbles \bea \mathfrak{T}_1, \dots, \mathfrak{T}_n,\eea one for each critical point $i \in \text{Crit}(W)$. They are ordered by $\text{Im}(-W)$ so that $i > j$ if $\text{Im}(W_i) < \text{Im}(W_j)$.

\begin{figure}%[here!]
\centering
\includegraphics[width=.8\textwidth]{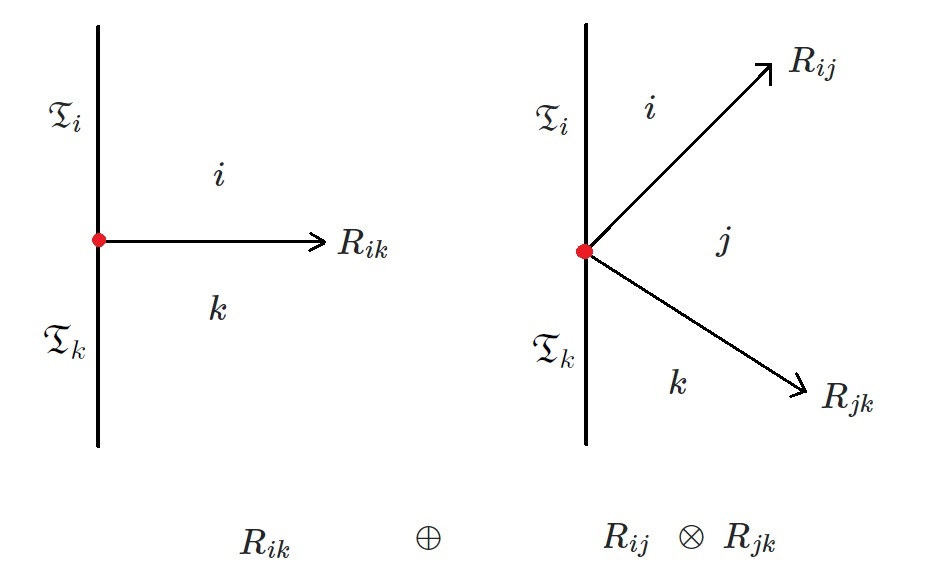}
\caption{Contribution of some half-plane fans to $\hp(\mft_i, \mft_k)$}
\label{hpfans}
\end{figure}

\paragraph{Morphisms} The morphisms are given as follows. In order to define \footnote{$\text{Hop}(A,B) := \text{Hom}^{\text{opp}}(A,B) = \text{Hom}(B,A)$} \bea \widehat{R}_{ij} := \hp(\mft_i, \mft_j) \eea we look at all half-plane fans with ``top" vacuum $i$ and ``bottom" vacuum $j$. To an edge separating $i$ and $j$ assign the vector space $R_{ij}$ and take the (ordered) tensor product along each edge. Thus to each half-plane fan $F_{ij}$ of this type we assign a vector space $R_{F_{ij}}$. The morphism space is then defined by taking direct sums over all $F_{ij}$ half-plane fans \bea \wR_{ij} = \bigoplus_{F_{ij}} R_{F_{ij}} .\eea See Figure \ref{hpfans} for an example of a morphism space where two fans contribute. Note that \bea \widehat{R}_{ii} = \hp(\mft_i, \mft_i) = \mathbb{Z}.\eea If there are no half-plane fans then \bea \widehat{R}_{ij} = 0 ,\eea so that the objects $\{\mathfrak{T}_1, \dots, \mathfrak{T}_n \}$ are an exceptional collection; the matrix of morphism spaces $\widehat{R}_{ij}$ is an upper-triangular matrix with $\mathbb{Z}$ on the diagonal.

\paragraph{Compositions} An associative composition law \bea \label{naivecomp} m_{ijk}: \wR_{ij} \otimes \wR_{jk} \rightarrow \wR_{ik} \eea is given simply by looking at whether $F_{jk}$ can be placed below $F_{ij}$ to form a fan $F_{ik}$. If so, we take the tensor product of the vectors in $R_{F_{ij}}$ and $R_{F_{jk}}$ to get a vector in $R_{F_{ik}}$. If not, we set it equal to zero. 

\paragraph{Differentials} Finally the differential $\widehat{d}_{ij}$ on \bea \widehat{R}_{ij} = R_{ij} \oplus( R_{ik} \otimes R_{kj}) \oplus \dots \eea will be inherited from the differentials on the complexes in the obvious way \bea \label{naivediff} \widehat{d}_{ij} = d_{ij} \oplus( d_{ik} \otimes 1 + 1 \otimes d_{kj}) \oplus \dots .\eea

\paragraph{Remark}The differential-graded algebra \bea \text{End}\big(\oplus_{i} \mft_i) = \bigoplus_{i,j} \widehat{R}_{ij}\eea in which the algebra multiplication is specified by the morphisms as defined above, as explained in Appendix \ref{app}, carries the same information as the category $\widehat{R}$ and so we often use the terms algebra and category interchangeably in what follows.

\subsection{Interior Amplitudes and Deformations of $\widehat{R}$} While $\widehat{R}^{\text{bare}}$ indeed gives $\widehat{\mu}$ as its matrix of Euler characters, the cohomology space $H^{\bullet}(\widehat{R}_{ij}, \widehat{d}_{ij})$ is not very physically meaningful. In particular, it is not isomorphic to the space of boundary BPS local operators at a $\mft_i$-$\mft_j$ brane junction, like we would want it to be. The reason for this is similar to the failure of our naive categorification: we have not taken into account all $\zeta$-instantons. In particular these $\zeta$-instantons will correct the differential \eqref{naivediff} and the composition law \eqref{naivecomp} described in the previous section. 

\paragraph{} The precise way to take $\zeta$-instantons into account again uses the interior amplitude $\beta$. Similar to how one can use taut webs with $n$ vertices to define $L_{\infty}$-maps, \bea \rho(\mathfrak{t}^{(n)}): S^n R_c \rightarrow R_c, \eea we can use taut half-plane webs with $p$ boundary vertices and $q$ bulk vertices to define maps \bea \rho(\mathfrak{t}_{\mathcal{H}}^{(p,q)}): (\widehat{R})^{\otimes p}\otimes (R_c)^{\otimes q} \rightarrow \widehat{R} \eea which satisfy the $LA_{\infty}$-axioms \cite{Gaiotto:2015aoa} (these are also known as the axioms of an open-closed homotopy algebra, see \cite{Kajiura:2004xu}). We now make use of the
\paragraph{Theorem} Suppose $(A,L)$ is an open-closed homotopy algebra with structure maps \bea m_{k,l}: A^{\otimes k} \otimes L^{\otimes l} \rightarrow A, \,\,\,\,\,\, k \geq 1, l \geq 0 \eea and suppose $\gamma \in L$ is a Maurer-Cartan element for the $L_{\infty}$ algebra $L$. Then the collection of maps \bea m_k[\gamma]: A^{\otimes k} \rightarrow A,\eea defined by \bea \label{newainf} m_{k}[\gamma](-, \dots, -) := \sum_{l \geq 0} \frac{1}{l!}m_{k,l}(- \dots, -, \gamma^{\otimes l}) \eea give a (new) $A_{\infty}$-structure on $A$.

\paragraph{} Thus we use the $\zeta$-instanton counting element $\beta$ to deform the dg-category $\widehat{R}^{\text{bare}}$ to an $A_{\infty}$-category denoted by $\widehat{R}[X,W]$. The deformed category $\widehat{R}[X,W]$ is proposed as the physical brane category of the Landau-Ginzburg model associated to the pair $(X,W)$. In particular, we correct the differential $\widehat{d}_{ij}$ to $\widehat{d}_{ij}[\beta]$ via \eqref{newainf} with $k=1$, so that the cohomology \bea H^{\bullet}(\widehat{R}_{ij}, \widehat{d}_{ij}[\beta])\eea is isomorphic to the space of $\frac{1}{2}$-boundary BPS local operators at a $(\mathfrak{T}_i, \mft_j)$-brane junction. In addition $k=2$ of \eqref{newainf} also modifies the bilinear composition \eqref{naivecomp}. As a result of \eqref{newainf} higher operations \bea \{m_k[\beta]\}_{k > 2} \eea are also introduced. Together these operations turn $\widehat{R}[X,W]$ into a genuine $A_{\infty}$-category. 

\section{Homotopy Equivalence of Brane Categories} \label{homotopy} The categorical wall-crossing constraint is formulated as follows. 

\paragraph{Categorical Wall-Crossing Constraint} Suppose $W$ and $W'$ are superpotentials on different sides of a wall of marginal stability. Then the $\beta$-deformed thimble categories on either side of the wall are homotopy equivalent \bea \widehat{R}[X,W] \simeq \widehat{R}'[X,W']\eea as $A_{\infty}$-categories.

\paragraph{} We now relate our categorical wall-crossing formulas with the categorical wall-crossing constraint. First we construct the left and right $\{i,j,k\}$-subcategories. As an instructive first check, we verify that the canonical representatives indeed give homotopy equivalent categories. Finally we unpack the axioms for $A_{\infty}$ equivalence and show how the general statement follows. 

\subsection{Left Configuration} \label{leftconfig} Let us first construct the $\{i,j,k\}$ sub-algebra of $\widehat{R}$ for the configuration on the left of Figure \ref{cvform}. The soliton complexes are \bea (R_{ij},d_{ij}), (R_{ik}, d_{ik}), (R_{jk}, d_{jk}).\eea Because there are no half-plane fans with more than one edge emanating from the boundary, the morphism spaces are simply \bea \widehat{R}_{ij} &=& R_{ij}, \\ \widehat{R}_{jk} &=& R_{jk}, \\  \widehat{R}_{ik} &=& R_{ik}.\eea In the undeformed algebra, there are no non-trivial multiplications. 

\begin{figure}
\centering
\includegraphics[width=.3\textwidth]{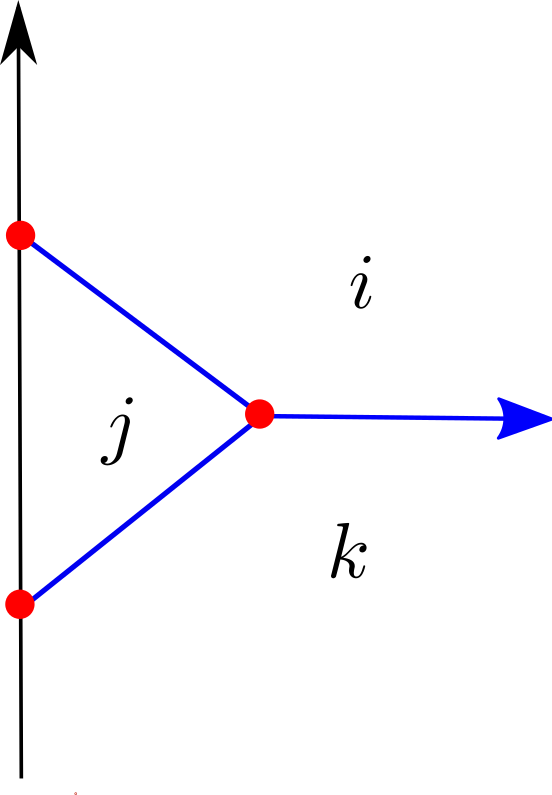}
\caption{A taut half-plane web which contributes to a non-trivial bilinear product, by inserting the interior amplitude $\beta$ in the bulk vertex.}
\label{hpweb1}
\end{figure}

\paragraph{} Now consider the interior amplitude component \bea \beta_{ikj} \in R_{ik} \otimes R_{kj} \otimes R_{ji} ,\eea and consider the $\beta$-deformed algebra $\widehat{R}(X,W)$. In $\widehat{R}(X,W)$ we see that the taut half-plane web shown in Figure \ref{hpweb1} now gives rise to a non-trivial morphism \bea M[\beta_{ikj}]: \widehat{R}_{ij} \otimes \widehat{R}_{jk} \rightarrow \widehat{R}_{ik} \eea given precisely by \eqref{cont} applied to $\beta_{ikj}$.  The differential $\widehat{d}$ remains uncorrected. 

\paragraph{} The only $A_{\infty}$ axiom to check is that  \bea d_{ik}(M[\beta_{ikj}]( r_{ij}, r_{jk}) ) = M[\beta_{ikj}] (dr_{ij}, r_{jk}) \pm M[\beta_{ikj}](r_{ij}, dr_{jk})\eea which follows from $\beta_{ikj}$ being an interior amplitude component. 

\subsection{Right Configuration} \label{rightconfig}

\begin{figure}
\centering
\includegraphics[width=.3\textwidth]{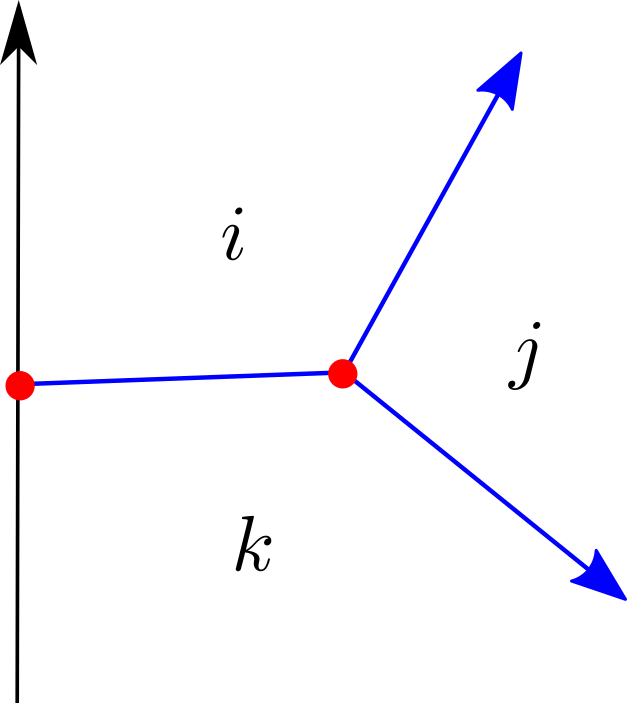}
\caption{A taut half-plane web which contributes to an off-diagonal element in the differential by inserting the interior amplitude $\beta'$ in the bulk vertex.}
\label{hpweb2}
\end{figure}

Suppose the BPS chain complexes on the right configuration are \bea (R'_{ij}, d'_{ij}), (R'_{jk}, d'_{jk}), (R'_{ik}, d'_{ik}).\eea There are now two half-plane fans of type $ik$, shown in Figure \ref{hpfans} with one and two edges emanating from the boundary vertex respectively. This gives that the morphism spaces are \bea \widehat{R}'_{ij} &=& R'_{ij}, \\ \widehat{R}'_{jk} &=& R'_{jk}, \\  \widehat{R}'_{ik} &=& R'_{ik} \oplus (R'_{ij} \otimes R'_{jk}). \eea Denote the interior amplitude on the right configuration to be \bea \beta'_{ijk} \in R'_{ij} \otimes R'_{jk} \otimes R'_{ki}.\eea %The deformed algebra $\widehat{R}'[\beta']$ now has a corrected differential $$\widehat{d}'[\beta']: \widehat{R}'_{ik} \rightarrow \widehat{R}'_{ik}$$ given by the inherited differentials on the diagonals, whereas the off-diagonal element is defined by the half-plane web of Figure \ref{hpweb2} by inserting $\beta_{ijk}$ in the bulk vertex. 
Writing an element of $\wR'_{ik}$ as a column vector $\begin{pmatrix}
r'_{ik} \\ r'_{ij} r'_{jk}
\end{pmatrix}$ the differential on $\wR'_{ik}$ is of the form \bea \widehat{d}'_{ik}[\beta'] &=& \begin{pmatrix}
d'_{ik} && 0 \\
M'[\beta'_{ijk}] && d'_{ijk}
\end{pmatrix} \eea where \bea M'[\beta'_{ijk}]: R'_{ik} \rightarrow R'_{ij} \otimes R'_{jk} \eea is a degree $+1$ map defined by Figure \ref{hpweb2}. Nilpotence of $\widehat{d}'_{ik}[\beta']$ holds if \bea d'_{ijk} M'[\beta'_{ijk}] +M'[\beta'_{ijk}]  d'_{ik}  = 0,\eea therefore we may equivalently view $M'[\beta'_{ijk}]$ as a chain map \bea M'[\beta'_{ijk}]: R'_{ik}[1] \rightarrow R'_{ij} \otimes R'_{jk} \eea and we can rewrite \bea \wR'_{ik} = \text{Cone}\big(M'[\beta'_{ijk}]: R'_{ik}[1] \rightarrow R'_{ij} \otimes R'_{jk} \big).\eea The only non-trivial multiplication map is \bea i: \wR'_{ij} \otimes \wR'_{jk} \rightarrow \wR'_{ik}\eea given by inclusion. The $A_{\infty}$ axiom says that $i$ is a chain map with respect to $d'_{ijk}$, the product rule differential on $R'_{ij} \otimes R'_{jk}$ and $\widehat{d}'_{ik} [\beta'] =  d_{M'} $ the mapping cone differential on $\widehat{R}'_{ik}$. 

\subsection{Canonical Representatives Satisfy Wall-Crossing Constraint}
\paragraph{} In this section we show that the canonical representatives \eqref{cat1}, \eqref{cat2}, \eqref{cat3}, \eqref{intprime} satisfy the categorical wall-crossing constraint.

\paragraph{Claim:} Suppose the primed complexes \bea (R'_{ij}, d'_{ij}), (R'_{jk}, d'_{jk}), (R'_{ik}, d'_{ik})\eea and interior amplitude \bea \beta'_{ijk} \in R'_{ij} \otimes R'_{jk} \otimes R'_{kj} \eea are given as in \eqref{cat1}, \eqref{cat2}, \eqref{cat3} and \eqref{intprime}. Then there is a functor  \bea T: \widehat{R} \rightarrow \widehat{R}'\eea which defines a quasi-isomorphism of $A_{\infty}$-categories. We call $T$ the wall-crossing functor.

\paragraph{Proof:} By virtue of the categorical wall-crossing statement, we have the primed morphism spaces \bea \widehat{R}'_{ij} &=& R_{ij}, \\ \widehat{R}'_{jk} &=& R_{jk}, \\ \widehat{R}'_{ik} &=& R_{ik} \oplus \big(R_{ij} \otimes R_{jk}\big)[-1]\oplus \big(R_{ij} \otimes R_{jk}\big).\eea The differentials deformed by the interior amplitude component $\beta'_{ijk}$ are of the form \bea \widehat{d}'_{ik}[\beta'] &=& \widehat{d}_{ik}, \\ \widehat{d}'_{jk}[\beta'] &=& \widehat{d}_{jk}, \\ \label{cyldiff} \widehat{d}'_{ik}[\beta']&=&   \begin{pmatrix}
d_{ik} && M[\beta_{ikj}] && 0 \\
0 && d^{[-1]}_{ijk} && 0 \\
M'_1[\beta'_{ijk}] && M'_2[\beta'_{ijk}] && d_{ijk}
\end{pmatrix} ,\eea where $M[\beta_{ikj}]$ was defined as before and \bea  M'_1[\beta'_{ijk}]&:& R_{ik} \rightarrow R_{ij} \otimes R_{jk}, \\ M'_2[\beta'_{ijk}]&:& \big(R_{ij} \otimes R_{jk}\big)[-1] \rightarrow R_{ij} \otimes R_{jk}\eea are the different components of the maps defined by Figure \ref{hpweb2} by inserting $\beta'$ in the bulk vertex.   The functor $T$ can then be defined as follows. On objects we simply have the identity map. On morphism spaces we define \bea T_1: \widehat{R}_{ij} \rightarrow \widehat{R}'_{ij}, \\ T_1 : \widehat{R}_{kj} \rightarrow \widehat{R}'_{kj}\eea as identity maps, whereas \bea T_1: \widehat{R}_{ik} \rightarrow \widehat{R}'_{ik}\eea is defined as inclusion, \bea T_1(r_{ik}) = \begin{pmatrix} r_{ik} \\0 \\0 \end{pmatrix} \eea Furthermore \bea T_2: \widehat{R}_{ij} \otimes \widehat{R}_{jk} \rightarrow \widehat{R}'_{ik} \eea is again defined to be inclusion, but into the summand with shifted degree, \bea T_2(r_{ij} r_{jk}) = \begin{pmatrix} 0 \\ (r_{ij}r_{jk})^{[-1]} \\ 0 \end{pmatrix}.\eea Indeed $(T_1,T_2)$ have degrees $(0,-1)$ respectively. The higher maps $T_n$ are set to be zero for $n \geq 3$. 

\paragraph{} First we have to show the axioms of an $A_{\infty}$-morphism are satisfied. Here there are just two axioms to check. At $n=1$ we have to check if $T_1$ is a chain map.  The only non-trivial check is on the $ik$-component of $T_1$ and it follows that we have a chain map from the form of the differential \eqref{cyldiff}. At $n=2$ we must check \bea \begin{split} T_1\big(m_2(r_{ij}, r_{jk}) \big) - m'_2\big(T_1(r_{ij}), T_1(r_{jk})\big) \\ = T_2(r_{ij}, d r_{jk}) \pm T_2(d r_{ij}, r_{jk}) \pm d'(T_2(r_{ij}, r_{jk})). \end{split}  \eea This follows from the following simplification for the expression of $\widehat{d}'_{ik}[\beta']$. The explicit form of $\beta'$  \bea \beta'_{ijk} = K^{-1}_{ij} K^{-1}_{jk}, \eea from \eqref{intprime}, implies that the off-diagonal maps $M_{1,2}'[\beta'_{ijk}]$ are \bea M'_1[\beta_{ijk}'] &=& 0 \\ M'_2[\beta_{ijk}'] &=& \text{id}. \eea Note that the identity map $M'_2$ has degree $+1$ due to the degree shift on the domain. Thus we can rewrite the differential as \bea \widehat{d}'_{ik}[\beta']&=&   \begin{pmatrix}
d_{ik} && M[\beta_{ikj}] && 0 \\
0 && d^{[-1]}_{ijk} && 0 \\
0 && \text{id} && d_{ijk}
\end{pmatrix} .\eea Using this expression for $d'$ on the right hand side, the axiom easily follows. Thus $T$ defines an $A_{\infty}$-functor.

\paragraph{} Finally, we must show that the wall-crossing functor $T$ is a quasi-isomorphism. Again this is non-trivial only on the $ik$-component. The simplification of $\widehat{d}'[\beta']$ in fact allows us to relate this to the mapping cylinder construction: similar to \eqref{cat3} one can recognize $\widehat{R}'_{ik}$ as the mapping cone of the projection map \bea \pi: R'_{ik}[1] = \text{Cone}\big(M[\beta_{ikj}] \big) [1] \rightarrow R_{ij} \otimes R_{jk}.\eea In other words we can rewrite \bea \widehat{R}'_{ik} = \text{Cyl}\big(M[\beta_{ikj}]\big).\eea 

\paragraph{} Applying the \textbf{Proposition} about mapping cylinders from Appendix \ref{hom} to $f = M[\beta_{ikj}]$ yields that $T$ is a quasi-isomorphism. $\qed$

\paragraph{Remark} Two $A_{\infty}$-algebras are homotopy equivalent if and only if they are quasi-isomorphic (this is a theorem of Prout\'e, \cite{Proute84}). We can thus say \bea \widehat{R}[X,W] \simeq \widehat{R}'[X,W'] \eea where $\simeq$ is meant to be understood as homotopy equivalence.

\subsection{Homotopy Equivalence $\implies$ Categorical WCF} Finally we come to the main claim. \paragraph{Claim}  The categorical wall-crossing constraint, namely the homotopy equivalence of $A_{\infty}$-categories \bea \wR[X,W] \simeq \wR[X,W']\eea implies the categorical wall-crossing formula \bea R'_{ij} &\simeq& R_{ij}, \\ R'_{jk} &\simeq& R_{jk}, \\ R'_{ik} &\simeq& \text{Cone}\big(M[\beta_{ikj}]: R_{ij} \otimes R_{jk} \rightarrow R_{ik} \big).\eea 

\paragraph{}Consider first the $A_{\infty}$ morphism \bea T: \widehat{R}[X,W] \rightarrow \widehat{R}'[X,W']. \eea This in particular means that there are chain maps \bea T_1 &:& \widehat{R}_{ij} \rightarrow \widehat{R}'_{ij}, \\ T_1 &:& \widehat{R}_{jk} \rightarrow \widehat{R}'_{jk}, \\ T_1 &:& \widehat{R}_{ik} \rightarrow \widehat{R}'_{ik} .\eea We showed in \ref{leftconfig}, \ref{rightconfig} that the hatted and un-hatted spaces coincide as chain complexes except for $\widehat{R}'_{ik}$ which is of the form \bea \widehat{R}'_{ik} = \text{Cone}(M'[\beta'_{ijk}]: R'_{ik}[1] \rightarrow R'_{ij} \otimes R_{jk}). \eea Therefore we have chain maps  \bea T_1 &:& R_{ij} \rightarrow R'_{ij}, \\ T_1 &:& R_{jk} \rightarrow R'_{jk}, \\ T_1 &:& R_{ik} \rightarrow \text{Cone}\big(M'[\beta'_{ikj}]\big) .\eea In addition the $A_{\infty}$-morphism $T$ provides a degree $-1$ map \bea T_2: \widehat{R}_{ij} \otimes \widehat{R}_{jk} \rightarrow \widehat{R}'_{ik} = \text{Cone} \big(M'[\beta'_{ikj}]\big) \eea such that the second $A_{\infty}$-morphism axiom, \eqref{secondaxiom}, which in the present case reads \bea \begin{split} T_1\big(M[\beta_{ikj}](r_{ij}, r_{jk}) \big) \pm M_2'\big(T_1(r_{ij}), T_1(r_{jk}) \big) \\ = \widehat{d}'_{ik}[\beta'] T_2(r_{ij}, r_{jk}) \pm T_2 (dr_{ij}, r_{jk}) \pm T_2(r_{ij}, dr_{jk}), \end{split}\eea holds.  We showed that $M_2'$ the bilinear multiplication \bea M_2': \widehat{R}'_{ij} \otimes \widehat{R}'_{jk} \rightarrow \text{Cone}\big(M'[\beta'_{ijk}] \big) \eea is given simply by the inclusion map $i$ in \ref{rightconfig}. We therefore see that the conceptual way to interpret this axiom is that it is saying that the square \bea \begin{CD} 
R_{ij} \otimes R_{jk} @> M[\beta_{ikj}] >> R_{ik} \\
@V T_1\otimes T_1 VV  @VV T_1 V  \\
R'_{ij} \otimes R'_{jk} @>i>> \text{Cone}\big(M'[\beta'_{ijk}]\big) \end{CD} \eea commutes up to homotopy  \footnote{Note that the the compositions are chain maps \bea \begin{CD} R_{ik} \otimes R_{jk} @> i\circ T_1 \otimes T_1 > T_1 \circ M[\beta] > \text{Cone}\big(M'[\beta'_{ijk}]),\end{CD} \eea } \bea i(T_1 \otimes T_1 ) \simeq T_1(M[\beta_{ikj}] ) \eea with $T_2$ providing the chain homotopy. This condition is precisely \eqref{diagram3}.

\paragraph{}Let the morphism in the other direction be \bea S: \widehat{R}'[X,W'] \rightarrow \widehat{R}[X,W]\eea which in particular says that we have chain maps \bea S_1 &:& R'_{ij} \rightarrow R_{ij}, \\ S_1 &:& R'_{jk} \rightarrow R_{jk} , \\ S_1 &:& \text{Cone} \big( M'[\beta'_{ijk}] \big) \rightarrow R_{ik} \eea that provide homotopy inverses to the $T_1$'s. $S$ also provides us with a degree $-1$ map \bea S_2: R'_{ij} \otimes R'_{ik} \rightarrow R_{ik} \eea that satisfies the second $A_{\infty}$ axiom which in this case says that the the square \bea \begin{CD} 
R'_{ij} \otimes R'_{jk} @>i>> \text{Cone}(M'[\beta'_{ijk}]) \\
@VS_1 \otimes S_1 VV @VVS_1V \\
R_{ij} \otimes R_{jk} @>>M[\beta_{ikj}]> R_{ik}
\end{CD} \eea commutes up to homotopy, with $S_2$ providing the chain homotopy. \bea \begin{CD} R'_{ij} \otimes R'_{jk} @> S_1 \circ i > M[\beta] \circ S_1 \otimes S_1 > R_{ik} \end{CD} . \eea In particular the existence of $(S_1, T_1)$ implies that \bea R_{ik} &\simeq& R'_{ik}, \\ R_{jk} &\simeq& R'_{jk}, \\ R_{ik} &\simeq& \text{Cone}\big(M'[\beta'_{ijk}]: R'_{ik}[1] \rightarrow R'_{ij} \otimes R'_{jk}\big),\eea which are precisely the homotopy equivalences \eqref{cat4}, \eqref{cat5}, \eqref{cat6} asserted in the categorical wall-crossing statement. The statement that these are homotopy equivalences follows from the definition of homotopy equivalence of $A_{\infty}$-algebras. Similarly the commutative square above is precisely \eqref{diagram4}. 

\paragraph{} Finally we use the Triangularity Lemma from Appendix \ref{hom}. 

\paragraph{} We found above that \bea R_{ik} \simeq \text{Cone}\big(M'[\beta'_{ijk}] : R'_{ik}[1] \rightarrow R'_{ij} \otimes R'_{jk} \big)\eea so an application of the Triangularity Lemma implies that \bea R'_{ik} \simeq \text{Cone}( S_1 \circ i : R'_{ij} \otimes R'_{jk} \rightarrow R_{ik}).\eea Next we recall that the $A_{\infty}$-axiom for $S_2$ implies that \bea S_1 \circ i \simeq M[\beta_{ikj} ] \circ (S_1 \otimes S_1) \eea and so their cones are homotopy equivalent. This gives \bea R'_{ik} \simeq \text{Cone}\big(M[\beta_{ikj}] \circ (S_1 \otimes S_1): R'_{ik} \otimes R'_{jk} \rightarrow R_{ik} \big).\eea Finally since \bea S_1 &:& R'_{ij} \rightarrow R_{ij}, \\ S_1&:& R'_{jk} \rightarrow R_{jk} \eea are individually homotopy equivalences, so is \bea S_1 \otimes S_1: R'_{ij} \otimes R'_{jk} \rightarrow R_{ij} \otimes R_{jk}.\eea Therefore the latter part has a trivial mapping cone and can be ``factored out" to conclude that \bea R'_{ik} \simeq \text{Cone} \big(M[\beta_{ikj}]: R_{ij} \otimes R_{jk} \rightarrow R_{ik} \big),\eea the result to be shown.

\section{The Fermion Degree of a $\zeta$-instanton} \label{maslov} Recall that a $\zeta$-instanton with boundary conditions labeled by the triple of solitons \bea \bphi = (\phi_{ik}, \phi_{kj}, \phi_{ji}) \eea that occupy the edges of an $ikj$ wall-crossing triangle contributes to the differential in a categorical wall-crossing process if and only if \bea F(\phi_{ik} \otimes \phi_{kj} \otimes \phi_{ji}) = 2. \eea Therefore it is quite important to determine the degree of a given gradient polygon.

\paragraph{} By definition the Fermion number is the index of the Dirac operator \bea \mathcal{D}_{\zeta}: \Gamma\big( \phi^*(TX) \big) \rightarrow \Gamma \big(\phi^*(TX) \big)\eea given by  \bea \mathcal{D}_{\zeta}  = \begin{pmatrix}
\delta^I_J D_{\bar{s}}^{(1,0)} && 0 \\
0 && \delta^{\bar{I}}_{\bar{J}} D_{s}^{(0,1)}
\end{pmatrix} - \begin{pmatrix}
0 && \frac{\zeta}{2} g^{I \bar{K}} D_{\bar{K}} \del_{\bar{J}} \ov{W} \\
\frac{\zeta^{-1}}{2} g^{\bar{I} K} D_K \del_J W && 0
\end{pmatrix}. \eea in the background of a $\zeta$-instanton $\phi$ with $\bphi$ boundary conditions. Clearly such an index will be difficult to compute if we work directly with $\mathcal{D}$ \footnote{Moreover the question of whether $\mathcal{D}$ is even Fredholm is a very delicate one, \cite{Carey}}. However  a Maslov index type construction, described in \cite{Kapranov:2014uwa}, gives a more geometric prescription to obtain a well-defined integer $d(\bphi)$ which is expected to agree with the index of $\mathcal{D}$ up to an overall shift. It would be interesting to prove the equality of $d(\bphi)$ with the index of $\mathcal{D}$, but this would take us too far afield in the present paper. We proceed assuming the equality holds and use the geometric prescription in what follows. The Maslov index construction also assumes that $X$ is equipped with a nowhere vanishing holomorphic volume form $\Omega$.

\paragraph{} Starting from a convex gradient polygon \bea \bphi = (\phi_{i_0 i_1}, \dots \phi_{i_n i_0}) \eea the Maslov index prescription gives us $d(\bphi) \in \mathbb{Z}$ as follows. The main step consists of assigning to the gradient polygon $\bphi$ a (homotopy class of a) loop in the Lagrangian Grassmannian of $X$, \bea \text{Lag}(TX) = \{(p, E)| p \in X,\,\,\, E \text{ Lagrangian subspace of } T_p X \},\eea constructed as follows.

\paragraph{} First to each soliton $\phi_{ij}$ we associate an open path $\gamma$ in $\text{Lag}(TX)$ simply by taking a point $p$ along the soliton trajectory and assigning to it the Lagrangian subspace \bea T_{p}L_{i}(\zeta_{ij})\subset T_p X\eea as the fiber. Let $\gamma_k$ denote the open path assigned to $\phi_{i_{k-1} i_{k}}$ in this way. One notices that the endpoint of $\gamma_k$ and the starting point of $\gamma_{k+1}$ have the same base point, the $k$th critical point $i_k$, but the Lagrangians fibers differ. The endpoint of $\gamma_k$ has fiber \bea \ell_k := T_{i_k} L_{i_{k-1}}(\zeta_{i_{k-1} i_k}) \eea whereas the starting point of $\gamma_{k+1}$ has the fiber \bea \ell_{k+1} := T_{i_{k}} L_{i_k}(\zeta_{i_{k} i_{k+1}}).\eea  $\ell_{k}, \ell_{k+1}$ are Lagrangians living in the same ambient space $T_{i_k} X$. Between any two Lagrangian subspaces $L_1, L_2$ in a symplectic vector space $V$, there is a canonical homotopy class of paths $\kappa_{L_1, L_2}$ in $\text{Lag}(V)$ that connects these points, known as the \textbf{symplectic bridge} \footnote{This is also known as the \textbf{canonical short path}, see for instance \cite{Auroux2013}.} connecting $L_1$ and $L_2$. For instance if $\text{dim}(V)=2$, the Lagrangians are specified by points $\theta_1, \theta_2$ in $\mathbb{RP}^1 \cong S^1$ and $\kappa_{\theta_1, \theta_2}$ is the circular arc going in the \underline{counter-clockwise} direction between these two angles. Therefore there is a well-defined way to connect the open path $\gamma_k$ to $\gamma_{k+1}$. Going around the gradient polygon by gluing adjacent open paths via symplectic bridges, one obtains a loop in $\text{Lag}(TX)$. 

\paragraph{} Next we need to define a winding number of the loop $\gamma$. Let $\bar \gamma$ be the loop in $X$ obtained by projecting $\gamma$ to $X$. Thus, if $\bar\gamma(t) = p \in X$ then $\gamma(t) \subset T_pX$ is a maximal Lagrangian subspace.  Let $2n$ denote the rank of $TX$ considered as a real vector bundle over $X$. Then $\gamma(t)$ is a real vector space of dimension $n$. The $n^{\text{th}}$ exterior product of this space is a real line associated to the point $p$. Now, recall that $TX$ can also be considered to be a complex vector bundle of rank $n$. Therefore, the $n^{\text{th}}$ exterior power of $TX$ as a complex vector bundle is a complex line associated to $p$. Indeed, this is the fiber of the canonical bundle at $p$, denoted $\mathcal{K}_p$. Note that $\Lambda^n \gamma(t) \subset \mathcal{K}_p$ is a real line 
inside a complex line. Finally we use $\Omega$ to trivialize the canonical bundle and therefore get a 
real line $\ell_p \subset \mathbb{C}$. That is, to the loop $\gamma:S^1 \to {\rm Lag}(TX)$ we associate a loop in   $\text{Lag}(\mathbb{C}) = \mathbb{RP}^1 \cong S^1$. All-in-all we get a map \bea \psi(\bphi): S^1 \rightarrow \text{Lag}(\mathbb{C}) \cong S^1. \eea  The integer $d(\bphi)$ is defined to be the winding number of $\psi(\bphi)$. The fermion number is then \bea F(\bphi) = d(\bphi) + 1.\eea

\paragraph{} We illustrate the computation of $d(\bphi)$ in some examples.

\subsection{Gradient Polygons in $\mathbb{C}$}\label{polygonmoduli} Suppose our target space is the complex plane, and say for simplicity that the solitons trace out straight lines so that the gradient polygon $\bphi = (\phi_{i_0 i_1}, \dots \phi_{i_n i_0})$ traces out the boundary of an $(n+1)$-gon. This boundary can be clockwise or counter-clockwise oriented and we analyze each case.

\begin{figure}
\centering
\includegraphics[width=1.0\textwidth]{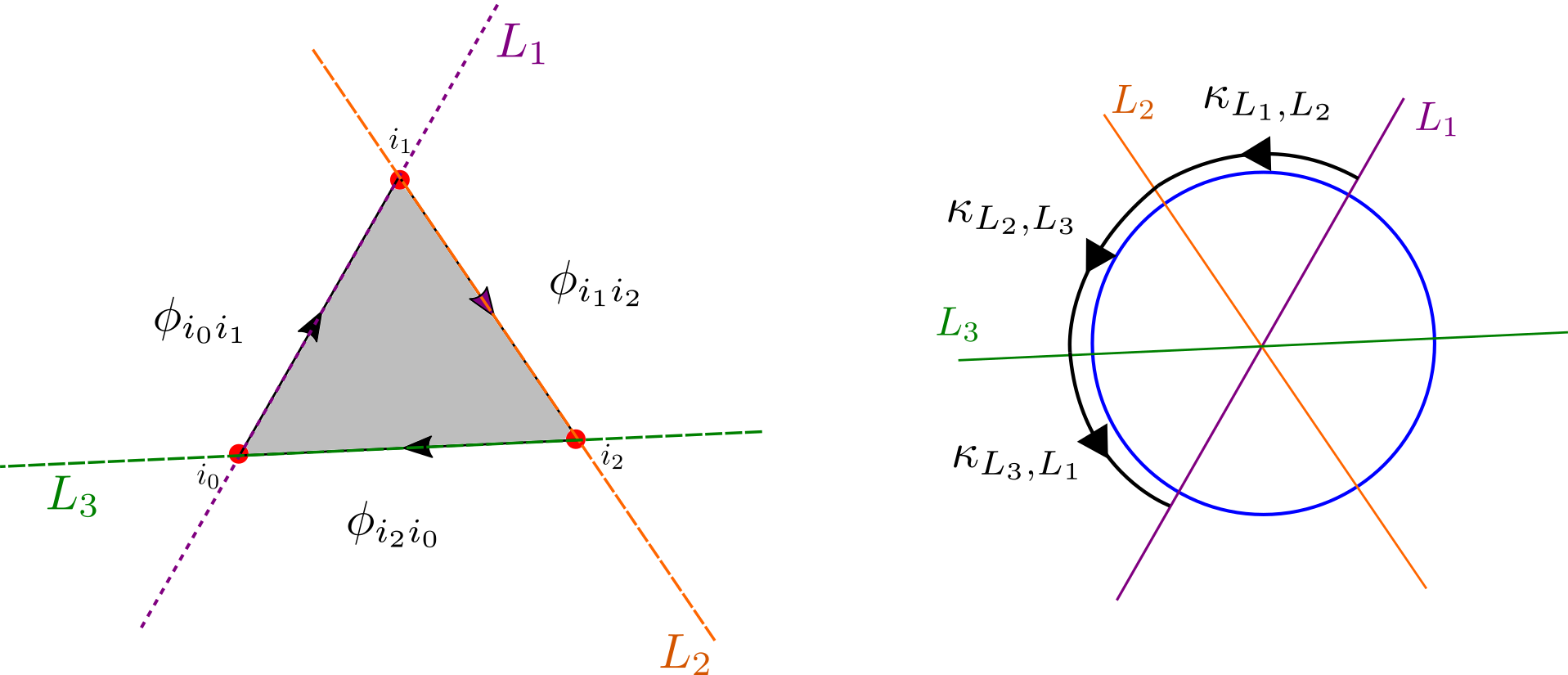}
\caption{The left shows the gradient polygon $\bphi = (\phi_{i_0 i_1}, \phi_{i_1 i_2}, \phi_{i_2 i_0})$ assumed to trace out straight lines on the complex plane. The dashed lines depict the Lagrangians tangent to these solitons. On the right we show the symplectic bridges $\kappa_{L_i, L_{i+1}}$ connecting these Lagrangians. The winding number of the total path in $\text{Lag}(\mathbb{C}) = S^1/\mathbb{Z}_2$ is $+1$, therefore $d(\phi) = 1$.}
\label{maslov1}
\end{figure}

\paragraph{} For the case of clockwise oriented boundaries, the tangent Lagrangian does not vary along the soliton. The symplectic bridge between $\phi_{i_{k-1} i_k}$ and $\phi_{i_{k} i_{k+1}}$ chooses to take the route that takes $\theta_k$ radians where $\theta_k$ is an internal angle of the polygon. Adding up these angles gives one a total winding number in $S^1/\mathbb{Z}_2$ of  \bea d(\bphi) &=& \frac{\big((n+1)-2\big)\pi}{\pi} \\ &=& n-1\eea where in the first equality we divide by $\pi$ (not $2\pi$) because of the $\mathbb{Z}_2$ quotient. See Figure \ref{maslov1} for the case of $n=2$.

\paragraph{} For counter-clockwise oriented (convex) polygons \footnote{We don't know any examples of $W(\phi)$ where this happens, although we don't see a reason why it cannot happen in principle. }, the symplectic bridge chooses to connect adjacent Lagrangians via the route that takes $\pi- \theta_k$ radians. This gives one \bea d(\bphi) = 2,\eea an index independent of $n$.

\paragraph{} That clockwise versus counterclockwise give such different answers might be a bit puzzling first, but its origin is clarified if one thinks about the analogous situation in Morse theory. Suppose that $\mathcal{M}(x_a, x_b)$ denotes the reduced moduli space of solutions of the gradient flow equation \bea \frac{d\phi^I}{dx} = g^{IJ} \frac{\del h}{\del \phi^J}, \eea between two critical points $x_a, x_b$ of $h$ with Morse indices $\mu_a, \mu_b$ \footnote{Not to be confused with the BPS index $\mu_{ij}$}. Then supposing $\mu_b > \mu_a$ we have \bea \text{dim} \mathcal{M}(x_a, x_b) = \mu_b - \mu_a - 1.\eea On the other hand, \bea \text{dim} \mathcal{M}(x_b, x_a) =  0.\eea $\mathcal{M}(x_b,x_a)$ is in fact empty, as a consequence of the ascending property of the gradient flow. Thus it should not be very surprising that the moduli space of $\zeta$-instantons is not very well-behaved under orientation reversal of a cyclic fan.

\subsection{Paths in $\mathbb{C}^*$} \label{cstarpath} Let's now consider a gradient polygon of solitons in the punctured complex plane $\mathbb{C}^*$ so that the total path winds around the origin. We choose the holomorphic volume form that trivializes $T \mathbb{C}^*$ to be \bea \Omega = \frac{dX}{X}. \eea One can show that a loop that winds around the origin, by virtue of this trivialization satisfies \bea d(\bphi) = 0.\eea This will be useful for the trigonometric Landau-Ginzburg models.

\subsection{Fermion Degrees for $\mathbb{Z}_N$-symmetric Models} We can use the observations above to determine  (integral part of) the fermion degrees of solitons in at least two interesting $\mathbb{Z}_N$-symmetric family of models. These are \begin{enumerate}
\item $W = \frac{1}{N+1} \phi^{N+1} - t \phi,$ the deformed $A_{N-1}$ model.
\item $W = \phi + \frac{1}{N-1} \phi^{-(N-1)},$ the $\mathbb{Z}_N$ invariant ``trigonometric" LG model.
\end{enumerate} Let's analyze each one.

\subsubsection{Deformed $A_{N-1}$-Model} \label{minimal} The model of a single chiral superfield $\phi$ with superpotential \bea W = \frac{1}{N+1} \phi^{N+1} - t \phi \eea is a well-studied one. The critical points are \bea \phi_k = t^{\frac{1}{N}}e^{\frac{2\pi i k}{N}} \eea for $k=0, 1, \dots, N-1$, with critical values \bea W_k = -\frac{N}{N+1} t^{\frac{N+1}{N}} e^{\frac{2\pi i k}{N}}.\eea It is well-known that there is a unique soliton $\phi_{ij}$ interpolating between each pair $(\phi_i, \phi_{j})$ of distinct critical points. Therefore \bea R_{ij} = \mathbb{Z} \langle \phi_{ij} \rangle.\eea The degree $F_{ij}$ of $\phi_{ij}$ is of the form \bea F_{ij} = n_{ij} + f_{ij} \eea where $n_{ij}$ is the integral part and $f_{ij}$ is the fractional part (for which we have a universal formula). It remains to determine $n_{ij}$.

\paragraph{} For this we use the constraint coming from the Maslov index: Let $\bphi = (\phi_{i_0 i_1}, \dots, \phi_{i_k i_0})$ be a convex gradient polygon. Then \bea n_{i_0 i_1} + n_{i_1 i_2} + \dots + n_{i_k i_0} = d(\bphi) + 1.\eea For the present model, we have that \bea (\phi_{i_0 i_1 }, \phi_{i_1 i_2}, \dots \phi_{i_k i_0} ) \eea is a gradient polygon if and only if $i_0 > i_1 > i_2 \dots > i_n$ up to cyclic reordering. In the complex plane the gradient polygon traces out a clockwise oriented closed path with $k$-segments, and thus the computation in \ref{polygonmoduli} implies \bea d(\phi_{i_0 i_1}, \phi_{i_1 i_2}, \dots, \phi_{i_k i_0}) = k-2.\eea We thus get the constraint \bea n_{i_0 i_1} + n_{i_1 i_2} + \dots + n_{i_k i_0} = k-1,\eea which is satisfied by a particularly simple solution: \bea n_{ij} &=& 1 \text{  for  } i>j, \\ n_{ij} &=& 0 \text{  for  } i<j.\eea By induction on $k$ we see the solution is unique up to shifts \bea n_{ij} \rightarrow n_{ij} + n_i - n_j.\eea Therefore we conclude that \bea R_{ij} &=& \mathbb{Z}[1]\, \text{  for  }i> j, \\ R_{ij} &=& \mathbb{Z}\, \text{  for  } i<j. \eea 

\subsubsection{Trigonometric Models} \label{trigmodels} We can do a similar analysis for the $\mathbb{Z}_N$-symmetric trigonometric Landau-Ginzburg models. These have target space $\mathbb{C}^*$ and superpotential \bea W = \phi + \frac{1}{N-1} \phi^{-(N-1)}.\eea The critical points are again located at the roots of unity \bea \phi_k = e^{2\pi i k/N}\eea for $k = 0, 1, \dots, N-1$ and the critical values are \bea W_k = \frac{N}{N-1} e^{\frac{2\pi i k}{N}} .\eea The soliton spectrum of this model is also known (this model is example $3$ in section $8.1$ of \cite{Cecotti:1992rm}): There is a unique soliton between each nearest neighbor pair $(\phi_i, \phi_{i+1}), (\phi_{i}, \phi_{i-1})$ and none between the other pairs. Therefore the only gradient polygon $\bphi$ with more than $2$ solitons consists of the full $N$-gon \bea \bphi = (\phi_{N-1, N-2}, \phi_{N-2, N-3}, \dots, \phi_{0,N-1} ).\eea The paths these solitons trace out in $\mathbb{C}^*$ consists of round arcs that together wind around the origin once in the clockwise direction. The computation of the Maslov index for paths in $\mathbb{C}^*$ allows us to conclude that $d(\bphi) = 0$ and therefore \bea n_{N-1,N-2} + n_{N-2, N-3} + \dots + n_{0,N-1} = 1. \eea We choose the solution \bea n_{i,i-1} &=& 0, \\ n_{i,i+1} &=& 1.\eea  Thus the non-zero BPS chain complexes with this solution read \bea R_{i,i+1} &=& \mathbb{Z} \langle \phi_{i,i+1} \rangle \cong \mathbb{Z}, \\ R_{i,i-1} &=& \mathbb{Z} \langle \phi_{i,i-1} \rangle \cong \mathbb{Z}[1]. \eea

\section{Examples} \label{examples} Finally let's illustrate categorical wall-crossing in a few examples. 

\subsection{Quartic LG Model} Let's return to the quartic Landau-Ginzburg model that was alluded to in the introduction. The target space is the complex plane $\mathbb{C}$ and the superpotential is \bea W = \frac{1}{4} \phi^4 - \frac{t_1}{2} \phi^2 - t_2 \phi.\eea Consider the point $(t_1, t_2) = (0,1)$ where the critical points are \bea \phi_1 = e^{-\frac{2\pi i }{3}},\,\,\,\, \phi_2 = 1, \,\,\,\, \phi_3 = e^{\frac{2\pi i}{3}}\eea with critical values \bea W_1 = -\frac{3}{4}e^{-\frac{2\pi i}{3}},\,\,\,\, W_2 = -\frac{3}{4}, \,\,\,\, W_3 = -\frac{3}{4} e^{\frac{2\pi i}{3}}.\eea The BPS chain complexes consist of \bea R_{12} &=& \mathbb{Z}\langle \phi_{12} \rangle, \\ R_{13} &=& \mathbb{Z}\langle \phi_{13} \rangle, \\ R_{23} &=& \mathbb{Z}\langle \phi_{23} \rangle, \eea where $\phi_{ij}$ is the unique soliton interpolating between $\phi_i$ and $\phi_j$. As discussed in \ref{minimal}, an assignment of degrees consistent with the Maslov index is that all three spaces are concentrated in degree zero. 

\begin{figure}
\centering
\includegraphics[width=0.5\textwidth]{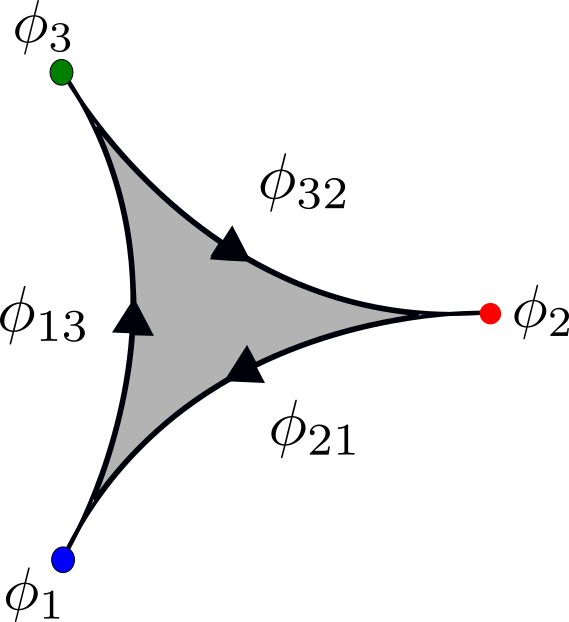}
\caption{Image of the $\zeta$-instanton with fan boundary conditions $\{1,3,2\}$ in the $X$-plane. It sweeps out a region bounded by the soliton paths.} 
\label{quarticinst}
\end{figure}

\paragraph{} Now we must count $\zeta$-instantons. Consider the cyclic fan $\{1,3,2\}$ which has degree $+2$. It is argued in papers on domain wall junctions \cite{Gibbons:1999np} that there is indeed a solution with no reduced moduli with these trivalent fan boundary conditions. Therefore we have \bea N(\phi_{13}, \phi_{32}, \phi_{21}) = 1 .\eea The image swept out by this instanton $\phi(\mathbb{C})$ is depicted in Figure \ref{quarticinst}.

\paragraph{} Crossing the wall of marginal stability we consider $(t_1, t_2) = (1, \epsilon)$ where $\epsilon$ is some small number. Categorical wall-crossing says that the chain complex is \bea R'_{13} = \mathbb{Z} \langle (\phi_{12} \phi_{23})^{[-1]} \rangle \oplus \mathbb{Z} \langle \phi_{13} \rangle.\eea The differential reads \bea d'_{13}\big((\phi_{12} \phi_{23})^{[-1]}\big) &=& \phi_{13}, \\ d'_{13}(\phi_{13}) &=& 0,\eea by virtue of the $\zeta$-instanton of Figure \ref{quarticinst}. Therefore the cohomology is trivial \bea H^{\bullet}(R'_{13}, d'_{13}) = 0.\eea Indeed this is the correct BPS Hilbert space on the other side of the wall. 

\subsection{Trigonometric LG Model} Next we consider the model with target space the complex cylinder $\mathbb{C}^*$ with coordinate $\phi$. The family of superpotentials we consider is \bea W = \phi + \lambda \phi^{-1} + \frac{1}{2} \phi^{-2}.\eea The model at $\lambda = 0$ is known in \cite{Cecotti:1992rm} as the Bullough-Dodd model and that's where we begin our analysis. Here we have the critical points \bea \phi_1 = e^{\frac{2\pi i }{3}}, \,\,\,\,\, \phi_2 = 1, \,\,\,\,\,\, \phi_3 = e^{-\frac{2\pi i}{3}}\eea with critical values $W_i = \frac{3}{2} X_i$. As discussed in \ref{trigmodels}, there is a single soliton between each pair of vacua and so the BPS chain complexes read \bea R_{12} &=& \mathbb{Z}\langle \phi_{12} \rangle, \\ R_{23} &=& \mathbb{Z} \langle \phi_{23} \rangle,\\ R_{13} &=& \mathbb{Z} \langle \phi_{12} \rangle. \eea As discussed in \ref{cstarpath}, consistent with the Maslov index is to choose these spaces to be concentrated in degree zero (the vacua have been relabeled compared to that section). Note that there's a crucial difference with the quartic Landau-Ginzburg model. The vector space associated to the cyclic fan $\{1,2,3\}$ is one-dimensional but now concentrated in degree $+1$. The interior amplitude must therefore be trivial \bea \beta = 0.\eea Therefore, there are no $\zeta$-instantons. 

\begin{figure}
\centering
\includegraphics[width=1.0\textwidth]{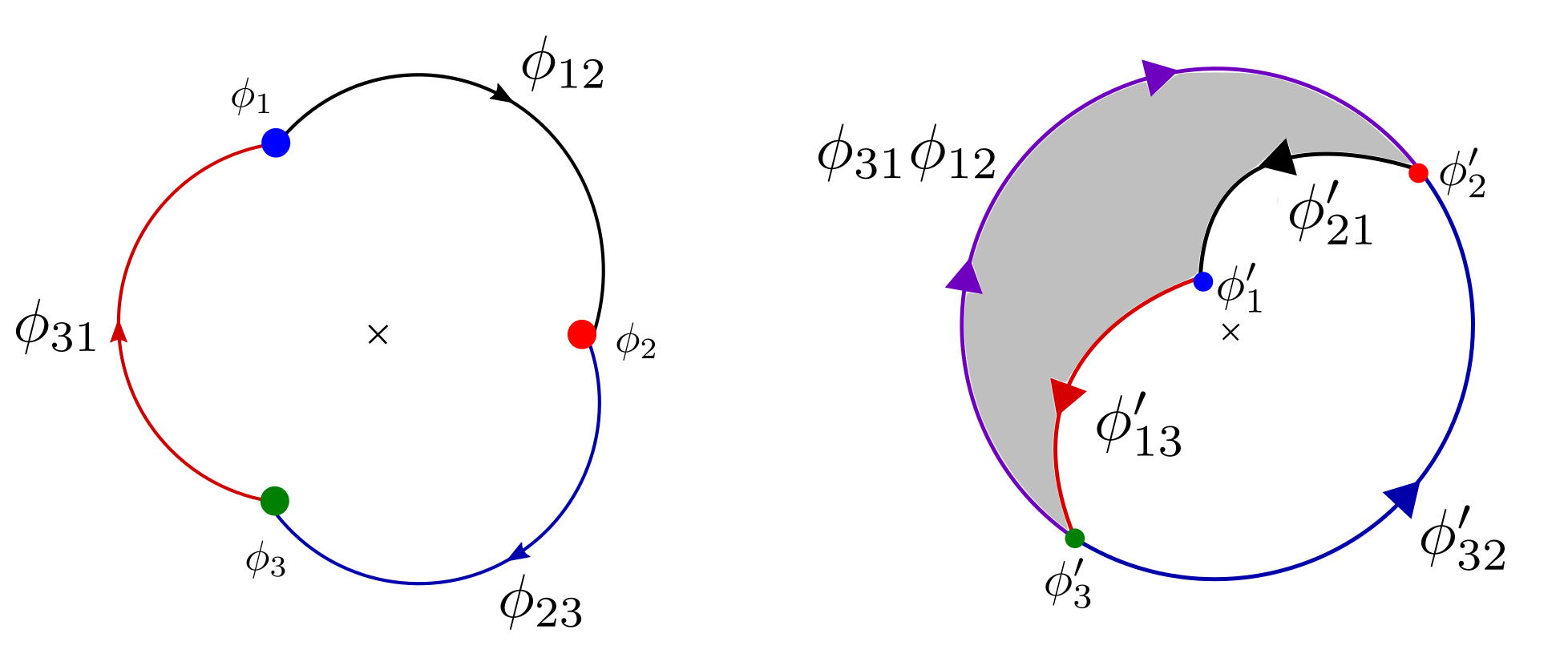}
\caption{Left: the solitons in the $\lambda = 0$ model. There is one between each pair of vacua. On the right we cross the wall of marginal stability and go to $\lambda = 2i$. There are now two solitons in the $32$ sector. We also gain a non-trivial $\zeta$-instanton contributing to the interior amplitude.} 
\label{trigsolitons}
\end{figure}

\paragraph{} The absence of $\zeta$-instantons with trivalent boundary conditions may also be geometrically argued as follows. The cyclic fan of solitons sweep out a path that winds around the origin. Were a $\zeta$-instanton to exist, its image would be a region bounded by this path. However, the latter region contains the singular point $\phi=0$, which means that the $\zeta$-instanton blows up at finite $(x,\tau)$. 

\paragraph{} We now vary $\lambda$ by taking it to be purely imaginary and increasing the magnitude from the $\mathbb{Z}_3$ symmetric point $\lambda = 0$. The wall of marginal stability is crossed at $\lambda \sim 1.5i$. $W_1$ passes through the line between $W_2$ and $W_3$. Therefore $R_{23}$ jumps. We have \bea R'_{23} &=& (R_{21} \otimes R_{13} \big)[-1] \otimes R_{23}, \\ &=& \mathbb{Z} \langle (\phi_{21} \phi_{13})^{[-1]} \rangle \oplus \mathbb{Z} \langle \phi_{23} \rangle, \\ &\cong& \mathbb{Z}^2. \eea Trivial $\beta$ implies that this is also the cohomology. We see that the $23$ sector has gained a bound state of the $21$ and $13$ sectors. 

\paragraph{} These two states post wall-crossing have a simple interpretation. When $\lambda$ is large the theory consists of the $\mathbb{CP}^1$ mirror along with a vacuum $W_1$ running away to infinity. The solitons between $2$ and $3$ are the solitons of this model.

\paragraph{} Categorical wall-crossing also predicts the interior amplitude after wall-crossing. Formula \eqref{intprime} says that the interior amplitude should be \bea \beta'_{132} = (\phi_{31} \phi_{12}) \otimes \phi_{21} \otimes \phi_{13}.\eea Indeed the geometry of solitons allows the region between the new soliton that appears, $\phi_{31} \phi_{12}$, between $3$ and $2$ and the the old solitons $\phi_{21}$ and $\phi_{13}$ to be filled up by a $\zeta$-instanton. See Figure \ref{trigsolitons}.

\subsection{Elliptic LG Model} Let the target space be $T^2_{\tau} \backslash \{0\}$ and \bea W = \wp(\phi, \tau).\eea  We study the wall-crossing properties as we vary $\tau$, the complex structure parameter of the torus\footnote{The moduli space of models is the stack $\mathbb{H}/\text{PSL}(2, \mathbb{Z})$ where $\mathbb{H}$ is the upper-half plane. The moduli space of models with marked vacua is $\mathbb{H}/\Gamma(2)$ where $\Gamma(2)$ is the level $2$ principal congruence subgroup of $\text{SL}(2,\mathbb{Z})$. See \cite{Bergamin:2019dhg} for further examples of this type.}. The critical points are the familiar half-periods \bea \{\frac{1}{2}, \frac{\tau}{2}, \frac{1+ \tau}{2}\} \text{ mod } \big(\mathbb{Z} \oplus \mathbb{Z} \tau \big)\eea with critical values being the elliptic constants \bea \{e_1(\tau), e_2(\tau), e_3(\tau)\}.\eea It is well-known \cite{Cecotti:1992rm, Cecotti:1992vy} that this model has precisely two solitons between each pair of critical points, independent of the value of $\tau$. On the other hand, there are still marginal stability walls. For example when $\tau$ is pure imaginary the $e_i(\tau)$ are all real and hence co-linear, so the imaginary axis and its $\text{PSL}(2, \mathbb{Z})$-images are marginal stability walls in the upper-half plane. The fact that there are two solitons in any chamber, is explained at the level of BPS indices by the equations \bea 2 &=& -2 + 2 \cdot 2, \text{ or }, \\ -2 &=& 2-2\cdot 2. \eea We will now see what happens at the level of chain complexes.

\begin{figure}
\centering
\includegraphics[width=0.8\textwidth]{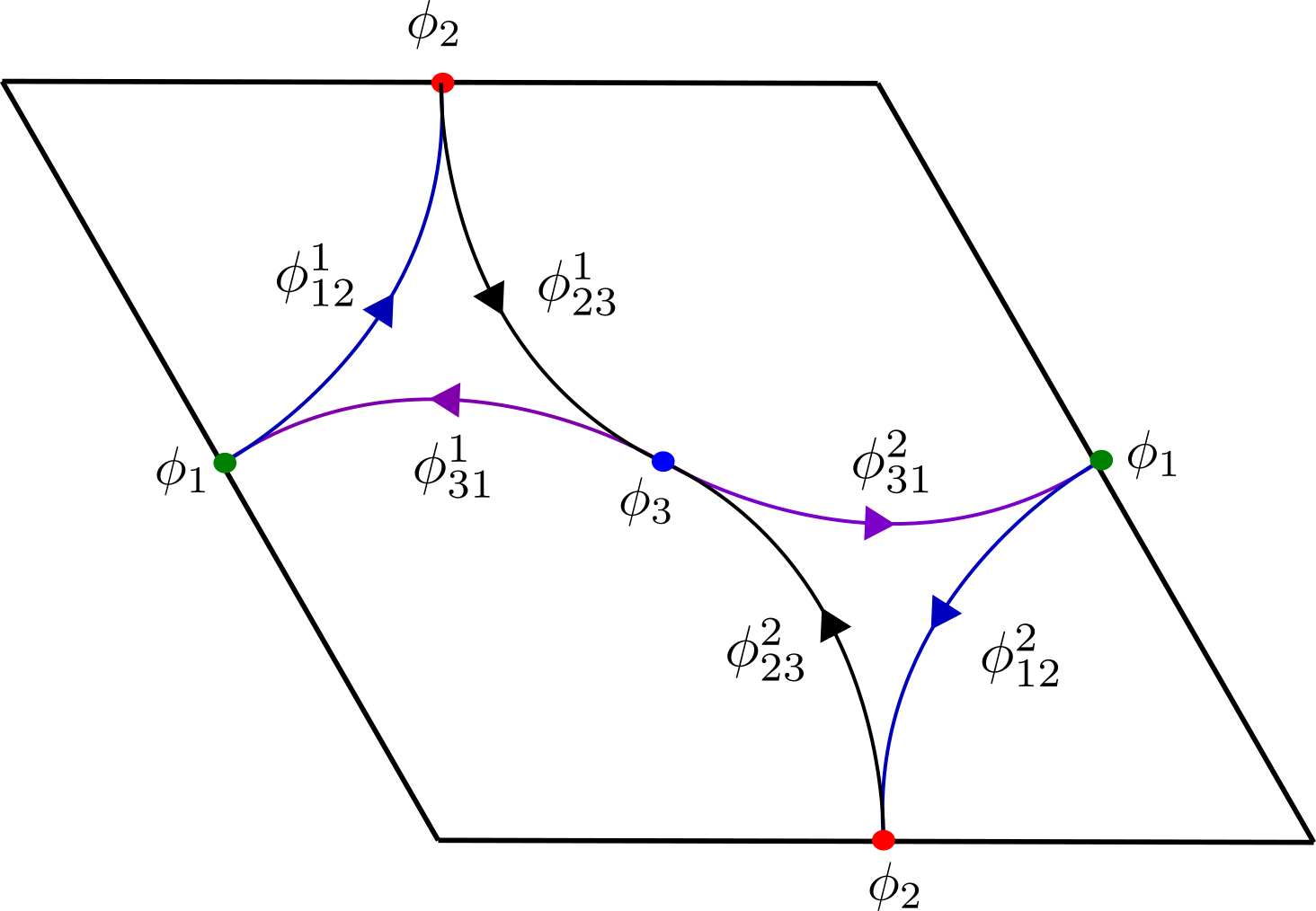}
\caption{BPS solitons in the $W = \wp(\phi,\tau)$ model with $\tau = e^{\frac{2\pi i }{3}}$. There are two solitons between each pair of vacua, the paths they trace out are depicted.} 
\label{ellipticsolitons}
\end{figure}

\paragraph{}First work at the $\mathbb{Z}_3$ symmetric point $\tau_0 = e^{\frac{2\pi i}{3}}$.  We set \bea \phi_1 = \frac{\tau_0}{2},\,\,\,\,\,\,\, \phi_2=\frac{1}{2},\,\,\,\,\,\,\, \phi_3= \frac{1+\tau_0}{2}.\eea The homogeneity property of $\wp(\phi,\tau)$ at the $\mathbb{Z}_3$ symmetric point implies that the critical values are proportional to the cubic roots of unity \bea W_1 =   W_0\, e^{\frac{2\pi i}{3}},\,\,\,\,\,\,\,W_2 =   W_0, \,\,\,\,\,\, W_3 =   W_0 \, e^{\frac{-2\pi i}{3}}, \eea where the proportionality constant is, according to \cite{DLMF}: \bea W_0 = \Bigg(\frac{\Gamma^3(\frac{1}{3})}{2^{\frac{1}{3}}2\pi} \Bigg)^2 .\eea  The chain complexes are \bea R_{12} &=& \mathbb{Z}\langle \phi_{12}^1 \rangle \oplus \mathbb{Z} \langle \phi_{12}^2  \rangle \cong \mathbb{Z}^{2}[1], \\ R_{13} &=& \mathbb{Z}\langle \phi_{13}^1 \rangle \oplus \mathbb{Z} \langle \phi_{13}^2 \rangle \cong \mathbb{Z}^{2}[1], \\ R_{23} &=& \mathbb{Z}\langle \phi_{23}^1 \rangle \oplus \mathbb{Z} \langle \phi_{23}^2 \rangle \cong \mathbb{Z}^2[1]  . \eea A computation similar to the deformed $A_{N-1}$-minimal models can be performed to conclude that these chain complexes are all concentrated in degree $+1$ and so all the individual differentials $d_{ij}$ vanish. The trajectories these solitons trace out on $T^2_{\tau_0}$ are depicted in Figure \ref{ellipticsolitons}.

\paragraph{} The anti-particles are associated to the BPS complexes \bea R_{21} &=& \mathbb{Z}\langle \phi_{21}^1 \rangle \oplus \mathbb{Z} \langle \phi_{21}^2  \rangle \cong \mathbb{Z}^{2}, \\ R_{31} &=& \mathbb{Z}\langle \phi_{31}^1 \rangle \oplus \mathbb{Z} \langle \phi_{31}^2 \rangle \cong \mathbb{Z}^{2}, \\ R_{32} &=& \mathbb{Z}\langle \phi_{32}^1 \rangle \oplus \mathbb{Z} \langle \phi_{32}^2 \rangle \cong \mathbb{Z}^2  . \eea The pairings $K_{12}, K_{13}, K_{23}$ are diagonal in this basis of solitons.

\begin{figure}
\includegraphics[width=1.1\textwidth]{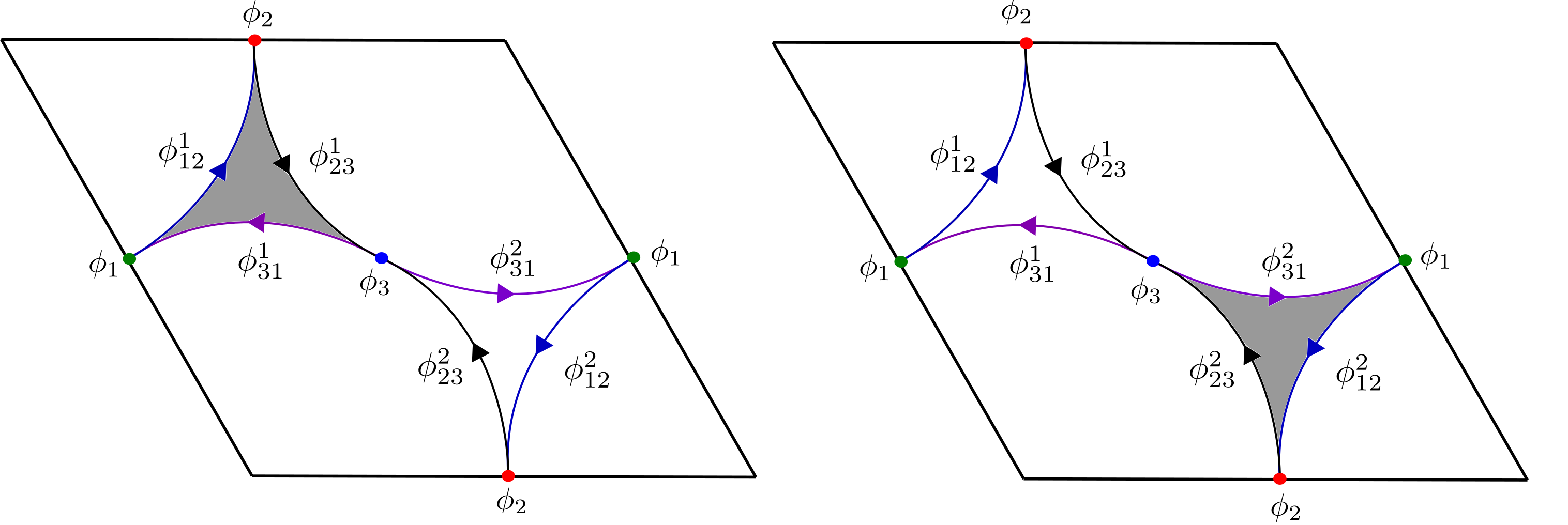}
\caption{$\zeta$-instantons in the $W = \wp(\phi,\tau)$ model with $\tau = e^{\frac{2\pi i }{3}}$.} 
\label{ellipticinstanton1}
\end{figure}

%\begin{figure}
%\centering
%\includegraphics[width=0.8\textwidth]{ellipticinstanton2.png}
%\caption{Another possible $\zeta$-instanton in the same model.} 
%\label{ellipticinstanton2}
%\end{figure}

\paragraph{} Let's now consider $\zeta$-instantons. The vector space corresponding to the cyclic fan \bea \{1,2,3\}, \eea $R_{12} \otimes R_{23} \otimes R_{31}$ is concentrated in degree $+2$ and so this model allows rigid instantons. There are eight possible gradient polygons $\bphi^{a,b,c} = (\phi^{a}_{12}, \phi^{b}_{23}, \phi^{c}_{31})$ for $a,b,c = 1,2$ which could a-priori be occupied. However, the model has additional flavor symmetries whose charges are associated with the winding numbers around the torus
\footnote{More precisely this symmetry doesn't come from translational invariance, since the pole in the superpotential distinguishes a point in the torus (there is a puncture at $X=0$). Nevertheless we can form a conserved current for each harmonic one-form $\alpha$ given by $j = *\phi^*(\alpha)$. }.
These symmetries reduce the number of possibilities as follows. Denoting $q_1, q_2$ the fugacities for the cycles that (half)-wind around the horizontal and $\tau$-direction respectively, the solitons have the following (exponentiated) winding numbers:  States in $R_{12}$ have winding numbers $q_1 q_2$ and $(q_1 q_2)^{-1}$, in $R_{23}$ they have $q_2,q_2^{-1}$, and in $R_{13}$ they have $q_1, q_1^{-1}$. On the other hand $\beta$ must have zero winding charge. This cuts down the allowed gradient polygons that can be occupied to \bea \bphi^1 &=& (\phi_{12}^1, \phi_{23}^1, \phi_{31}^1), \\ \bphi^2 &=& (\phi_{12}^2, \phi_{23}^2, \phi_{31}^2).\eea The simplest non-trivial guess is to posit that these polygons indeed support $\zeta$-instantons with degeneracies \bea N(\bphi^1) &=& 1, \\ N(\bphi^2) &=& 1. \eea Thus we predict the interior amplitude for this model is \bea \beta = \phi^1_{12} \otimes \phi^1_{23} \otimes \phi^1_{31} + \phi^2_{12} \otimes \phi^2_{23} \otimes \phi^2_{31}.\eea

\paragraph{} Assuming this is indeed the case, we now evolve from $\tau_0 = e^{2\pi i /3}$ to, a point of the form $\tau_1= i e^{-i\epsilon}$ with $\epsilon>0$. In doing so we must cross the wall at $\text{Re}(\tau) = 0$. In such a move, one can check (numerically for instance) that the point $W_3$ passes through the line connecting $W_1$ and $W_2$. Therefore the chain complexes $R_{13}, R_{32}$ remain the same as before \bea R'_{13} &=& \mathbb{Z}^2[1], \\ R'_{32} &=& \mathbb{Z}^2,\eea but $R_{12}$ can jump: \bea R'_{12} &=& \big(R_{13} \otimes R_{32}\big)[-1] \oplus R_{12} \\ &=& \big(\mathbb{Z} \langle \phi^1_{13}, \phi^2_{13} \rangle \otimes \mathbb{Z} \langle \phi^1_{32}, \phi^2_{32} \rangle \big)[-1] \oplus \mathbb{Z} \langle \phi^{1}_{12}, \phi^{2}_{12} \rangle. \eea  The first summand is concentrated in degree zero whereas the second factor is in degree one. The $\zeta$-instanton count imply that the differentials act as follows. \bea d'_{12}\big( (\phi^{1}_{13} \phi^{1}_{32})^{[-1]} \big) &=& \phi^1_{12}, \\ d'_{12} \big( (\phi^{1}_{13} \phi^{2}_{32})^{[-1]} \big) &=& 0,\\ d'_{12}\big( (\phi^{2}_{13} \phi^{1}_{32})^{[-1]} \big) &=& 0, \\ d'_{12}\big( (\phi^{2}_{13} \phi^{2}_{32})^{[-1]} \big) &=& \phi_{12}^2. \eea Thus the cohomology is \bea H^{\bullet}(R'_{12}, d'_{12}) = \mathbb{Z} \langle (\phi^{1}_{13} \phi^{2}_{32})^{[-1]}, (\phi^{2}_{13} \phi^{1}_{32})^{[-1]} \rangle ,\eea which is two-dimensional as expected. Categorical wall-crossing has allowed us to see that there has been a non-trivial reorganization of the BPS states in the $12$-sector: in particular their winding numbers jump. This was not visible at the level of ordinary BPS indices\footnote{Of course a \textit{refined} index could have still detected this. In particular upgrading $\mu_{ij}$ to a character valued index $\mu_{ij}(q_1, q_2)$ and applying Cecotti-Vafa does the job in this example. In general such a refinement might not always be available.}.

\section{Conclusions and Future Directions} \label{conclusion}

\paragraph{} There are various future directions that might be worth pursuing. While staying in the two-dimensional world, it is desirable to categorify more general wall-crossing statements. In particular the presence of twisted masses leads to interesting new phenomena. These new phenomena and how they affect the discussion of categorical wall-crossing will be the subject of a separate paper. Similarly, another interesting direction would be to categorify the beautiful formula of Kontsevich and Soibelman, perhaps by constructing the category of infrared line defects in four-dimensional $\mathcal{N}=2$ theories as a first step.

\begin{figure}[H]
\centering
\includegraphics[width=0.6\textwidth]{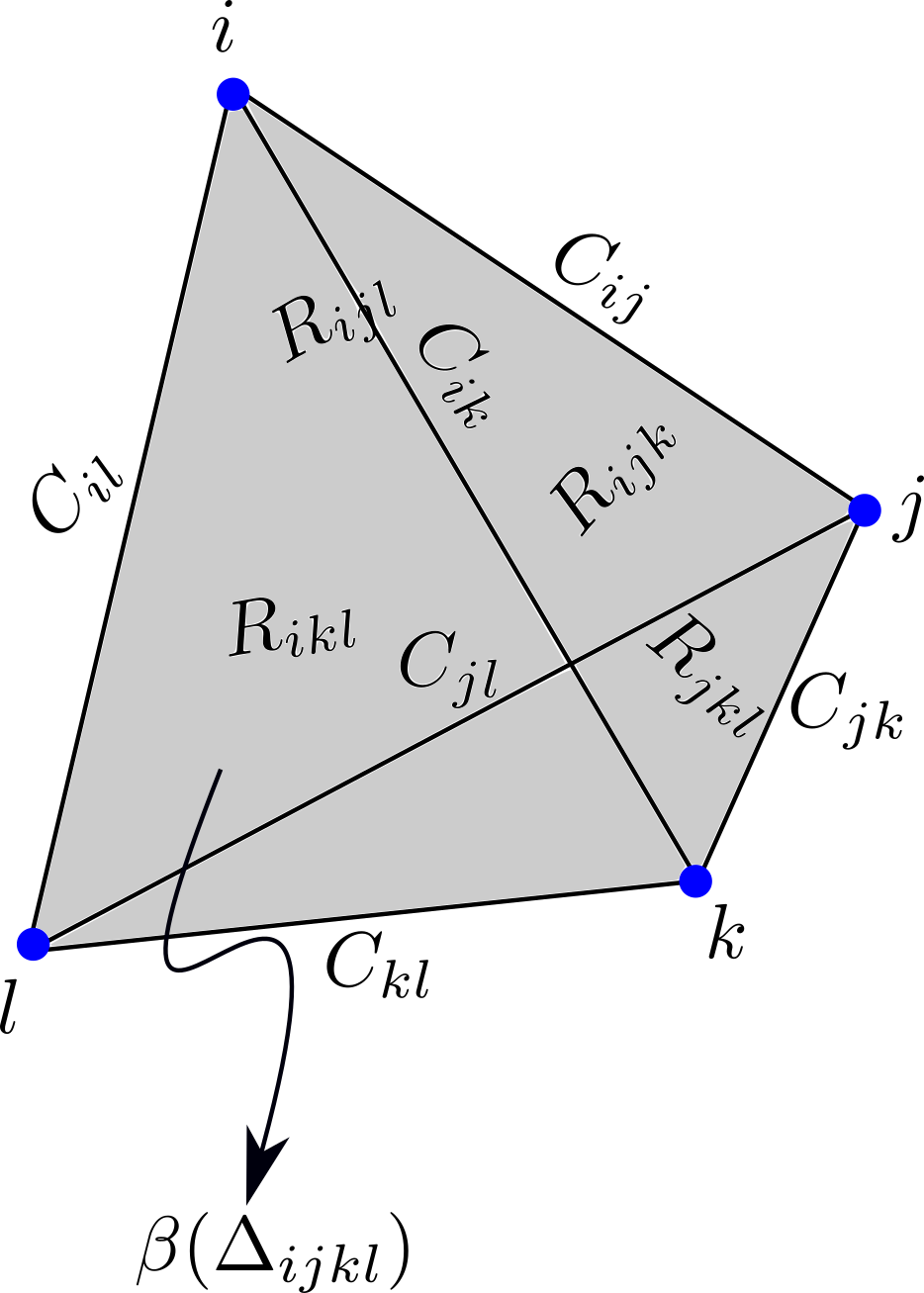}
\caption{A speculative wall-crossing simplex. $C_{ij}$ etc denote categories associated to edges, $R_{ijk}$ etc denote chain complexes associated to faces and $\beta(\Delta_{ijkl})$ denotes a collection of numbers associated to the interior. Wall-crossing would occur when $i$ passes through the base $jkl$ triangle and moves over to the other side.}  
\label{tetra}
\end{figure}

\paragraph{} In a more speculative direction one might wonder about the following. We were studying two-dimensional theories, both in spacetime and from the perspective of the $W$-plane. Edges between vacua in the latter were initially supported by BPS indices, which are integers, and in particular we can use these edges to form a wall-crossing triangle. Categorifying upgraded these integers to chain complexes, but a lesson we learned is that information about these chain complexes by themselves is not sufficient to describe categorical wall-crossing: they must be accompanied by integers associated to the interior of the wall-crossing triangle. In a higher-dimensional generalization of the formalism, let's say three dimensions, we can imagine having a tetrahedron, whose edges carry categories, faces carry chain complexes and whose interior carries the data of integers. See Figure \ref{tetra}. Wall-crossing would occur when the vertices of the tetrahedron lie on a common plane followed by the apex switching sides as viewed from the base. It would be interesting to spell out the wall-crossing structure of this hierarchy of categories, vector spaces and integers that lie on the various faces of the tetrahedron. Even more compelling would be to find a quantum field theoretic realization of such a higher-dimensional ``wall-crossing simplex."

\paragraph{} In the process of categorifying the simplest wall-crossing formula, we have been lead to an interesting blend of mathematics and physics. The physics of domain wall junctions and their moduli spaces allows one to construct canonical objects in homological algebra: the mapping cone and mapping cylinder. These mathematical objects allow us to compactly express the answer to the question we had initially asked. This is the very essence of physical mathematics.

\section*{Acknowledgements} GM would like to thank Tudor Dimofte, Davide Gaiotto, and Edward Witten for previous discussion and collaboration on categorical wall-crossing. AK thanks S. Cecotti for useful discussions and J. Cushing for a helpful lesson on Inkscape. We would like to thank especially Tudor Dimofte for organizing an outstanding virtual seminar dedicated to categorical wall-crossing and the web formalism. We also thank N. Sheridan for useful correspondence.  AK and GM are supported by the DOE under the grant DOE-SC0010008 to Rutgers.

\begin{appendix} 

\section{Some Basic Homological Algebra}\label{hom} The categorical wall-crossing formula is most cleanly stated using some standard homological algebra. We summarize the concepts we need below and refer the reader to \cite{Weibel} for further details. 

\paragraph{Homotopy Equivalence of Complexes} Two complexes $(C,d)$ and $(C',d')$ are said to be homotopy equivalent if there are chain maps $f: C \rightarrow C'$ and $g: C' \rightarrow C$ such that \bea g f &=& 1_C + \{d, s\}, \\ fg &=& 1_{C'} + \{d', s'\}, \eea for some degree $-1$ maps $s: C' \rightarrow C$ and $s': C\rightarrow C'$. $s$ and $s'$ are known as chain homotopies. 

\paragraph{Mapping Cone Recollection}  Given two chain complexes $(A^{\bullet}, d_A)$ and $(B^{\bullet}, d_B)$ along with a chain map \bea f: A^{\bullet} \rightarrow B^{\bullet}, \eea there is a canonical chain complex $\text{Cone}(f)$ defined as follows. The underlying space consists of \bea \text{Cone}(f) = B \oplus A[-1]. \eea Writing an element of $\text{Cone}(f)$ as a column vector \bea \begin{pmatrix}
 b \\ a^{[-1]}
\end{pmatrix} ,\eea the differential on $\text{Cone}(f)$ is \bea d[f] = \begin{pmatrix}
d_B && f \\
0 && -d_A
\end{pmatrix} .\eea $d[f]$ is nilpotent as a consequence of $f$ being a chain map. The projection map \bea \pi: \text{Cone}(f) \rightarrow A[-1], \eea and the inclusion map \bea i: B \rightarrow \text{Cone}(f), \eea are chain maps that fit into the exact sequence \bea \begin{CD} 0 @>>> B @>i>> \text{Cone}(f) @>\pi>> A[-1] @>>> 0. \end{CD} \eea

\paragraph{Mapping Cylinder Recollection} Suppose we are in the setting of the mapping cone of a morphism $f: A \rightarrow B$, i.e consider  $\text{Cone}(f)$. Note that the projection map \bea \pi: \big(\text{Cone}(f)\big)[1] \rightarrow A \eea is a chain map. The $\textbf{mapping cylinder}$ of $f$ is then by definition \bea \text{Cyl}(f) := \text{Cone}(\pi). \eea More explicitly, we can write \bea \text{Cyl}(f) = B \oplus A[-1] \oplus A \eea The differential on $\text{Cyl}(f)$ reads \bea d = \begin{pmatrix}
d_B && f && 0 \\
0 && -d_A && 0 \\
0 && \text{id} && d_A
\end{pmatrix}. \eea The following is standard in homological algebra and topology (for instance see Lemma 1.5.6 in Weibel \cite{Weibel} ).

\paragraph{Proposition} Suppose $(A,d_A),( B, d_B) $ are chain complexes and $f: A \rightarrow B$ is a chain map. Then $B$ and $\text{Cyl}(f)$ are canonically homotopy equivalent. The map $i: B \rightarrow \text{Cyl}(f)$ is given by inclusion and its homotopy inverse $j: \text{Cyl}(f) \rightarrow B$ is given by \bea j \begin{pmatrix}
b \\ a^{[-1]} \\ a'
\end{pmatrix} = b + f(a'). \eea

\paragraph{Remark} The mapping cone and mapping cylinder constructions have their origins in topology. If $f: (X, p_*) \rightarrow (Y,q_*)$ is a continuous map of topological spaces we can define topological spaces \bea \text{Cyl}(f) &=& (X \times I) \cup Y / (x,1) \sim f(x), \\ \text{Cone}(f) &=& \text{Cyl(f)}/ (x,0) \sim  p_*.\eea These spaces are related to the previous constructions as follows. If $C_*(X), C_*(Y)$ denote the singular chain complexes of $X$ and $Y$, then \bea C_*(\text{Cyl}(f) ) &\cong& \text{Cyl} \big(f_*: C_*(X) \rightarrow C_*(Y) \big), \\ C_*(\text{Cone}(f)) &\cong& \text{Cone}\big(f_*: C_*(X) \rightarrow C_*(Y) \big)   , \eea $f_*$ being the induced map on complexes.

\paragraph{Triangularity Lemma:} Let $A,B,C$ be chain complexes and $f: A \rightarrow B$ be a chain map. Suppose that \bea \begin{CD} C \simeq \text{Cone}(f: A \rightarrow B) \end{CD} .\eea Then we can construct chain maps \bea g&:& B \rightarrow C, \\ h&:& C[1] \rightarrow A \eea such that \bea  A[-1] &\simeq& \text{Cone}(g: B \rightarrow C) , \\  B &\simeq& \text{Cone}(h: C[1] \rightarrow A) .\eea

\paragraph{} The maps $g$ and $h$ can be written down explicitly. We set \bea g = u \circ i \eea where $u: \text{Cone}(f) \rightarrow C$ is one of the maps provided by homotopy equivalence and  $i: B \rightarrow \text{Cone}(f)$ is the inclusion map (also a chain map). Similarly \bea h = \pi \circ v \eea where $v: C \rightarrow \text{Cone}(f) $ is the homotopy inverse of $u$ and $\pi: \text{Cone}(f) \rightarrow A[-1]$ is the projection map (also a chain map). These maps may be remembered from the commutative diagram \bea \begin{CD} @. @.  C  @. @. \\
@. @. @VvVV @. @. \\
0 @>>> B @>i>> \text{Cone}(f) @>\pi>>  A[-1] @>>> 0 \\ @. @. @VuVV @. @.  \\ @. @. C @. @. \end{CD}.  \eea  

\section{$A_{\infty}$ Algebras and Morphisms} \label{app} This appendix serves as a reminder of some elementary formulas in $A_{\infty}$ theory. We refer the reader to the (unpublished) book of Kontsevich-Soibelman \cite{Kontsevich:Book}, Keller's notes \cite{Keller}, and appendix A of \cite{Gaiotto:2015aoa} for more details. 

\paragraph{$A_{\infty}$-algebra} Given a graded vector space $A$, denote by $T^{\bullet}(A)$ the tensor algebra of $A$, and $T_+^{\bullet}(A)$ the positive part of the tensor algebra: \bea T^{\bullet}(A) = \oplus_{n \geq 0} A^{\otimes n}, \\ T^{\bullet}_+(A) = \oplus_{n \geq 1} A^{\otimes n}.\eea $A$ is called an $A_{\infty}$-algebra if there is a square-zero, degree one derivation,\footnote{Meaning $\delta$ is both a derivation of the tensor algebra $\delta(X^a X^b) = \delta X^a X^b \pm X^a \delta X^b$, and a differential, a degree one map such that $\delta^2 = 0$. } \bea \delta: T^{\bullet}_+(A^{*}[1]) \rightarrow T_+^{\bullet}(A^{*}[1]) .\eea Extracting Taylor coefficients amounts to a collection maps \bea m_n : A^{\otimes n} \rightarrow A \eea of degree $2-n$ satisfying the $A_{\infty}$-associativity axioms: for each $d \geq 1$ we have \bea \begin{split} \sum_{\substack{k+l = d+1 \\  1 \leq i \leq k} }(-1)^{d_1 + \dots d_{i-1} -i+1} m_k(a_1, \dots, a_{i-1},  m_l(a_{i}, \dots, a_{i+l-1}), a_{i+l}, \dots, a_{d})\\ = 0. \end{split}\eea  $a_i$ is a homogeneous element and $d_i = \text{deg}(a_i)$.

\paragraph{$A_{\infty}$-morphism} Given two $A_{\infty}$-algebras \bea (A_1, \{m_n\}) \text{ and } (A_2, \{\mu_n\}) \eea an $A_{\infty}$-morphism \bea f: A_1 \rightarrow A_2 \eea is an algebra homomorphism (respects tensor algebra structure) \bea f: T^{\bullet}_+(A_2^{*}[1]) \rightarrow T^{\bullet}_+(A_{1}^{*}[1])\eea that is also a chain map: namely $f$ is degree $0$ map satisfying \bea f\delta_2=\delta_1 f.\eea  Again expanding out Taylor coefficients we get a collection of maps  \bea f_n : (A_{1})^{\otimes n} \rightarrow A_2 \eea of degree $1-n$ satisfying the $A_{\infty}$-morphism axioms \bea \label{ainfmorphisms} \begin{split} \sum_{\substack{k+l = d+1 \\ 1 \leq i \leq k} }(-1)^{d_1 + \dots d_{i-1} -i+1} f_k(a_1, \dots, a_{i-1}, m_l(a_i, \dots, a_{i+l-1}), a_{i+l}, \dots, a_d) = \\ \sum_{\substack{n_1 + \dots + n_k = d \\ k \geq 1}} \mu_k (f_{n_1}(a_1, \dots, a_{n_1}), \dots, f_{n_k}(a_{d-n_k + 1}, \dots, a_d) ). \end{split}\eea The $d=1$ relation is \bea \mu_1( f_1(a_1)) = f_1 ( m_1(a_1)) \eea which simply says that $f_1$ is a chain map. \\ \\ The $d=2$ relation is \bea \label{secondaxiom}\begin{split}f_1(m_2(a_1, a_2)) \pm \mu_2(f_1(a_1), f_1(a_2)) = \\ f_2(m_1(a_1), a_2) \pm f_2(a_1, m_1(a_2) ) \pm \mu_1(f_2(a_1, a_2)) \end{split} \eea where the precise signs can be restored via \eqref{ainfmorphisms}. This says that the diagram \bea \begin{CD} A_1^{\otimes 2} @>f_1 \otimes f_1>> A_2^{\otimes 2} \\
@Vm_2VV  @VV \mu_2 V \\
A_1 @>>f_1> A_2 ,\end{CD} \eea commutes up to homotopy, with $f_2$ providing the chain homotopy.

\paragraph{Quasi-isomorphism of $A_{\infty}$-algebras} An $A_{\infty}$-morphism $\{f_n\}_{n \geq 1}$ is said to be a quasi-isomorphism if $f_1: (A_1,m_1) \rightarrow (A_2,\mu_1)$ is a quasi-isomorphism of chain complexes.

\paragraph{Homotopy Equivalence of $A_{\infty}$-algebras} \label{ainf} Two $A_{\infty}$-morphisms $f, g: A_1 \rightarrow A_2$, between $A_{\infty}$-algebras are said to be homotopic $ f \simeq g$, if there is a degree $-1$ map \bea S: T^{\bullet}_+(A^*_2[1]) \rightarrow T^{\bullet}_+(A^*_1[1]) \eea such that \bea f - g = S \delta_2 + \delta_1 S.\eea That is $S$ provides a homotopy between the parent maps $f,g$ of the tensor algebra. $A_1$ and $A_2$ are said to be homotopy equivalent $A_{\infty}$ algebras if there are $A_{\infty}$-morphisms $f: A_1 \rightarrow A_2$ and $g: A_2 \rightarrow A_1$ such that the compositions in either direction are homotopic to the identities on the tensor algebras: \bea g \circ f &\simeq& 1_{T_+ A^*_2}, \\ f \circ g & \simeq & 1_{T_+A^*_1}. \eea In particular, $(A_1,m_1)$ and $(A_2, \mu_1)$ are homotopy equivalent chain complexes.

 % We say that $A_1$ and $A_2$ are homotopy equivalent if there are $A_{\infty}$-morphisms $f: A_1 \rightarrow A_2$ and $g: A_2 \rightarrow A_1$ such that $(f_1, g_1)$ induce homotopy equivalences of $(A_1,m_1)$ and $(A_2, m_2)$ viewed as chain complexes.
 
\paragraph{$L_{\infty}$-algebra} A graded vector space $L$ is called an $L_{\infty}$-algebra if there is a derivation differential $$ \delta: S^{\bullet}_+(L^{*}[2]) \rightarrow S^{\bullet}_+(L^*[2]) .$$ Extracting coefficients gives us that we have a collection of maps \bea \lambda_n : L^{\otimes n} \rightarrow L\eea of degree $3-2n$ which are graded symmetric, and satisfy the $L_{\infty}$-associativity axioms: for each $d \geq 1$ we have \bea \sum_{\substack{k+l = d+1 \\ \sigma \in \text{Sh}_2(k-1,l)}} \epsilon(\sigma, \vec{\ell}\,) \lambda_k(\lambda_l( \ell_{\sigma(1)}, \dots, \ell_{\sigma(l)}), \ell_{\sigma(l+1)}, \dots, \ell_{\sigma(d)} ) = 0. \eea In the above $\sigma \in \text{Sh}_2(k,l)$ denotes a permutation $\sigma \in S_{k+l}$ such that \bea \sigma(1) < \dots < \sigma(k), \,\,\,\, \sigma(k+1) < \dots < \sigma(k+l).\eea

\paragraph{$L_{\infty}$-morphism} Given \bea (L_1, \{\lambda_n\}) ,\,\,\,\,\,\, (L_2, \{\kappa_n\}) \eea two $L_{\infty}$-algebras an $L_{\infty}$-morphism $f: L_1 \rightarrow L_2$ is an algebra homomorphism \bea f: S^{\bullet}_+(L_2^{*}[2]) \rightarrow S^{\bullet}_+(L_1^{*}[2])\eea that is also a chain map with respect to the $L_{\infty}$-structures. Extracting coefficients we get a collection of maps \bea f_n: (L_1)^{\otimes n} \rightarrow L_2 \eea of degree $1-n$ satisfying axioms for an $L_{\infty}$-morphism: for each $d \geq 1$ \bea \begin{split} \sum_{\substack{k+l = d+1 \\ \sigma \in \text{Sh}_2(k-1,l)}} \epsilon(\sigma, \vec{\ell}\,) f_k(\lambda_l( \ell_{\sigma(1)}, \dots, \ell_{\sigma(l)}), \ell_{\sigma(l+1)}, \dots, \ell_{\sigma(d)} ) = \\ \sum_{\substack{n_1 + \dots + n_k  \\ \sigma \in \text{Sh}_k(n_1, \dots, n_k) \\ k \geq 1}} \frac{1}{k!} \epsilon'(\sigma)\kappa_n( f_{n_1}( \ell_{\sigma(1)}, \dots \ell_{\sigma(n_1)}), \dots f_{n_k}(\ell_{\sigma(d-n_k+1)}, \dots, \ell_{\sigma(d)})),\end{split} \eea and $\epsilon(\sigma, \vec{\ell})$ and $\epsilon'(\sigma)$ are suitable signs.

\paragraph{Quasi-isomorphism of $L_{\infty}$-algebras} An $L_{\infty}$-morphism  $\{f_n\}_{n \geq 1}$ from $(L_1, \{\lambda_n\})$ and $(L_2, \{\kappa_n\})$ is said to be a quasi-isomorphism if \bea f_1 : (L_1, \lambda_1) \rightarrow (L_2, \kappa_2) \eea is a quasi-isomorphism of chain complexes.

\paragraph{Maurer-Cartan elements of $L_{\infty}$-algebras} A Maurer-Cartan element $\gamma$ of an $L_{\infty}$ algebra $(L, \{\lambda_n\})$  is a degree two element that solves the $L_{\infty}$ Maurer-Cartan equation \bea \sum_{n \geq 1} \frac{1}{n!} \lambda_n (\gamma, \dots, \gamma) = 0. \eea An infinitesimal gauge transformation of a Maurer-Cartan element $\gamma$ is written as \bea \delta_{\epsilon} \gamma = \sum_{n \geq 1} \frac{1}{n!} \lambda_n(\gamma^{\otimes (n-1)}, \epsilon)\eea where $\epsilon$ is any degree one element of $L$. Indeed one checks that $\gamma + \delta_{\epsilon} \gamma$ solves the Maurer-Cartan equation to first order in $\epsilon$. 

\paragraph{Terminology: Algebras vs Categories} In the bulk text of this paper we have often used the terms ``algebra" and ``category" interchangeably. This is justified because we can go between the two in a precise manner. Following the discussion in chapter 6 of \cite{Kontsevich:Book}, given a linear category with a finite object set $S$, we can define a unital algebra to be \bea A = \oplus_{r,s \in S} \text{Hom}(r,s), \eea with the unit being the direct sum of identity compositions and multiplications given by compositions of morphisms. Conversely, if a unital algebra $A$ is equipped with commuting idempotents $\{\Pi_i\}_{i \in I}$ such that $1_A = \oplus_i \Pi_i$, then we can construct a category $\mathcal{C}$ by setting the object set to be $I$ and letting \bea \text{Hom}(i,j) = \Pi_i A \,\Pi_j. \eea 

\section{$N_{ij} \in \{0,1\}$ for $W \in \mathbb{C}[X]$} \label{polynomialsolitons} We give a proof of the assertion that a Landau-Ginzburg model with target $\mathbb{C}$ and $W(X)$ a Morse polynomial has at most a single soliton between any pair of critical points. For this we consider the relative homology group \bea V = H_1(\mathbb{C}, \text{Re}(\zeta^{-1}W) \rightarrow \infty; \mathbb{Z}) \eea where $\zeta$ is a phase not equal to any of the critical phases. $V$ is easily constructed. Supposing that the degree of $W$ is $n$, we divide the complex plane $\mathbb{C}$ into $2n$ wedges of equal angle $\frac{2\pi}{n}$ and shade alternating regions $R_1, \dots, R_n$. A basis for $V$ is provided by cycles $\gamma_{a,a+1}$ that connect $R_a$ and $R_{a+1}$ for $a=1, \dots, n-1$. On the other hand, Picard-Lefschetz theory says that the homology class of the Lefschetz thimbles $L_i(\zeta)$ for $i = 1, \dots, n-1$ critical points of $W$ must also form a $\mathbb{Z}$-module basis for $V$. In particular this implies that if $L_i(\zeta)$ connects $R_a, R_b$ and $L_j(\zeta)$ connects $R_c, R_d$ then $\{a,b\} \neq \{c,d\}$ since otherwise they will be multiples of each other by $\pm 1$ in homology, and thus linearly dependent elements of $V$. Considering a point $p$ on the $\zeta$-ray emanating from $W_i$ far out enough, $W^{-1}(p) \cap L_i(\zeta)$ is a pair of points lying in distinct regions $R_a, R_b$ which are connected by $L_i$. Therefore $|L_i(\zeta_{ji} e^{-i\epsilon}) \cap L_j (\zeta_{ji} e^{i \epsilon})| $ is at most one, concluding the proof.

\end{appendix}


\begin{thebibliography}{99}

\bibitem[ADJM]{Andriyash:2010qv}
E.~Andriyash, F.~Denef, D.~L.~Jafferis and G.~W.~Moore,
 ``Wall-crossing from supersymmetric galaxies,''
JHEP \textbf{01}, 115 (2012)
doi:10.1007/JHEP01(2012)115
[arXiv:1008.0030 [hep-th]].

\bibitem[Aur]{Auroux2013}
D.~Auroux,
``A beginner's introduction to Fukaya categories,"
[arXiv:1301.7056 [math.SG]].

\bibitem[BC]{Bergamin:2019dhg}
R.~Bergamin and S.~Cecotti,
``FQHE and $tt^{*}$ geometry,''
JHEP \textbf{12}, 172 (2019)
doi:10.1007/JHEP12(2019)172
[arXiv:1910.05022 [hep-th]].

\bibitem[CGGLPFZ]{Carey}
A.~Carey, F.~Gesztesy, H.~Grosse, G.~Levitina, D.~Potapov, F.~Sukochev, D.~Zanin,
``Trace Formulas for a Class of non-Fredholm Operators: A Review,"
Reviews in Mathematical Physics 28, No. 10 (2016) 1630002
[arXiv:1610.04954 [math.AP]].

\bibitem[CHT]{Carroll:1999wr}
S.~M.~Carroll, S.~Hellerman and M.~Trodden,
``Domain wall junctions are 1/4 - BPS states,''
Phys. Rev. D \textbf{61}, 065001 (2000)
doi:10.1103/PhysRevD.61.065001
[arXiv:hep-th/9905217 [hep-th]].

\bibitem[Cec]{CecottiLectures}
S.~Cecotti,
``Trieste Lectures on Wall–Crossing Invariants,"
\url{https://people.sissa.it/~cecotti/ictptext.pdf}

\bibitem[CFIV]{Cecotti:1992qh}
S.~Cecotti, P.~Fendley, K.~A.~Intriligator and C.~Vafa,
``A New supersymmetric index,''
Nucl. Phys. B \textbf{386}, 405-452 (1992)
doi:10.1016/0550-3213(92)90572-S
[arXiv:hep-th/9204102 [hep-th]].

\bibitem[CV1]{Cecotti:1992rm}
S.~Cecotti and C.~Vafa,
``On classification of N=2 supersymmetric theories,''
Commun. Math. Phys. \textbf{158}, 569-644 (1993)
doi:10.1007/BF02096804
[arXiv:hep-th/9211097 [hep-th]].

\bibitem[CV2]{Cecotti:1992vy}
S.~Cecotti and C.~Vafa,
``Ising model and N=2 supersymmetric theories,''
Commun. Math. Phys. \textbf{157}, 139-178 (1993)
doi:10.1007/BF02098023
[arXiv:hep-th/9209085 [hep-th]].

\bibitem[DM]{Denef:2007vg}
F.~Denef and G.~W.~Moore,
 ``Split states, entropy enigmas, holes and halos,''
JHEP \textbf{11}, 129 (2011)
doi:10.1007/JHEP11(2011)129
[arXiv:hep-th/0702146 [hep-th]].

\bibitem[DLMF]{DLMF}
``Digital Library of Mathematical Functions,"
\url{https://dlmf.nist.gov/23.5}

\bibitem[Dor]{Dorey:1998yh}
N.~Dorey,
``The BPS spectra of two-dimensional supersymmetric gauge theories with twisted mass terms,''
JHEP \textbf{11}, 005 (1998)
doi:10.1088/1126-6708/1998/11/005
[arXiv:hep-th/9806056 [hep-th]].

\bibitem[GMN1]{Gaiotto:2008cd}
D.~Gaiotto, G.~W.~Moore and A.~Neitzke,
``Four-dimensional wall-crossing via three-dimensional field theory,''
Commun. Math. Phys. \textbf{299}, 163-224 (2010)
doi:10.1007/s00220-010-1071-2
[arXiv:0807.4723 [hep-th]].

\bibitem[GMN2]{Gaiotto:2010be}
D.~Gaiotto, G.~W.~Moore and A.~Neitzke,
 ``Framed BPS States,''
Adv. Theor. Math. Phys. \textbf{17}, no.2, 241-397 (2013)
doi:10.4310/ATMP.2013.v17.n2.a1
[arXiv:1006.0146 [hep-th]].

\bibitem[GMN3]{Gaiotto:2011tf}
D.~Gaiotto, G.~W.~Moore and A.~Neitzke,
``Wall-Crossing in Coupled 2d-4d Systems,''
JHEP \textbf{12}, 082 (2012)
doi:10.1007/JHEP12(2012)082
[arXiv:1103.2598 [hep-th]].

\bibitem[GMW]{Gaiotto:2015aoa}
D.~Gaiotto, G.~W.~Moore and E.~Witten,
``Algebra of the Infrared: String Field Theoretic Structures in Massive ${\cal N}=(2,2)$ Field Theory In Two Dimensions,''
[arXiv:1506.04087 [hep-th]].

\bibitem[GMWSh]{Gaiotto:2015zna}
D.~Gaiotto, G.~W.~Moore and E.~Witten,
``An Introduction To The Web-Based Formalism,''
[arXiv:1506.04086 [hep-th]].

\bibitem[GT]{Gibbons:1999np}
G.~W.~Gibbons and P.~K.~Townsend,
``A Bogomolny equation for intersecting domain walls,''
Phys. Rev. Lett. \textbf{83}, 1727-1730 (1999)
doi:10.1103/PhysRevLett.83.1727
[arXiv:hep-th/9905196 [hep-th]].

\bibitem[HIV]{Hori:2000ck}
K.~Hori, A.~Iqbal and C.~Vafa,
``D-branes and mirror symmetry,''
[arXiv:hep-th/0005247 [hep-th]].

\bibitem[MS1]{Hori:2003ic}
K.~Hori, S.~Katz, A.~Klemm, R.~Pandharipande, R.~Thomas, C.~Vafa, R.~Vakil and E.~Zaslow,
``Mirror symmetry.''

\bibitem[INOS]{Ito:2000mt}
K.~Ito, M.~Naganuma, H.~Oda and N.~Sakai,
``An Exact solution of BPS junctions and its properties,''
Nucl. Phys. B Proc. Suppl. \textbf{101}, 304-313 (2001)
doi:10.1016/S0920-5632(01)01515-8
[arXiv:hep-th/0012182 [hep-th]]

\bibitem[KS]{Kajiura:2004xu}
H.~Kajiura and J.~Stasheff,
``Homotopy algebras inspired by classical open-closed string field theory,''
Commun. Math. Phys. \textbf{263}, 553-581 (2006)
doi:10.1007/s00220-006-1539-2
[arXiv:math/0410291 [math.QA]].

\bibitem[KKS]{Kapranov:2014uwa}
M.~Kapranov, M.~Kontsevich and Y.~Soibelman,
``Algebra of the infrared and secondary polytopes,''
Adv. Math. \textbf{300}, 616-671 (2016)
doi:10.1016/j.aim.2016.03.028
[arXiv:1408.2673 [math.SG]].

\bibitem[Kel]{Keller}
B.~Keller,
``Introduction to A-infinity algebras and modules,"
[arXiv:math/9910179 [math.RA]].

\bibitem[KoSo1]{Kontsevich:2008fj}
M.~Kontsevich and Y.~Soibelman,
``Stability structures, motivic Donaldson-Thomas invariants and cluster transformations,''
[arXiv:0811.2435 [math.AG]].

\bibitem[KoSo2]{Kontsevich:2009xt}
M.~Kontsevich and Y.~Soibelman,
 ``Motivic Donaldson-Thomas invariants: Summary of results,''
[arXiv:0910.4315 [math.AG]].

\bibitem[KoSo3]{KontsevichReview}
M.~Kontsevich and Y.~Soibelman, 
``Lectures on motivic Donaldson-Thomas invariants and wall-crossing formulas."

\bibitem[KoSo4]{Kontsevich:Book}
M.~Kontsevich and Y.~Soibelman,
``Deformation Theory. I."

\bibitem[LY]{Lee:1998nv}
K.~M.~Lee and P.~Yi,
``Dyons in N=4 supersymmetric theories and three pronged strings,''
Phys. Rev. D \textbf{58}, 066005 (1998)
doi:10.1103/PhysRevD.58.066005
[arXiv:hep-th/9804174 [hep-th]].

\bibitem[M1]{MooreLectures}
G.~Moore, 
``2010 PiTP Lectures," 
\url{http://www.physics.rutgers.edu/~gmoore/PiTP-LectureNotes.pdf},  
\url{https://video.ias.edu/pitp-2010}

\bibitem[M2]{GMSlides}
G.~Moore,
``Framed BPS States In Four And Two Dimensions,"
\url{http://www.physics.rutgers.edu/~gmoore/StringMath2016L.pdf},
\url{http://www.physics.rutgers.edu/~gmoore/StringMath2016-Aspen.pdf},
\url{http://www.physics.rutgers.edu/~gmoore/Hamburg2019-Short.pdf}.

\bibitem[N]{NeitzkeLectures}
A.~Neitzke,
``PCMI Lectures on BPS States and Spectral Networks."

\bibitem[Pio]{Pioline:2011gf}
B.~Pioline,
 ``Four ways across the wall,''
J. Phys. Conf. Ser. \textbf{346}, 012017 (2012)
doi:10.1088/1742-6596/346/1/012017
[arXiv:1103.0261 [hep-th]].

\bibitem[Pro]{Proute84} A. Prout\'e, 
``Alg\`ebres diff\'erentielles fortement homotopiquement associatives", 
Th\`ese d'Etat, Universit\'e Paris VII, 1984.

\bibitem[Seib]{Seiberg:1994bp}
N.~Seiberg,
``The Power of holomorphy: Exact results in 4-D SUSY field theories,''
[arXiv:hep-th/9408013 [hep-th]].

\bibitem[SW]{Seiberg:1994rs}
N.~Seiberg and E.~Witten,
 ``Electric - magnetic duality, monopole condensation, and confinement in N=2 supersymmetric Yang-Mills theory,''
Nucl. Phys. B \textbf{426}, 19-52 (1994)
doi:10.1016/0550-3213(94)90124-4
[arXiv:hep-th/9407087 [hep-th]].

\bibitem[Seid]{Seidel}
P. Seidel
``Fukaya categories and Picard-Lefschetz theory," Zurich Lectures in Advanced Mathematics. European Mathematical Society (EMS), Zürich

\bibitem[Weib]{Weibel}
C.~Weibel,
``An Introduction to Homological Algebra,"
Cambridge Studies in Advanced Mathematics.

\bibitem[Wit1]{Witten:1981nf}
E.~Witten,
``Dynamical Breaking of Supersymmetry,''
Nucl. Phys. B \textbf{188}, 513 (1981)
doi:10.1016/0550-3213(81)90006-7

\bibitem[Wit2]{Witten:1982im}
E.~Witten,
``Supersymmetry and Morse theory,''
J. Diff. Geom. \textbf{17}, no.4, 661-692 (1982)

\end{thebibliography}
\end{document}